\documentclass[a4paper,12pt,twoside,openright]{UGthesis}
\usepackage{amssymb}
\usepackage[swedish, english]{babel}
\usepackage[latin1]{inputenc}
\usepackage{enumerate}
\usepackage{epic}
\usepackage{fancyheadings}
\usepackage[dvips]{graphicx}
\usepackage{ifthen}
\usepackage{subfigure}
\usepackage{textcomp}
\usepackage{textfit}

\usepackage[errorshow]{tracefnt}

\newcommand{\we}{\wedge}
\newcommand{\dxmu}{dx^\mu}
\newcommand{\dxnu}{dx^\nu}
\newcommand{\la}{\lambda}
\newcommand{\La}{\Lambda}
\newcommand{\si}{\sigma}
\newcommand{\ta}{\tau}

\newcommand{\ka}{\kappa}
\newcommand{\om}{\omega}

\newcommand{\da}{\dagger}

\newcommand{\com}[1]{\lbrack #1 \rbrack}
\newcommand{\expo}{\textrm{exp}}

\newcommand{\non}{\nonumber}

\newcommand{\Lagr}{\mathcal{L}}
\newcommand{\D}{\mathfrak{D}}
\newcommand{\V}{\mathcal{V}}
\newcommand{\bein}{\textsf{e}}
\newcommand{\g}{\textbf{g}}

\newcommand{\pa}{\partial}

\newcommand{\pand}{\partial_\nu}

\newcommand{\beq}{\begin{equation}}
\newcommand{\eeq}{\end{equation}}
\newcommand{\beqa}{\begin{eqnarray}}
\newcommand{\eeqa}{\end{eqnarray}}

\newcommand\ffam{\sffamily}
\newcommand\fser{\bfseries}
\newcommand\fsh{\upshape}

\newcommand\blankpage{\thispagestyle{empty}\mbox{}\newpage}

\newcommand\rctr{\renewcommand{\theenumi}{\roman{enumi}}}

\newcommand\benu{\begin{enumerate}}
\newcommand\eenu{\end{enumerate}}
\newcommand\bit{\begin{itemize}}
\newcommand\eit{\end{itemize}}

\newcommand{\chrf}[3]{\left\{ \hspace{-2mm}
\begin{array}{ccc}
#1 \\ [-1.2mm]#2 ~ #3
\end{array}
\hspace{-2mm} \right\}}

\newcommand{\eqnlab}[1]{\label{eqn:#1}}
\newcommand{\figlab}[1]{\label{fig:#1}}
\newcommand{\tablab}[1]{\label{tab:#1}}
\newcommand{\eqnref}[1]{(\ref{eqn:#1})}
\newcommand{\figref}[1]{\ref{fig:#1}}
\newcommand{\tabref}[1]{\ref{tab:#1}}
\newcommand{\Eqnref}[1]{Eq.~(\ref{eqn:#1})}
\newcommand{\Figref}[1]{Fig.~\ref{fig:#1}}
\newcommand{\Tabref}[1]{Table~\ref{tab:#1}}
\newcommand{\Eqsref}[1]{Eqs.~(\ref{eqn:#1})}
\newcommand{\Figsref}[1]{Figs.~\ref{fig:#1}}
\newcommand{\Tabsref}[1]{Tables~\ref{tab:#1}}

\newcommand{\chlab}[1]{\label{ch:#1}}

\newcommand{\chref}[1]{\ref{ch:#1}}

\newcommand{\Chref}[1]{Chapter~\ref{ch:#1}}

\RequirePackage{hyperref}

\begin{document}

%
%



\pagenumbering{roman}
\setcounter{page}{1}

%
%

\thispagestyle{empty}


\begin{center}
  {\fsh\ffam\fser Thesis for the degree of Master of Science in
    Engineering Physics}
\end{center}

\vspace*{0.5cm}

\begin{center}
{\upshape\sffamily\bfseries\LARGE Algebraic Structures in M-theory}
\end{center}

\vspace*{2mm}
\begin{center}
        \rule{110mm}{2pt}
\end{center}

\vspace*{4mm}
\begin{center}
  {\fsh\ffam\fser\Large Ling Bao}\\
\end{center}
\vfill

\begin{center}
\includegraphics[width=9cm]{pics/ChalmGUmarke.epsi} 
\end{center}

\vfill
\begin{center}
        {\ffam\fsh Department of Theoretical Physics\\*[1mm]
        Chalmers University of Technology and Göteborg
        University\\*[2mm]
        Göteborg, Sweden 2004}
\end{center}

%
%

\mbox{}\thispagestyle{empty}\newpage
\vspace*{165mm}


\blankpage

%
%

\thispagestyle{empty}
\begin{center}
        {\ffam 
        {\fser\Large Algebraic Structures in M-theory}\\
\vspace{3mm}
        {\normalsize Ling Bao  \\
        Department of Theoretical Physics \\
        Chalmers University of Technology and Göteborg University \\
        SE-412 96 Göteborg, Sweden} \\[7mm]}
\end{center}

\centerline{\ffam\fser Abstract}
\medskip
\normalsize
\noindent In this thesis we reformulate the bosonic sector of eleven dimensional
supergravity as a simultaneous nonlinear realisation based on the
conformal group and an enlarged affine group called G$_{11}$. The
vielbein and the gauge fields of the theory appear as space-time
dependent parameters of the coset representatives. Inside the
corresponding algebra g$_{11}$ we find the Borel subalgebra of e$_7$,
whereas performing the same procedure for the Borel subalgebra of
e$_8$ we have to add some extra generators. The presence of these new
generators lead to a new formulation of gravity, which includes both a
vielbein and its dual. We make the conjecture that the final symmetry
of M-theory should be described by a coset space, with the global and
the local symmetry given by the Lorentzian Kac-Moody algebra e$_{11}$
and its subalgebra invariant under the Cartan involution,
respectively. The pure gravity itself in $D$ dimensions is argued to
have a coset symmetry based on the very extended algebra
A$^{+++}_{D-3}$. The tensor generators of a very extended algebra are
divided into representations of its maximal finite dimensional
subalgebra. The space-time translations are thought to be introduced
as weights in a basic representation of the Kac-Moody algebra. Some
features of the root system of a general Lorentzian Kac-Moody algebra
are discussed, in particular those related to even self-dual lattices.

\vfill

\newpage

%
%
\thispagestyle{plain}
\vspace*{4cm}

\centerline{\ffam\fser\Large Acknowledgments}
\medskip
\smallskip

\normalsize
\noindent\begin{center}
In the ancient Crete, Ariadne's clew guided Theseus through Minotaur's
maze and helped him escape the dreadful death. 
\end{center}

\vspace{0.5 cm}

\noindent I would like to express my deepest gratitudes to my
supervisor and source of inspiration throughout this thesis, professor
Bengt E.W. Nilsson, a fantastic world has been opened before my
eyes. Many thanks to all the members of the Elementary Particle Theory
and Mathematical Physics group for surviving my endless
questions. Special thanks to my roommate Ulrik Svensson for those
invaluable discussions. Finally I would like to thank my dear family,
you have always given me the courage through the sunshine and the
rain.

%
%
\cleardoublepage
\tableofcontents
\pagestyle{empty}

\cleardoublepage
\pagestyle{fancy}
\renewcommand{\chaptermark}[1]{\markboth{Chapter \thechapter\ \ \ #1}{#1}}
\renewcommand{\sectionmark}[1]{\markright{\thesection\ \ #1}}
\lhead[\fancyplain{}{\sffamily\thepage}]%
  {\fancyplain{}{\sffamily\rightmark}}
\rhead[\fancyplain{}{\sffamily\leftmark}]%
  {\fancyplain{}{\sffamily\thepage}}
\cfoot{}
\setlength\headheight{14pt}

\rctr

\setcounter{page}{1}
\pagenumbering{arabic}

%
%

\chapter{Introduction}
\chlab{introduction}

A central theme in fundamental physics during the twentieth century
was the search for ways to bring together gravity and quantum
mechanics, that is, to find a gravitational theory at the Planck
length. Until 1995 the best candidates for the \emph{theory of
  everything} were the ten-dimensional superstrings. Although the
superstrings succeeded to shine new light on the issue of quantum
gravity, one of the major problems was that the number of different
self-consistent superstring theories was not one, but five. Also the
required ten-dimensional space-time is somewhat startling, since
supersymmetry permits a supersymmetric field theory in eleven
dimensions. The discoveries of the supermembrane and the
superfivebrane then made it clear that the superstring theories cannot
be the whole story. In the last few years just before the new
millennium, Edward Witten opened a new window revealing a mysterious
theory called M-theory. It is a non-perturbative theory in eleven
dimensions, where the various superstrings correspond to different
extremal regions of the coupling constants arising after
compactification and related by dualities to one another. At the low
energy limit, M-theory is described by eleven dimensional
supergravity.

This thesis work deals with the possible symmetries of M-theory,
having eleven dimensional supergravity as the starting point. The
symmetry groups appear manifestly by using a technique called
nonlinear realisation. Typically, one has a Lie group as the global
symmetry group, while the local symmetry is its maximal compact
subgroup. Certain features of eleven dimensional supergravity and its
dimensionally reduced theories seem to point towards a rank eleven
Lorentzian Kac-Moody algebra named e$_{11}$.

The outline of this thesis is the following:
\textbf{\Chref{nonlinear}} deals with nonlinear realisations. First
the general theory is reviewed, detailed calculations are then carried
out for the scalar fields of type IIB supergravity and the bosonic
part of eleven dimensional supergravity. The correct field equations
of the latter are obtained by simultaneous nonlinear realisation based
on a group called G$_{11}$ and the conformal group, both having the
Lorentz group as the local subgroup.

In \textbf{\Chref{algebra}} some basic properties of Lorentzian
Kac-Moody algebras are presented, with the focus being mainly on the
root system. Conditions are found for a Lorentzian Kac-Moody algebra
to include a real principal so(2,1) subalgebra. Also, a section is
devoted to the so called very extension of a finite dimensional Lie
algebra g, which is obtained from g by adding three more simple roots
in a specific way.  The very extended Lie algebras related to
self-dual lattices are of special interest, e.g., e$_{11}$, m$_{19}$
and k$_{27}$.

The conjecture about e$_{11}$ being a symmetry of M-theory is made in
\textbf{\Chref{symmetry}}. It is shown that the eleven dimensional
supergravity contains the finite dimensional Lie groups of exceptional
type as symmetries. The concept of level is introduced, and the root
space of e$_{11}$ is divided into tensor representations of its
subalgebra A$_{10}$. The translation generators are found to be
included in a basic representation of e$_{11}$. An alternative
formulation of gravity, involving a field dual to the vielbein, is
also presented in this chapter.

Conclusions are made in \textbf{\Chref{conclusion}}, together with
some brief words about relevant issues not covered in this thesis.

There are some loosely attached subjects included as appendices,
providing short introductions to differential forms, the conformal
group, general coordinate transformations, real principal subalgebras
and the Lorentzian self-dual lattices $\Pi_{D-1;1}$. Last but not
least, there is a summary about field transformations under the
conformal group.

\chapter{Nonlinear realisations}
\chlab{nonlinear}

When the eleven dimensional supergravity is dimensionally reduced on a
$n$-torus to (11-$n$) dimensions, for $n=1,\ldots,8$, the scalars in
the resulting theory can be associated with the coset space formed by
a non-compact symmetry group E$_n$ divided by its maximal compact
subgroup F$_n$, see \Tabref{coset}.
\begin{table}[h!]
  \begin{center}
    \begin{tabular}{l|ll}
      Dimension & E$_n$         & F$_n$ \\ \hline 
      11        & 1             & 1 \\ 
      10, IIB   & SL(2)         & SO(2) \\ 
      10, IIA   & SO(1,1)/Z$_2$ & 1 \\ 
      9         & GL(2)         & SO(2) \\ 
      8         & E$_3$ $\sim$ SL(3)$\times$SL(2)   & U(2) \\
      7         & E$_4$ $\sim$ SL(5)   & USp(4) \\
      6         & E$_5$ $\sim$ SO(5,5) & USp(4)$\times$USp(4) \\
      5         & E$_6$         & USp(8) \\
      4         & E$_7$         & SU(8) \\
      3         & E$_8$         & SO(16)
    \end{tabular}
    \caption{\textit{Coset spaces of the supergravities.}}
    \tablab{coset}
  \end{center}
\end{table}
The local subgroup F$_n$ can be used to choose the coset
representatives of E$_n$/F$_n$ to belong to the Borel subgroup of
E$_n$, i.e., exponentials of the Cartan and the positive root
generators in E$_n$. The number of scalar fields is then just the
number of generators in the Borel subalgebra. Dimensional reduction of
eleven dimensional supergravity to two dimensions and one dimension is
thought to be theories invariant under the affine extension of E$_8$,
denoted E$_9$, and the hyperbolic group E$_{10}$, respectively. It has
been proposed that the eleven dimensional supergravity itself has some
kind of exceptional geometry. The local subgroups F$_n$ for
$n=1,\ldots,8$ are the subgroups invariant under the Cartan
involution, and for the cases of $n=9,10$ one needs to introduce an
involution in order to define possible local subgroups. The gauge
fields and the scalars transform non-trivially under the E$_n$ group,
whereas gravity is inert under this global internal symmetry group.
Our goal is to treat the gauge fields and gravity on the same footing
as nonlinear realisations.

We will in this chapter first look at the general theory of nonlinear
realisations with a coset space symmetry. More detailed calculations
are then carried out for the scalar part of type IIB supergravity and
the bosonic sector of eleven dimensional supergravity. In the latter
case we will have to make simultaneous realisation of the conformal
group and the G$_{11}$ group, which is the affine group IGL(11)
extended with some extra generators.

The technique of nonlinear realisations can also be applied to the
bosonic sector of IIA and IIB supergravities, the closed bosonic
string as well as the branes in M-theory, by using different coset
symmetries simultaneously realised with the conformal group. To
include supersymmetry in the eleven dimensional supergravity, we have
to consider simultaneous realisation of the groups IGL(11$|$32) and
OSp(1$|$64), where OSp(1$|$64) is replacing the conformal group
\cite{West:2000ga}.

\section{General theory}

There is a complete duality between sets of fields, which transform
linearly while being subject to certain nonlinear constraints, and
equivalent sets of unconstrained fields which transform according to
nonlinear realisation. The bridge between these two sets of fields are
some massless Goldstone bosons called preferred fields, which
generally enter nonlinearly. If nonlinear realisations are introduced
by considering linear realisations of the preferred fields together
with a constraint, the constraint implies that the vacuum state in the
theory must be non-invariant, and the symmetry is spontaneously
broken. The non-vanishing of some vacuum expectation values indicates
the vacuum asymmetry. If some long-range vector fields are present and
the currents of the spontaneously broken symmetries are coupled to a
gauge field of the Yang-Mills type, then the symmetry breaking
manifests itself not through the appearance of Goldstone bosons, but
in the generation of mass for some of the components of the gauge
field.

Consider a general Lie group G with some specified subgroup H. Suppose
that the linear irreducible representations of G
\beq
\eqnlab{linear_transf}
\Psi \rightarrow \Psi' = D(g_0)\Psi, \hspace{0.5 cm} g_0 \in G
\eeq
and their decomposition into linear irreducible representations of H
are known. $D({*})$ denote the representations of the group, to which
the field $\Psi$ belongs, and $g_0$ is a group element of G. We want
to find covariant objects undergoing spontaneous symmetry breaking
from G to H. A representative $g$ of the coset G/H is called a
reducing matrix, and will obey the following requirements
\benu
\item $g$ belongs also to the group G, thus for any finite-dimensional
  representation $g_0 \rightarrow D(g_0)$ the functional $D(g)$ shall
  be well-defined. The number of independent fields $\phi_a$ needed to
  parametrise $g$ is equal to or less than the dimensionality of G.
  $\phi_a$ are called the preferred fields.

\item Under the action of the group G, the reducing matrix transforms
  as  
\beq
\eqnlab{reducing_transf}
g \rightarrow g' = g_0 g h^{-1}(\phi,g_0); \hspace{0.5 cm} g_0 \in G
  \textrm{ , } h(\phi,g_0) \in H. 
\eeq

\item Under the operation of the subgroup H, the reducing matrix
  transforms as 
\beq
g \rightarrow h g h^{-1}.
\eeq
\eenu
For any finite-dimensional representation, the transformation
  \eqnref{reducing_transf} will become $D(g) \rightarrow
  D(g_0)D(g)D(h^{-1})$. We can project the nonlinear realisations out
  of the linear ones by defining
\beq
\eqnlab{nonlin_field}
\psi \equiv D(g^{-1})\Psi,
\eeq
which transforms under G as 
\beq
\eqnlab{nonlin_transf}
\psi \rightarrow \psi' = D(g'^{-1})\Psi' = D(h)\psi.
\eeq

We define the reducing matrix as
\beq
\eqnlab{reducing_matrix}
g = e^{\phi_a K^a},
\eeq
where $K^a$ are the set of infinitesimal generators of G that are not
contained in the algebra of H. The matrix representation of $g$
contains at least one column $\chi$ transforming linearly
\beq
D(g)\chi \rightarrow D(g')\chi = D(g_0)D(g)\chi,
\eeq
i.e., $\chi$ is a singlet under H and $D(h^{-1})\chi = \chi$.
Expressing the components of $\chi$ in terms of the preferred fields,
we can calculate transformations of the preferred fields under
\Eqnref{reducing_transf} and the local elements $h$ by $h(\phi,g_0) =
g'^{-1}g_0g$.

The covariant derivatives are found using the outline given in
\cite{Salam:1969rq}. The ordinary derivative transforms as
\beq
\pa_\mu\psi \rightarrow D(h)\pa_\mu\psi + \pa_\mu D(h)\psi,
\eeq
using which we define a new operator
\beq
\Delta_\mu\psi \equiv D(g^{-1})\pa_\mu\Psi = \pa_\mu\psi +
(D(g^{-1})\pa_\mu D(g))\psi 
\eeq
with the transformation rule
\beq
\eqnlab{delta_transf}
\Delta_\mu\psi \rightarrow D(h)\Delta_\mu\psi.
\eeq
Note that $\Psi$ transforms only globally according to
\Eqnref{linear_transf}. The operator $g^{-1}\pa_\mu g$ belongs to the
infinitesimal algebra of G, since
\beq
g^{-1}(x)g(x+\delta x) = 1 + \delta x^\mu (g^{-1}\pa_\mu g) + \ldots
\eeq
for any infinitesimal transformation of G. We can then write
$g^{-1}\pa_\mu g = (g^{-1}\pa_\mu g)_i s^i$, where $s^i$ is a chosen
basis for the algebra and $(g^{-1}\pa_\mu g)_i$ are the coefficients.
In a specific representation of the field $\Psi$, the relation is
$D(g^{-1})\pa_\mu D(g) = (g^{-1}\pa_\mu g)_i S^i$ with $S^i$ being the
representations of the basis. Using \Eqnref{reducing_transf} we find
\beq
\eqnlab{igdg_transf}
g^{-1}\pa_\mu g \rightarrow h(g^{-1}\pa_\mu g)h^{-1} + h\pa_\mu
h^{-1},
\eeq
where the inhomogeneous term belongs to the algebra of the subgroup H.
We can thus divide $(g^{-1}\pa_\mu g)_i$ into a set transforming
covariantly and another set containing the inhomogeneity
\beq
\eqnlab{igdg_basisdecomp}
g^{-1}\pa_\mu g = \Gamma_{\mu\alpha}m^\alpha + \la
D_\mu\phi_a n^a.
\eeq
The preferred fields used to parametrise the reducing matrix are
denoted $\phi_a$, $m^\alpha$ is a basis for the subalgebra
corresponding to H and $n^a$ are the rest of the basis for the total
algebra of the group G. It should be noted that $m^\alpha$ and $n^a$
are transforming independently. \Eqnref{igdg_basisdecomp} defines the
field quantities $\Gamma_{\mu\alpha}$ and $D_\mu\phi_a$, with the
latter to be interpreted as the covariant derivatives of the preferred
fields. By doing so, the inhomogeneous term in the transformation will
only affect $\Gamma_{\mu\alpha}$. Putting \Eqnref{igdg_basisdecomp}
into the definition of $\Delta_\mu\psi$ yields
\beq
\Delta_\mu\psi = \pa_\mu\psi + \Gamma_{\mu\alpha}M^\alpha\psi + \la
D_\mu\phi_a N^a\psi, 
\eeq
where $M^\alpha$ and $N^a$ are the representations of the generators
$m^\alpha$ and $n^a$, respectively. Since the left hand side and the
last term on the right hand side of the above equation are covariant,
we find the covariant derivative of an arbitrary field under the coset
\beq
D_\mu\psi = \pa_\mu\psi + \Gamma_{\mu\alpha}M^\alpha\psi.
\eeq

If the group G is space-time dependent
\beq
\Psi(x) \rightarrow D(g_0)\Psi(x), \hspace{0.5 cm} g_0 = g_0(x) \in G,
\eeq
the transformation of the ordinary derivative on a linear field will
be
\beq
\pa_\mu \Psi \rightarrow D(g_0)\pa_\mu\Psi(x) + \pa_\mu D(g_0)\Psi(x),
\eeq
where the second term vanishes for a space-time independent
group. Using a basis $S^i$ for the generators, the transformation law
becomes 
\beq
\pa_\mu \Psi \rightarrow D(g_0)[\pa_\mu + (g_0^{-1}\pa_\mu g_0)_i S^i]
\Psi(x).
\eeq
Introduce now the gauge fields $A_\mu = A_{\mu i}s^i$, which
transform as
\beq
A_\mu \rightarrow g_0 A_\mu g_0^{-1} + \frac{1}{f}g_0 \pa_\mu
g_0^{-1}, 
\eeq
where $f$ is a constant. The covariant derivative for the linear field
is then
\beq
\D_\mu\Psi = (\pa_\mu + fA_{\mu i} S^i)\Psi
\eeq
transforming as $\D_\mu\Psi \rightarrow D(g_0)\D_\mu\Psi$. We can
define the antisymmetric field strength associated with the gauge
field
\beq
F_{\mu\nu} = \pa_\mu A_\nu - \pa_\nu A_\mu + f\com{A_\mu,A_\nu},
\eeq
which transforms as $F_{\mu\nu} \rightarrow g_0 F_{\mu\nu} g_0^{-1}$.

Introduce the nonlinear fields $\psi=D(g^{-1})\Psi$. As for the
space-time independent case, the covariant derivatives should be
contained in the quantity
\beq
\Delta_\mu\psi = D(g^{-1})(\pa_\mu + fA_{\mu i} S^i)\Psi = (\pa_\mu +
fB_{\mu i} S^i)\psi.
\eeq
The gauge field $B_\mu$ is defined as
\beq
B_\mu \equiv g^{-1} A_\mu g + \frac{1}{f}g^{-1}\pa_\mu g,
\eeq
and fulfills
\beq
B_\mu \rightarrow hB_\mu h^{-1} + \frac{1}{f}h\pa_\mu h^{-1}.
\eeq
The corresponding field strength is
\beq
B_{\mu\nu} = \pa_\mu B_\nu - \pa_\nu B_\mu + f\com{B_\mu,B_\nu} =
g^{-1}F_{\mu\nu}g
\eeq
satisfying $B_{\mu\nu} \rightarrow hB_{\mu\nu}h^{-1}$. It is now easy
to verify that $\Delta_\mu\psi$ satisfies \Eqnref{delta_transf}, and
we write $\Delta_\mu\psi = D_\mu\psi + \la(D_\mu\phi_a)N^a\psi$. In
conclusion, the covariant derivatives of an arbitrary nonlinear field
and a preferred field are
\beq
\begin{array}{rcl}
D_\mu \psi &=& (\pa_\mu + f B_{\mu\alpha}M^\alpha)\psi \\
\la D_\mu \phi_a &=& fB_{\mu a},
\end{array}
\eeq
respectively.

Using the covariant objects ($\psi$, $D_\mu\psi$, $D_\mu\phi_a$ and
$B_{\mu\nu})$ we can construct general Lagrangians. Since the
preferred fields themselves are not transforming covariantly, we
cannot have any mass term ($\phi_a^{*}\phi^a$) for these, i.e., they
are massless bosons. On the other hand, if the group G is space-time
dependent, we can have a mass term for the gauge fields $B_\mu$ and
the preferred fields $\phi_a$ will be absorbed by the gauge fields.
$B_{\mu \alpha}$ is then still massless, while $B_{\mu a}$ has a
well-defined mass $f/\la$.

\section{Type IIB supergravity}

The kinetic terms of the scalar fields in the type IIB supergravity
are given in Einstein frame \cite{Argurio:1998cp} by
\beq
\Lagr_{IIB} = -\frac{1}{2} \left( \pa_{\mu}\phi\pa^{\mu}\phi +
  e^{2\phi}\pa_{\mu}\chi\pa^{\mu}\chi \right),
\eeq
where $\phi$ and $\chi$ are real scalar fields.

The scalar part of the total Lagrangian possesses a SU(1,1)/U(1) coset
space symmetry, with SU(1,1) being the non-compact global symmetry and
U(1) its maximal compact subgroup. This invariance can be shown by
performing three consecutive coordinate transformations and thus
writing the Lagrangian in a manifestly invariant form. The first
coordinate transformation is to combine the real scalar fields into a
complex field
\beq
\la = \chi + ie^{-\phi} \hspace{1 cm} , \hspace{1 cm} \bar{\la} = \chi
- ie^{-\phi}.
\eeq
The Lagrangian becomes 
\beq
\Lagr_{IIB} = \frac{2}{(\la-\bar{\la})^2} \pa_{\mu}\la\pa^{\mu}\bar{\la}.
\eeq
A Möbius transformation is then performed
\beq
z = \frac{i-\la}{i+\la} \hspace{1 cm} , \hspace{1 cm} \bar{z} =
\frac{-i-\bar{\la}}{-i+\bar{\la}}
\eeq
yielding
\beq
\Lagr_{IIB} = -2 \frac{\pa_{\mu}z\pa^{\mu}\bar{z}}{(1-z\bar{z})^2}.
\eeq
Introducing two complex scalars, $\phi_0$ and $\phi_1$, satisfying
\beq
z = \frac{\phi_1}{\phi_0} \hspace{1 cm} \textrm{and} \hspace{1 cm}
|\phi_0|^2 - |\phi_1|^2 = 1
\eeq
brings the Lagrangian to the form
\beq
\eqnlab{scalarlagr}
\Lagr_{IIB} = -2 |\phi_0\pa_{\mu}\phi_1 - \phi_1\pa_{\mu}\phi_0|^2.
\eeq
We can now define a gauge field $A_{\mu}$ in terms of the complex
scalars $\phi_0$ and $\phi_1$ by
\beq
A_{\mu} \equiv \frac{i}{2} \left[(\phi_0\pa_{\mu}\bar{\phi_0} -
\phi_1\pa_{\mu}\bar{\phi_1}) - (\bar{\phi_0}\pa_{\mu}\phi_0 -
\bar{\phi_1}\pa_{\mu}\phi_1)\right],
\eeq
and the Lagrangian finally becomes
\beq
\Lagr_{IIB} = -2 ( |\pa_{\mu}\phi_1 - iA_{\mu}\phi_{1}|^2 -
|\pa_{\mu}\phi_0 - iA_{\mu}\phi_{0}|^2 ).
\eeq
From the construction of the gauge field, we conclude that $A_\mu$
will not give rise to some new independent degrees of freedom.

To see the symmetry of the Lagrangian, we write the fields $\phi_0$
and $\phi_1$ as components of a group element $g$ of SU(1,1)
\beq
g = \pmatrix{\phi_0 & \bar{\phi_1} \cr
             \phi_1 & \bar{\phi_0}}.
\eeq
This is an element of the group SU(1,1) since the complex scalar
fields satisfy the condition $|\phi_0|^2 - |\phi_1|^2 = 1$. For the
Lagrangian to be invariant under the coset group SU(1,1)/U(1), i.e., to
incorporate the local U(1) gauge invariance, we have to define a
covariant derivative
\beq
D_{\mu}g = \pa_{\mu}g - ig \pmatrix{1 & 0 \cr 0 & -1} A_{\mu},
\eeq
It is now straightforward to show that the Lagrangian can be written as 
\beq
\eqnlab{grouplagr}
\Lagr_{IIB} = -\textrm{Tr}{(g^{-1}D_{\mu}g)(g^{-1}D^{\mu}g)}.
\eeq
If one uses the usual partial derivatives instead of the covariant
derivatives in \Eqnref{grouplagr}, one will only get a Lagrangian with
manifestly global SU(1,1) invariance.

The group element $g$ transforms under SU(1,1)/U(1) as
\beq
g \rightarrow g' = g_0gh^{-1},
\eeq
where $g_0$ is a global SU(1,1) transformation whereas $h$ is a local
U(1) transformation. The field $A_{\mu}$ will transform as 
\beq
A_{\mu} \rightarrow A'_{\mu} = hA_{\mu}h^{-1} - i \pmatrix{1 & 0 \cr 0
  & -1} h\pa_{\mu}h^{-1},
\eeq
the local U(1) group element $h$ are represented here by $(2\times2)$
matrices. The resulting Lagrangian is
\beq
\begin{array}{rcl}
\Lagr_{IIB} \rightarrow \Lagr_{IIB}' &=& -
\textrm{Tr}{(g'^{-1}(D_{\mu}g)')(g'^{-1}(D^{\mu}g)')} \\
&=& - \textrm{Tr}\left( [h(g^{-1}D_{\mu}g)h^{-1}]
  [h(g^{-1}D^{\mu}g)h^{-1}] \right)\\
&=& \Lagr_{IIB},
\end{array}
\eeq
i.e., the Lagrangian is invariant.

\section{Eleven dimensional supergravity}

It is known since a long time that gravity can be formulated as a
simultaneous nonlinear realisation \cite{Isham:1971dv}. This is
because the group of general coordinate transformations is the closure
of the conformal group and the the group of affine transformations
\cite{Ogievetsky:1973ik,Ivanov:1985nu}. A summary of the conformal
generators and its closure with the affine group can be found in
Appendix~\chref{conformal} and \chref{closure}. We will in this
section show explicitly how the bosonic sector of eleven dimensional
supergravity can be expressed as a nonlinear realisation. But first we
will review shortly the field equations of the bosonic sector of
eleven dimensional supergravity.

\subsection{The bosonic gauge field equations}

The bosonic sector of the Lagrangian of eleven dimensional
supergravity is given by \cite{Cremmer:1978km}
\beq
\Lagr_{11} = \frac{e}{4\ka^2}R(\Omega(e)) -
\frac{e}{48}F_{\mu_1\ldots\mu_4}F^{\mu_1\ldots\mu_4} +
\frac{2\ka}{12^4}\varepsilon^{\mu_1\ldots\mu_{11}}F_{\mu_1\ldots\mu_4}
F_{\mu_5\ldots\mu_8}A_{\mu_9\mu_{10}\mu_{11}},
\eeq
where $e = \sqrt{-\textrm{det}(g_{\mu\nu})}$, $R$ is the curvature
scalar, $A_{\mu_1\mu_2\mu_3}$ is the rank three gauge field and
$F_{\mu_1\ldots\mu_4}$ the corresponding field strength. We use the
space-time signature $\eta_{ab}=\textrm{diag}(-1,1,1,\ldots,1)$. Note
that by defining $A'_{\mu\nu\rho}=\ka A_{\mu\nu\rho}$, the constant
$\ka$ appears only as an overall factor $\ka^{-2}$ in front of the
whole action. The field equations can be obtained by varying the
action with respect to the metric $g_{\mu\nu}$ and the gauge field
$A_{\mu_1\mu_2\mu_3}$, respectively. The calculations are greatly
simplified if we write the action in terms of differential forms. (See
Appendix~\chref{pform} for an introduction to the differential forms.)
We define the following $p$-forms
\beq
\begin{array}{rcl}
\eqnlab{formdef}
d &=& \dxmu\pa_{\mu} \\ 
A_3 &=& \frac{1}{3!}\dxmu \we \dxnu \we dx^{\rho} \, A_{\mu\nu\rho} \\
F_4 &=& \frac{1}{4!}\dxmu \we \dxnu \we dx^{\rho} \we dx^{\si} \,
F_{\mu\nu\rho\si} \\
{*}F_4 &=& \frac{e^{-1}}{4!7!} dx^{\mu_1} \we \ldots \we dx^{\mu_7}
\varepsilon_{\mu_1\ldots\mu_{11}} \, F^{\mu_8\ldots\mu_{11}}.
\end{array}
\eeq
The three and the four forms are related as
\beq
F_4 = dA_3 \hspace{1 cm} \Rightarrow \hspace{1 cm} F_{\mu\nu\rho\si} =
4\pa_{[\mu}A_{\nu\rho\si]},
\eeq
which is invariant under the gauge transformation
\beq
\eqnlab{gauge_transf}
A_3 \rightarrow A'_3 = A_3 + d\Lambda_2 \hspace{1 cm} \Rightarrow
\hspace{1 cm} F_4 \rightarrow F'_4 = dA_3 + d^2\Lambda_2 = F_4.
\eeq
More generally the field strength is invariant under a transformation
$A'_3 = A_3 + \delta A_3$, such that $d(\delta A_3) = 0$. 

The Bianchi identity for the four form $F_4$ is given by
\beq
dF_4 = d^2A_3 = 0 \hspace{1 cm} \Rightarrow \hspace{1 cm}
\pa_{[\ta}F_{\mu\nu\rho\si]} = 0. 
\eeq
Using the differential forms defined in \Eqnref{formdef} we can write
the action of the eleven dimensional supergravity as
\beq
\eqnlab{elevenaction}
S = \frac{1}{4\ka^2} \int d^{11}x \, \sqrt{-g}R(\Omega(e)) +
\frac{1}{2}\int F_4 \we *F_4 + \frac{\ka}{3}\int F_4 \we F_4 \we A_3.
\eeq
By varying the action with respect to $A_3$ we find the equation of
motion for the gauge field to be
\beq
\begin{array}{rcl}
\eqnlab{eomA}
d*F_4 + \ka F_4 \we F_4 &=& 0 \hspace{1 cm} \textrm{or} \\
\pa_{\mu_1}F^{\mu_1\ldots\mu_4} + \frac{\ka e^{-1}}{4 \cdot
  12^2}\varepsilon^{\mu_2\ldots\mu_4\nu_1\ldots\mu_8}F_{\nu_1\ldots\nu_4}
  F_{\nu_5\ldots\nu_8} &=& 0.  
\end{array}
\eeq
Since only the field strength is present in the field equation, the
equation of motion is invariant under the gauge transformation in
\Eqnref{gauge_transf}.

We can also introduce a dual gauge field $A_6$ and its corresponding
field strength $F_7$ obeying
\beq
\begin{array}{rcl}
F_7 &=& dA_6 + A_3 \we F_4 \hspace{2 cm} \textrm{or} \\
F_{\mu_1\ldots\mu_7} &=& 7\pa_{[\mu_1}A_{\mu_2\ldots\mu_7]} +
35A_{[\mu_1\ldots\mu_3}F_{\mu_4\ldots\mu_7]}.
\end{array}
\eeq
The Bianchi identity for the seven form $F_7$ will now automatically
yield \Eqnref{eomA} provided that
\beq
\begin{array}{rcl}
\eqnlab{dualA}
*F_4 &=& -\ka F_7 \hspace{2 cm} \textrm{or} \\
F_{\mu_1\ldots\mu_4} &=& \frac{\ka e^{-1}}{7!}
\varepsilon_{\mu_1\ldots\mu_{11}}F^{\mu_5\ldots\mu_{11}},
\end{array}
\eeq
which plays the role of the equation of motion in this dual
formulation of supergravity. Note that unlike \Eqnref{eomA},
\Eqnref{dualA} is a linear equation. 

\subsection{The G$_{11}$ realisation}

Consider the Lie algebra with the non-vanishing commutation relations
\beq
\eqnlab{IGL11}
\com{K^a_{\phantom{a}b},K^c_{\phantom{c}d}} = \delta^c_b
K^a_{\phantom{a}d} - \delta^a_d K^c_{\phantom{c}b} 
\hspace{0.5 cm} ; \hspace{0.5 cm}
\com{K^a_{\phantom{a}b},P_c} = -\delta^a_c P_b
\eeq
together with
\beq
\eqnlab{G11}
\begin{array}{rcl}
\com{K^a_{\phantom{a}b},R^{c_1c_2c_3}} &=&
3\delta^{[c_1}_bR^{|a|c_2c_3]} = \delta^{c_1}_bR^{ac_2c_3} +
\delta^{c_2}_bR^{ac_3c_1} + \delta^{c_3}_bR^{ac_1c_2} \\
\com{K^a_{\phantom{a}b},R^{c_1\dots c_6}} &=&
6\delta^{[c_1}_bR^{|a|c_2 \dots c_6]} \\
\com{R^{c_1c_2c_3},R^{c_4c_5c_6}} &=& 2R^{c_1\ldots c_6} \\
\com{R^{c_1\ldots c_6},R^{d_1\ldots d_6}} &=& 0 \\
\com{R^{c_1c_2c_3},R^{d_1\ldots d_6}} &=& 0.
\end{array}
\eeq
The generators $K^a_{\phantom{a}b}$ and $P_a$ are precisely those of
the affine group IGL(11) (see Appendix~\chref{closure}), whereas
$R^{c_1c_2c_3}$ and $R^{c_1\ldots c_6}$ can be identified as part of
the gl(32) automorphism algebra of the eleven dimensional
supersymmetry algebra \cite{West:2000ga}. We use G$_{11}$ to denote
the group defined by \Eqnref{IGL11} and \Eqnref{G11}, while g$_{11}$
is used when referring to the corresponding algebra. Note that
g$_{11}$ is not a finite dimensional semisimple Lie algebra, later in
\Chref{symmetry} we will identify it as part of the Borel subalgebra
of the very extended Kac-Moody algebra e$_{11}$.

Take the group G$_{11}$ as the global symmetry group and its Lorentz
subgroup SO(10,1) as the local symmetry group, i.e., we investigate a
theory with G$_{11}$/SO(10,1) space-time symmetry. The generators of
the Lorentz group are the antisymmetric part of the generators
$K^a_{\phantom{a}b}$ of the conformal group
\beq 
\eqnlab{Lorentz_gen}
J^a_{\phantom{a}b} = K^a_{\phantom{a}b} - K_b^{\phantom{b}a}.  
\eeq
Following the general procedure, we take a representative $g$ of the
coset space as 
\beq
\eqnlab{eleven_affine_coset}
g = e^{x^{f}P_{f}}e^{m_a^{\phantom{a}b}K^a_{\phantom{a}b}} \expo
\left(\frac{1}{3!}A_{c_1c_2c_3}R^{c_1c_2c_3} +
  \frac{1}{6!}A_{c_1\ldots c_6}R^{c_1\ldots c_6} \right),
\eeq
with $m_a^{\phantom{a}b}$, $A_{c_1c_2c_3}$ and $A_{c_1\ldots c_6}$
being the coordinate dependent Goldstone fields. Since the
antisymmetric part of $K^a_{\phantom{a}b}$ belongs to the local
symmetry algebra, only the symmetric part of $K^a_{\phantom{a}b}$ is
needed to write a simplest representative for the coset. However, we
will not assume that $m_a^{\phantom{a}b}$ is symmetric; a choice that
will turn out to make the identification with general relativity
easier. The translations are linearly realised, and thus we do not
associate any preferred fields with them.

We require the theory to be invariant under 
\beq
\eqnlab{coset_transf}
g \rightarrow g' = g_0gh^{-1},
\eeq
where $g_0$ is a constant element of G$_{11}$ and $h$ is a local
element of SO(10,1). The manifestly coset invariant Lagrangian takes
then the form $\Lagr = \textrm{Tr}(\V_\mu \V^\mu)$, with the Cartan
form $\V$ defined as \cite{West:2000ga}
\beq
\eqnlab{Cartan_form}
\V \equiv g^{-1}dg - \om = \dxmu\,\V_\mu .
\eeq
The Lorentz connection $\om$ is a one-form
\beq
\eqnlab{Lorentz_connect}
\om \equiv \frac{1}{2}\dxmu\,\om_{\mu a}^{\phantom{\mu a}b}
J^a_{\phantom{a}b},
\eeq
where $J^a_{\phantom{a}b}$ are the Lorentz generators. Note that the
Lorentz connection is a quadratic matrix, when writing out the matrix
indices explicitly \eqnref{Lorentz_connect} becomes
$\om^a_{\phantom{a}b} \equiv \frac{1}{2}\dxmu\,\om_{\mu
  c}^{\phantom{\mu c}d}(J^c_{\phantom{c}d})^a_{\phantom{a}b}$.

To find out how \eqnref{Cartan_form} transforms under
\eqnref{coset_transf}, we will first investigate the transformation
law of the Lorentz connection \eqnref{Lorentz_connect}. We write the
vielbein as a one-form $\bein^a = \dxmu\,
\bein_{\mu}^{\phantom{\mu}a}$, and it transforms only under local
Lorentz transformations according to
\beq
\eqnlab{bein_transf}
\bein^a \rightarrow \bein'^a = h^a_{\phantom{a}b} \bein^b.
\eeq
We can then define the torsion, which is nothing but the covariant
derivative of the vielbein
\beq
\eqnlab{torsion}
T^a \equiv D\bein^a = d\bein^a + \om^a_{\phantom{a}b}\bein^b,
\eeq
and express the Lorentz connection in terms of the vielbein by letting
the torsion vanish
\beq
\eqnlab{connect_vielbein}
\begin{array}{rcl}
\pa_{[\mu}\bein_{\nu]}^{\phantom{\nu]}a} &+&
\om_{[\mu\phantom{a}|b|}^{\phantom{[\mu}a}
\bein_{\nu]}^{\phantom{\nu]}b} = 0 \hspace{2 cm} \Rightarrow \\
\om_{\mu b c}(\bein) &=&
\frac{1}{2}(\bein^\rho_{\phantom{\rho}b}\pa_\mu\bein_{\rho c} - 
\bein^\rho_{\phantom{\rho}c}\pa_\mu\bein_{\rho b}) - 
\frac{1}{2}(\bein^\rho_{\phantom{\rho}b}\pa_\rho\bein_{\mu c} -
\bein^\rho_{\phantom{\rho}c}\pa_\rho\bein_{\mu b}) \\
& & - \frac{1}{2}(\bein^\la_{\phantom{\la}b}
\bein^\rho_{\phantom{\rho}c} \pa_\la\bein_{\rho a} -
\bein^\la_{\phantom{\la}c}\bein^\rho_{\phantom{\rho}b}
\pa_\la\bein_{\rho a}) \bein_\mu^{\phantom{\mu}a}.
\end{array}
\eeq
The Lorentz connection is antisymmetric in its last two indices, and
using this fact together with \Eqsref{Lorentz_gen} and
\eqnref{Lorentz_connect} we get
\beq
\om = \dxmu\,\om_{\mu a}^{\phantom{\mu a}b} K^a_{\phantom{a}b}.
\eeq
By definition the covariant derivative of $\bein^a$ (i.e., the torsion)
has to transform as the vielbein itself
\beq
\eqnlab{covd_bein_transf}
T^a \rightarrow T'^a = h^a_{\phantom{a}b}T^b.
\eeq
Using \Eqsref{bein_transf}, \eqnref{torsion} and
\eqnref{covd_bein_transf} we find the transformation of the Lorentz
connection to be
\beq
\eqnlab{connect_transf}
\begin{array}{rcl}
\om &\rightarrow& \om' = h\om h^{-1} + hdh^{-1} \hspace{4 cm}
\textrm{or} \\
\om^a_{\phantom{a}b} &\rightarrow& \om'^a_{\phantom{a}b} =
h^a_{\phantom{a}c} \om^c_{\phantom{c}d} (h^{-1})^d_{\phantom{d}b} +
h^a_{\phantom{a}c}d(h^{-1})^c_{\phantom{c}b}.
\end{array}
\eeq

Having found the transformation of the Lorentz connection, we are
ready to look at the transformation of the Cartan form $\V$
\beq
\eqnlab{Cartan_transf}
\begin{array}{rcl}
\V \rightarrow \V' &=& (g')^{-1}dg' - \om' \\
&=& (g_0gh^{-1})^{-1}d(g_0gh^{-1}) - (h\om h^{-1} + hdh^{-1}) \\
&=& h(g^{-1}dg - \om)h^{-1} = h \V h^{-1}.
\end{array}
\eeq
By now it should clear that the Lorentz connection is included in the
Cartan form to make it covariant and hence the Lagrangian invariant
under \eqnref{coset_transf}
\beq
\Lagr \rightarrow \Lagr' = \textrm{Tr}(h \V_\mu h^{-1} h \V^\mu
h^{-1}) = \Lagr,
\eeq
where we have omitted all the matrix indices. 

Motivated by the knowledge that the Cartan form in
\Eqnref{Cartan_form} will lead to an invariant Lagrangian, we can
write it out using the coset representative given in
\eqnref{eleven_affine_coset}. The constituents of the Cartan form will
then be identified with the covariant objects that appeared in the
equations of motion.
\beq
\eqnlab{affine_calc}
\begin{array}{rcl}
\V &=& g^{-1}dg - \om \\
   
   &=& \dxmu\,g^{-1}\pa_\mu e^{x^f P_f}
   e^{m_a^{\phantom{a}b}K^a_{\phantom{a}b}} \expo
   \left(\frac{A_{c_1c_2c_3}R^{c_1c_2c_3}}{3!} +
   \frac{A_{c_1\ldots c_6}R^{c_1\ldots c_6}}{6!} \right) - \om \\
   
   &=& \dxmu\,g^{-1} e^{x^{f}P_{f}} [\pa_\mu + (\pa_\mu x^c)P_c]
   e^{m_a^{\phantom{a}b}K^a_{\phantom{a}b}} \\
   & & \times \expo \left(\frac{A_{c_1c_2c_3}R^{c_1c_2c_3}}{3!} +
   \frac{A_{c_1\ldots c_6}R^{c_1\ldots c_6}}{6!} \right) - \om \\ 
   
   &=& \dxmu\,g^{-1} e^{x^{f}P_{f}}
   e^{m_a^{\phantom{a}b}K^a_{\phantom{a}b}} [\pa_\mu + \pa_\mu
   m_c^{\phantom{c}d}K^c_{\phantom{c}d} +
   (e^m)_{\mu}^{\phantom{\mu}c}P_c] \\
   & & \times \expo \left(\frac{1}{3!}A_{c_1c_2c_3}R^{c_1c_2c_3} +
   \frac{1}{6!}A_{c_1\ldots c_6}R^{c_1\ldots c_6} \right) - \om \\
   
   &=& \dxmu\,g^{-1}g\{ (e^m)_{\mu}^{\phantom{\mu}a}P_a +
   (\pa_\mu m_a^{\phantom{a}b} - \om_{\mu a}^{\phantom{\mu a}b})
   K^a_{\phantom{a}b} \\ 
   & & + \frac{1}{3!}[\pa_\mu A_{c_1c_2c_3} + 3\pa_\mu
   m_{[c_1}^{\phantom{[c_1}b}A_{|b|c_2c_3]}]R^{c_1c_2c_3} \\
   & & + \frac{1}{6!}[\pa_\mu A_{c_1\ldots c_6} + 6\pa_\mu
   m_{[c_1}^{\phantom{[c_1}b}A_{|b|c_2\ldots c_6]} \\
   & & + 20(\pa_\mu A_{c_1c_2c_3} + 3\pa_\mu
   m_{[c_1}^{\phantom{[c_1}b}A_{|b|c_2c_3]}) A_{c_4c_5c_6}]
   R^{c_1\ldots c_6} \} 
\end{array}
\eeq
In calculating \eqnref{affine_calc} we have used that all the
generators commute with the coordinates $x^\mu$. We have also
frequently used the Baker-Hausdorff lemma \cite{Fuchs:1997jv}
\beq
\eqnlab{baker_hausdorff}
\begin{array}{rcl}
Xe^Y &=& e^Y \left\{ X + \com{X,Y} + \frac{1}{2!}\com{\com{X,Y},Y} +
     \ldots \right. \\
     & & \left. + \frac{1}{n!}\com{\ldots\com{\com{X,Y},Y},\ldots,Y} +
    \ldots \right\}.
\end{array}
\eeq

We will now make the following definitions
\beqa
(\bein^{-1}\pa_\mu\bein)_a^{\phantom{a}b} &=& \pa_\mu m_a^{\phantom{a}b}
\\
\bein_{\mu}^{\phantom{\mu}a} &\equiv& (e^m)_{\mu}^{\phantom{\mu}a} \\
\Omega_{\mu a}^{\phantom{\mu a}b} &\equiv& \pa_\mu m_a^{\phantom{a}b} -
\om_{\mu a}^{\phantom{\mu a}b} =
(\bein^{-1}\pa_\mu\bein)_a^{\phantom{a}b} - \om_{\mu a}^{\phantom{\mu
    a}b} \\
\tilde{D}_\mu A_{c_1c_2c_3} &\equiv& \pa_\mu A_{c_1c_2c_3} +
3(\bein^{-1}\pa_\mu\bein)_{[c_1}^{\phantom{[c_1}b}A_{|b|c_2c_3]} \\
\non \tilde{D}_\mu A_{c_1\ldots c_6} &\equiv& \pa_\mu A_{c_1\ldots c_6} +
6(\bein^{-1}\pa_\mu\bein)_{[c_1}^{\phantom{[c_1}b}A_{|b|c_2\ldots
  c_6]} \\
& & - 20A_{[c_1c_2c_3}\tilde{D}_{|\mu|}A_{c_4c_5c_6]},
\eeqa
where we identify $\bein_{\mu}^{\phantom{\mu}a}$ with the vielbein.
Putting these definitions into \Eqnref{affine_calc} we finally get
\beq
\eqnlab{affine_Cartan}
\V_\mu = \bein_{\mu}^{\phantom{\mu}a}P_a + \Omega_{\mu
  a}^{\phantom{\mu a}b}K^a_{\phantom{a}b} + \frac{1}{3!}\tilde{D}_\mu
  A_{c_1c_2c_3}R^{c_1c_2c_3} + \frac{1}{6!}\tilde{D}_\mu A_{c_1\ldots
  c_6}R^{c_1\ldots c_6}. 
\eeq
From \Eqnref{affine_Cartan} we can read off the covariant derivatives
of the Goldstone fields $m_a^{\phantom{a}b}$, $A_{c_1c_2c_3}$ and
$A_{c_1\ldots c_6}$
\beqa
\Omega_{ab}^{\phantom{ab}c} &=&  \bein^{\mu}_{\phantom{\mu}a}\Omega_{\mu
  b}^{\phantom{\mu b}c} = \bein^{\mu}_{\phantom{\mu}a}
(\bein^{-1}\pa_\mu\bein)_b^{\phantom{b}c} - \om_{ab}^{\phantom{ab}c}
\\
\eqnlab{covd_A3}
\tilde{D}_a A_{c_1c_2c_3} &=&
\bein^{\mu}_{\phantom{\mu}a}\tilde{D}_\mu A_{c_1c_2c_3} \\
\eqnlab{covd_A6}
 \tilde{D}_a A_{c_1\ldots c_6} &=&
 \bein^{\mu}_{\phantom{\mu}a}\tilde{D}_\mu A_{c_1\ldots c_6},
\eeqa
where $\bein_\mu^{\phantom{\mu}a}\bein^{\nu}_{\phantom{\nu}a} =
\delta^\nu_\mu$. These covariant derivatives are not field strengths,
since they are not antisymmetrised. Only after a simultaneous
nonlinear realisation with the conformal group will the field
strengths be recovered. Note that we have defined
$\om_{ab}^{\phantom{ab}c} \equiv
\bein^{\mu}_{\phantom{\mu}a}\om_{\mu b}^{\phantom{\mu b}c}$.

Performing a local Lorentz transformation
\beq
\eqnlab{Lorentz_transf}
g \rightarrow hgh^{-1} \hspace{1 cm} \textrm{with} \hspace{1 cm}
h^a_{\phantom{a}b} = \left( e^{\frac{1}{2}\varsigma_c^{\phantom{c}d}(x)
    J^c_{\phantom{c}d}} \right)^a_{\phantom{a}b}
\eeq
we find that $\bein_{\mu}^{\phantom{\mu}a}$ really transforms in
agreement with \Eqnref{bein_transf}, and also the interpretation of
all the contracted indices in $\V_\mu$ as tangent space indices is
justified. A matter field $B$ will transform under
\eqnref{coset_transf} according to $B \rightarrow B' = D(h)B$, where
$D$ is the representation of the Lorentz group depending on the
specific type of the matter field $B$. The covariant derivative of the
matter field is then given by
\beq
\eqnlab{affine_covd}
\tilde{D}_a B \equiv \bein^\mu_{\phantom{\mu}a}\pa_\mu B +
\frac{1}{2} \om_{ab}^{\phantom{ab}c} S^b_{\phantom{b}c} B,
\eeq
where $S^a_{\phantom{a}b}$ are the representations of the generators
of the Lorentz group associated with $B$.

\subsection{The conformal realisation}

Let us now turn our attention to the conformal group. The conformal
transformations in space-time include translations, Lorentz
transformations, dilations and special conformal transformations. The
generators and their commutation relations can be found in
Appendix~\chref{conformal}. Here we will first use a slightly
different approach to obtain the generators, and then perform the
nonlinear realisation based on these. We assume the dimension to be
$(10,1)$, but the results can be directly generalised to arbitrary
$(d-1,1)$ dimensions.

As mentioned in Appendix~\chref{conformal}, the conformal group in
$(10,1)$ dimensions is equal to the Lorentz group in $(11,2)$
dimensions. A linear representation of O(11,2) acts on a field as
\beq
\eqnlab{conf_linear_transf}
\Psi(\zeta) \rightarrow \Psi'(\zeta) =
D(\Lambda)\Psi(\Lambda^{-1}\zeta), 
\eeq
where $\zeta_A$ denotes the 13-dimensional coordinate vector and
$\Lambda$ is a pseudo-orthogonal transformation on these coordinates
\beq
\zeta_A \rightarrow \zeta'_A = \Lambda_A^{\phantom{A}B}\zeta_B.
\eeq
We will in this subsection use the signature
$\eta_{AB}=\textrm{diag}(1,-1,\ldots,-1,1)$. The notation $D(*)$ is
used to denote the matrix representation of a group element or a
generator associated with the field $\Psi(\zeta)$.

The generators of the Lorentz group are as usual defined through the
representations of infinitesimal transformations $\Lambda = 1 +
\epsilon$ as
\beq
D(\La) = 1 + \epsilon^{AB}J_{AB}.
\eeq
The generators $J_{AB}$ are given by
\beq
(J_{AB})_{CD} = \eta_{AC}\eta_{BD} - \eta_{AD}\eta_{BC}
\eeq
in the 13-dimensional self-representation and obey the commutation
relations
\beq
\com{J_{AB},J_{CD}} = (\eta_{BC}J_{AD} - \eta_{BD}J_{AC} +
\eta_{AD}J_{BC} - \eta_{AC}J_{BD}).
\eeq
These generators can also be realised using the coordinates in 13
dimensions 
\beq
\eqnlab{13dimLorentz}
J_{AB} \equiv \pa_{AB} =  \zeta_A\pa_B - \zeta_B\pa_A.
\eeq

We can now identify the generators of the conformal group in (10,1)
dimensions with different parts of $J_{AB}$ \cite{Salam:1969qk}
\beq
\eqnlab{conf_gen_6dim}
\begin{array}{llll}
\textrm{Lorentz~generators:} & J_{\mu\nu} & & \cr
\textrm{Translation~generators:} & P_\mu & = & J_{13\mu} + J_{12\mu} \cr
\textrm{Special~conformal~generators:} & K_\mu & = & J_{13\mu} -
J_{12\mu} \cr 
\textrm{Dilation generator:} & D & = & J_{13,12},
\end{array}
\eeq
where the index $\mu$ runs from 0 to 10. Introduce a new set of
independent coordinate variables
\beq
\eqnlab{coord_change}
x_\mu = \frac{\zeta_\mu}{\zeta_{12} + \zeta_{13}}, \hspace{0.5 cm} \ka
= \zeta_{12} + \zeta_{13}, \hspace{0.5 cm} \la = \zeta_{12} -
\zeta_{13}. 
\eeq
Calculating the transformations of $\zeta_A$ under $e^{\epsilon X}$,
with $X$ being the generators in \eqnref{conf_gen_6dim}, verifies that
$x_\mu$ transforms indeed as a coordinate vector in 11 dimensions.
Putting \Eqnref{coord_change} into \Eqnref{13dimLorentz} we find the
generators of the conformal group in terms of $x_\mu$, $\ka$ and
$\la$. It is possible to eliminate the variable $\la$ by requiring
$\zeta^2 = 0$, since this condition is invariant under 13-dimensional
Lorentz transformations. This will allow us to set $\la = \ka x^2$,
and we finally arrive at
\beq
\begin{array}{rcl}
\pa_{\mu\nu} &=& x_\mu\pa_\nu - x_\nu\pa_\mu \\
\pa_{13\mu} + \pa_{12\mu} &=& \pa_{\mu} \\
\pa_{13\mu} - \pa_{12\mu} &=& 2x_\mu\left(x^\nu\pa_\nu -
  \ka\frac{\pa}{\pa\ka}\right) - x^2\pa_\mu \\ 
\pa_{13,12} &=& x^\nu\pa_\nu - \ka\frac{\pa}{\pa\ka},
\end{array}
\eeq
which are precisely the generators of the conformal group given in
\eqnref{conf_gen}. We define the degree of homogeneity as $l \equiv
\ka\pa/\pa\ka$ with the coordinates $x^\mu$ having degree zero.  For a
homogeneous field $\Phi(\zeta)$ of degree $l$, the dependence of $\ka$
will factor out as $\ka^l$ on the hypercone $\zeta^2 = 0$, see
reference \cite{Salam:1969qk}.

Let us consider the reducing matrix\footnote{For the conformal
  algebra, there is no difference between the algebra index $a$ and
  the space-time index $\mu$, hence in this subsection both $a$ and
  $\mu$ are raised and lowered using the metric given in
  \eqnref{conf_metric}.}
\beq
\eqnlab{eleven_conf_coset}
g = e^{-x^a P_a}e^{\phi^b K_b}e^{\si D},
\eeq
having the conformal group as the global and the Lorentz group as the
local symmetry, respectively. The preferred fields $\si$ and $\phi^a$
are coordinate dependent, and we do not associate any field with the
translation generators. As will become obvious later, the factor
$e^{-x^a P_a}$ is included to give a simpler form for the covariant
derivatives. Note also that we have chosen a coset representative,
which does not contain any contribution from the local Lorentz
subgroup ($J^a_{\phantom{a}b}$). The reducing matrix transforms under
the coset according to
\beq
\eqnlab{eleven_conf_transf}
g(\zeta) \rightarrow g'(\zeta') = \Lambda g(\Lambda^{-1}\zeta)
h^{-1}(\Lambda^{-1}\zeta,\Lambda),
\eeq
where $\Lambda$ corresponds to a global conformal transformation in
(10,1) dimensions, $h$ is a local Lorentz transformation and $\zeta' =
\Lambda\zeta$. The nonlinear field $\psi$ is then defined by
\beq
\psi(\zeta) \equiv D(g^{-1})\Psi(\zeta),
\eeq
where $\Psi(\zeta)$ is the linear field defined in
\Eqnref{conf_linear_transf}. The field $\psi(\zeta)$ transforms
irreducibly under the Lorentz group as
\beq
\psi(\zeta) \rightarrow \psi'(\zeta') = D(h)\psi(\Lambda^{-1}\zeta). 
\eeq

The transformations of the preferred fields $\phi_\mu$ and $\si$ can
be found if we require that $(g)_A^{\phantom{A}12}$ and
$(g)_A^{\phantom{A}13}$ transform as true 13-dimensional vectors, i.e.,
$h(\zeta, \Lambda)$ acts only in the subspace \{0,\ldots,10\}. Using
these columns we can define two independent linear fields
\beq
\begin{array}{rcl}
\eqnlab{phi_matrix}
\Phi_A &\equiv& \frac{1}{2}[g_A^{\phantom{A}12} - g_A^{\phantom{A}13}]
  \\ 
  & & \\
  &=& \pmatrix{(\phi_\mu + x_\mu\phi^2)e^{-\si} \cr 
  \frac{1}{2}(1+2x\cdot\phi+x^2\phi^2+\phi^2)e^{-\si} \cr
  -\frac{1}{2}(1+2x\cdot\phi+x^2\phi^2-\phi^2)e^{-\si} }
\end{array}
\eeq
and
\beq
\begin{array}{rcl}
\eqnlab{psi_matrix}
\Psi_A \equiv \frac{1}{2}[g_A^{\phantom{A}12} + g_A^{\phantom{A}13}]
  = \pmatrix{x_\mu e^{\si} \cr 
  \frac{1}{2}(1+x^2)e^{\si} \cr
  \frac{1}{2}(1-x^2)e^{\si} }.
\end{array}
\eeq

We are now ready to find the covariant derivatives. It should be
stressed that the notation $h_A^{\phantom{A}B}$ means the
13-dimensional matrix representation, whereas the notation $D(h)$
means the representation associated with the field $\Psi(\zeta)$. Let
us first define the operator $\Delta_{AB}\psi$ as
\beq
\eqnlab{d1}
\Delta_{AB}\psi \equiv D(g^{-1})\pa_{AB}\Psi,
\eeq
where $\Psi$ is an arbitrary linear field and $\pa_{AB}$ is defined in
\eqnref{13dimLorentz}. Under the coset, this operator transforms as
\beq
\eqnlab{d1_transf}
\Delta_{AB}\psi \rightarrow D(g'^{-1})\pa_{AB}\Psi' =
\Lambda_A^{\phantom{A}A'}\Lambda_B^{\phantom{B}B'}
D(h)\Delta_{A'B'}\psi, 
\eeq
where we have used \Eqsref{conf_linear_transf} and
\eqnref{eleven_conf_transf}. To avoid $\Lambda_A^{\phantom{A}B}$ in
the transformation laws we define a new operator
\beq
\eqnlab{d2}
\Delta_{(AB)}\psi \equiv (g^{-1})_A^{\phantom{A}A'}
(g^{-1})_B^{\phantom{B}B'}\Delta_{A'B'}\psi.
\eeq
Using the transformation laws \eqnref{eleven_conf_transf} and
\eqnref{d1_transf} we find
\beq
\eqnlab{d2_transf}
\Delta_{(AB)}\psi \rightarrow h_A^{\phantom{A}A'}
h_B^{\phantom{B}B'} D(h)\Delta_{(A'B')}\psi.
\eeq
To simplify the calculations one observes that
\beq
\pa_{AB} = \frac{1}{2} (e^{-x\cdot P})_A^{\phantom{A}A'}
(e^{-x\cdot P})_B^{\phantom{B}B'} (K^\mu\pa_\mu + 2lD)_{A'B'},
\eeq
putting which into \Eqsref{d1} and \eqnref{d2} yields
\beq
\eqnlab{d2_calc}
\begin{array}{rcl}
\Delta_{(AB)}\psi &=& (g^{-1})_A^{\phantom{A}C_1}
(g^{-1})_B^{\phantom{B}C_2} D(g^{-1}) \\
 & & \times \left[ \frac{1}{2} (e^{-x\cdot
     P})_{C_1}^{\phantom{C_{1}}C_3} (e^{-x\cdot
     P})_{C_2}^{\phantom{C_2}C_4} (K^\mu\pa_\mu + 2lD)_{C_3C_4}
 \right] \Psi \\ 
&=& \frac{1}{2}D(g^{-1}) \{
e^{-\si}(\pa_\mu + 2l\phi_\mu)(K^\mu)_{AB} + 2lD_{AB} \}\Psi.
\end{array}
\eeq
It is now clear that the expression \eqnref{d2_calc} gets much
simplified by putting a factor $e^{-x\cdot P}$ in the definition of
$g$. We separate \eqnref{d2_calc} into irreducible parts according to
\beq
\begin{array}{rcl}
\Delta_{(\mu\nu)}\psi &=& 0 \\
(\Delta_{(13\mu)} + \Delta_{(12\mu)})\psi &=& D(g^{-1}) e^{-\si}
(\pa_\mu + 2l\phi_\mu) \Psi \\
(\Delta_{(13\mu)} - \Delta_{(12\mu)})\psi &=& 0 \\
\Delta_{(13,12)}\psi &=& -l\cdot D(g^{-1})\Psi = -l\psi.
\end{array}
\eeq
It is then natural to find the covariant derivatives inside the term
\beq
\eqnlab{d3}
\Delta_\mu\psi \equiv D(g^{-1}) e^{-\si} (\pa_\mu + 2l\phi_\mu) D(g)
\psi, 
\eeq
which transforms as
\beq
\eqnlab{d3_transf}
\begin{array}{rcl}
\Delta_\mu\psi \rightarrow \left[ h_{13}^{\phantom{13}A}
h_{\mu}^{\phantom{\mu}B} + h_{12}^{\phantom{12}A}
h_{\mu}^{\phantom{\mu}B} \right] D(h) \Delta_{(AB)}\psi 
= h_{\mu}^{\phantom{\mu}\nu} D(h) \Delta_\nu\psi(\Lambda^{-1}\zeta).
\end{array}
\eeq
To calculate this transformation we have used the fact that
$h_A^{\phantom{A}B}$ only acts on the $(\mu\nu)$ subspace and takes
the form of a 11-dimensional Lorentz transformation.

To obtain the covariant derivatives in 11 dimensions we write
\eqnref{d3} as
\beq
\eqnlab{d3_calc}
\begin{array}{rcl}
\Delta_\mu \psi &=& e^{-\si}(\pa_\mu + 2l\phi_\mu)\psi +
D(g^{-1})e^{-\si}\pa_\mu D(g)\psi \\
&=& e^{-\si}(\pa_\mu + 2l\phi_\mu)\psi +
\frac{1}{2}(g^{-1}e^{-\si}\pa_\mu
g)^A_{\phantom{A}B}S_A^{\phantom{A}B}\psi, 
\end{array}
\eeq
where $S_A^{\phantom{A}B}$ generates infinitesimal SO(11,2)
transformations on $\Psi(\zeta)$. The Cartan form is given by
\beq
\eqnlab{conf_calc}
\begin{array}{rcl}
\V &=& g^{-1}dg \\
   
   &=& \dxmu\,g^{-1}\pa_\mu e^{-x^{a}P_{a}}
   e^{\phi^b K_b} e^{\si D} \\
   
   &=& \dxmu\,g^{-1} e^{-x^{a}P_{a}} [\pa_\mu - (\pa_\mu x^c) P_c]
   e^{\phi^b K_b} e^{\si D} \\
      
   &=& \dxmu\,g^{-1} e^{-x^{a}P_{a}} e^{\phi^b K_b} [\pa_\mu -
   (\pa_\mu x^c) P_c + \pa_\mu \phi^c K_c - 2\phi_\mu D \\
   & & + 2\delta_\mu^c\phi^d J_{cd} - 2\phi_\mu \phi^c K_c +
   \phi^c\phi_c\delta_\mu^d K_d] e^{\si D} \\ 
   
   &=& \dxmu\,\{ -e^\si\delta_\mu^a P_a + e^{-\si}(\pa_\mu
   \phi^a + \phi^b\phi_b\delta_\mu^a - 2\phi_\mu\phi^a)K_a \\
   & & + (\pa_\mu\si - 2\phi_\mu)D + (\delta_\mu^a\phi^b -
   \delta_\mu^b\phi^a)J_{ab} \}, \\
\end{array}
\eeq
where the identity element $g^{-1}g$ has been omitted. Observe that
$\V_\mu\V^\mu$, with the Cartan form defined without the spin
connection, is not invariant under the coset transformations
\eqnref{eleven_conf_transf}. The covariant parts of
$g^{-1}e^{-\si}\pa_\mu g$ are
\beq
\begin{array}{rcl}
(g^{-1}e^{-\si}\pa_\mu\Phi)_A &=& \frac{1}{2}[(g^{-1}e^{-\si}\pa_\mu
  g)_A^{\phantom{A}12} - (g^{-1}e^{-\si}\pa_\mu g)_A^{\phantom{A}13}]
  \\ 
  & & \\
  &=& \pmatrix{ e^{-2\si}(\pa_\mu\phi_\nu +
  \phi^\ta \phi_\ta \eta_{\mu\nu} - 2\phi_\mu\phi_\nu) \cr 
  -e^{-\si}(\pa_\mu \si - 2\phi_\mu) \cr
  e^{-\si}(\pa_\mu \si - 2\phi_\mu) }
\end{array}
\eeq
and
\beq
\begin{array}{rcl}
(g^{-1}e^{-\si}\pa_\mu\Psi)_A &=& \frac{1}{2}[(g^{-1}e^{-\si}\pa_\mu
  g)_A^{\phantom{A}12} + (g^{-1}e^{-\si}\pa_\mu g)_A^{\phantom{A}13}]
  \\ 
  & & \\
  &=& \pmatrix{g_{\mu\nu} \cr 
  e^{-\si}(\pa_\mu\si - 2\phi_\mu) \cr
  e^{-\si}(\pa_\mu\si - 2\phi_\mu) }.
\end{array}
\eeq
We use these to define the covariant derivatives of the fields $\si$
and $\phi_\mu$. The remaining part
\beq
\begin{array}{rcl}
(g^{-1}e^{-\si}\pa_\mu g)_{\la\rho} &=& e^{-\si}(\delta_\mu^\nu\phi^\ta
- \delta_\mu^\ta\phi^\nu)(J_{\nu\ta})_{\la\rho} \\
&=& e^{-\si}(\delta_\mu^\nu\phi^\ta - \delta_\mu^\ta\phi^\nu)
(\eta_{\nu\la}\eta_{\ta\rho} - \eta_{\nu\rho}\eta_{\ta\la}) \\
&=& 2e^{-\si}(\eta_{\mu\la}\phi_{\rho} - \eta_{\mu\rho}\phi_\la),
\end{array}
\eeq
becomes covariant together with the other terms in \Eqnref{d3_calc}
\beq
\begin{array}{rcl}
\D_\mu \psi &\equiv& e^{-\si}(\pa_\mu + 2l\phi_\mu)\psi +
\frac{1}{2}\cdot 2e^{-\si}(\delta_\mu^\la\phi_{\rho} -
\eta_{\mu\rho}\phi^\la)S_\la^{\phantom{\la}\rho}\psi \\
&=& e^{-\si}(\pa_\mu + 2l\phi_\mu \psi +
2S_\mu^{\phantom{\mu}\rho}\phi_\rho) \psi.
\end{array}
\eeq
We have used that the generators $S_{\mu\nu}$ obey $S_{\mu\nu} =
-S_{\nu\mu}$. 

In summary, the covariant derivatives of the various fields in
11 dimensions are
\beqa
\eqnlab{covd_sigma}
\D_\mu \si &=& e^{-\si}(\pa_\mu \si - 2\phi_\mu) \\
\eqnlab{covd_phi}
\D_\mu \phi_\nu &=& e^{-2\si}(\pa_\mu\phi_\nu + \phi^\ta \phi_\ta
\eta_{\mu\nu} - 2\phi_\mu\phi_\nu) \\
\eqnlab{conf_covd}
\D_\mu \psi &=& e^{-\si}(\pa_\mu + 2l\phi_\mu +
2S_\mu^{\phantom{\mu}\rho}\phi_\rho) \psi,
\eeqa
where $\psi(\zeta)$ is homogeneous of degree $l$, whereas the fields
$\si$ and $\phi_\mu$ are both of degree zero.

We can now calculate the transformations of the preferred fields, or
more generally how an arbitrary field transforms under the coset, with
the conformal group as the global symmetry and the Lorentz group as
the local symmetry. The detailed calculations can be found in
Appendix~\chref{transformation}. The transformations of the preferred
fields with special conformal transformations as the global
transformations are
\beq
\begin{array}{rcl}
\si'(x') &=& \si(x) + \ln|\det \left( \frac{\pa x'}{\pa x}
\right)|^{-1/4} \\
\phi'_\mu(x') &=& \frac{\pa x^\nu}{\pa x'^\mu} \left\{ \phi_\nu(x) +
    \frac{1}{2}\frac{\pa}{\pa x^\nu}\left[ \ln|\det \left( \frac{\pa
    x'}{\pa x} \right)|^{-1/4} \right] \right\},
\end{array}
\eeq
where $x'^\mu = \frac{x^\mu+b^\mu x^2}{1+2b \cdot x+b^2x^2}$ and
$\left( \ln|\det \left( \frac{\pa x'}{\pa x} \right)|^{-1/4} \right) =
1+2b \cdot x+b^2x^2$. Thus we see that the preferred fields transform
with great similarity to the gauge fields, and it is motivated to
try assign geometric meanings to these fields.

We would like to create a metric tensor and a connection using only
$\si$ and $\phi_\mu$, i.e., without introducing any new dynamical
variable. The easiest choice of a metric tensor is
\beq
\eqnlab{conf_metric}
g_{\mu\nu} = \eta_{\mu\nu} e^{2\si(x)} \hspace{0.5 cm} \Rightarrow
\hspace{0.5 cm} g^{\mu\nu} = \eta^{\mu\nu} e^{-2\si(x)},
\eeq
using the transformation laws for $\si$ given in
Appendix~\chref{transformation} we can verify that $g_{\mu\nu}$ indeed
transforms as a rank two tensor. Having the metric, the Christoffel
symbol will be
\beq
\begin{array}{rcl}
\chrf{\ta}{\mu}{\nu} &=& \frac{1}{2}g^{\ta\rho}(\pa_\mu g_{\nu\rho} +
\pa_\nu g_{\mu\rho} - \pa_\rho g_{\mu\nu}) \\
&=& \delta^\ta_\mu\pa_\nu\si + \delta^\ta_\nu\pa_\mu\si -
\eta_{\mu\nu}\eta^{\ta\rho}\pa_\rho\si. 
\end{array}
\eeq
We can also define an alternative connection using the preferred
fields $\phi_\mu$
\beq
\Gamma^\ta_{\mu\nu} = C(\delta^\ta_\mu\phi_\nu +
\delta^\ta_\nu\phi_\mu\ - g_{\mu\nu}\phi^\ta).
\eeq
The constant $C$ can be determined by requiring the connection to
transform as
\beq
\Gamma^\ta_{\mu\nu} \rightarrow \Gamma'^\ta_{\mu\nu} = \frac{\pa
  x'^{\ta}}{\pa x^{\rho_1}} \frac{\pa x^{\rho_2}}{\pa x'^{\mu}}
\frac{\pa x^{\rho_3}}{\pa x'^{\nu}} \Gamma^{\rho_1}_{\rho_2\rho_3} -
\frac{\pa^2 x'^\ta}{\pa x^{\rho_1}\pa x^{\rho_2}} \frac{\pa
  x^{\rho_1}}{\pa x'^{\mu}} \frac{\pa x^{\rho_2}}{\pa x'^{\nu}},
\eeq
which sets $C = 2$ to yield 
\beq
\eqnlab{conf_connect}
\Gamma^\ta_{\mu\nu} = 2(\delta^\ta_\mu\phi_\nu +
\delta^\ta_\nu\phi_\mu\ - g_{\mu\nu}\phi^\ta).
\eeq
Comparing the expression for the Christoffel symbol with
$\Gamma^\ta_{\mu\nu}$ leads to
\beq
\eqnlab{si_phi}
\chrf{\ta}{\mu}{\nu} = \Gamma^\ta_{\mu\nu} \hspace{0.5 cm} \Rightarrow
\hspace{0.5 cm} \pa_\mu \si = 2\phi_\mu.
\eeq
If the connection is not equal to the Christoffel symbol, then the
geometry is non-Riemannian. This means that the covariant derivative
of the metric tensor with respect to this connection will not
vanish. Whenever there exist two connections, we can form a general
connection
\beq
\begin{array}{rcl}
& & \chrf{\ta}{\mu}{\nu} + f\left( \chrf{\ta}{\mu}{\nu} -
  \Gamma^\ta_{\mu\nu} \right) \\
& & \\
&=& \chrf{\ta}{\mu}{\nu} + f\left[
  \delta^\ta_\mu(\pa_\nu\si-2\phi_\nu) +
  \delta^\ta_\nu(\pa_\mu\si-2\phi_\mu) -
  g_{\mu\nu}g^{\ta\rho}(\pa_\rho\si-2\phi_\rho) \right],
\end{array}
\eeq
where $f$ is an arbitrary scalar function. From the nonlinear
realisation we know that the second part of this general connection
transforms as a tensor, since $\D_\mu \si = \pa_\mu\si-2\phi_\mu$ is
the covariant derivative of $\si$. We can thus really put $\pa_\mu\si
= 2\phi_\mu$ and use the Christoffel symbol as the connection, this
choice is the same as to let the torsion vanish. However, we will
continue the calculations with a general $\Gamma^\ta_{\mu\nu}$, i.e.,
assume $\phi_\mu$ to be unrelated to $\si$.

The covariant derivative of an arbitrary tensor under general
coordinate transformations is \cite{Isham:1971dv}
\beq
\eqnlab{gen_covd}
\D_\mu T = \pa_\mu T - F^\nu_{\phantom{\nu}\ta} \Gamma^\ta_{\mu\nu} T, 
\eeq
where $F^\nu_{\phantom{\nu}\ta}$ defines the representation of
GL(11,$\mathbb R$) to which $T$ belongs, and satisfy the commutation
relations
\beq
\com{F^\mu_{\phantom{\mu}\nu},F^\ta_{\phantom{\ta}\rho}} =
\delta^\ta_\nu F^\mu_{\phantom{\mu}\rho} - \delta^\mu_\rho
  F^\ta_{\phantom{\ta}\nu}.
\eeq
We see that these generators are precisely $K^a_{\phantom{a}b}$
introduced in \eqnref{IGL11}. To get the covariant derivatives under
conformal transformations, the parts in $F^\mu_{\phantom{\mu}\nu}$
corresponding to Lorentz transformations and dilations are projected
out by our specific choice of the connection $\Gamma^\ta_{\mu\nu}$. We
take the eigenvalue of $F^\nu_{\phantom{\nu}\nu}$ to be $(-l)$.
Putting \Eqnref{conf_connect} into \eqnref{gen_covd} we find
\beq
\begin{array}{rcl}
\D_\mu \psi &=& \pa_\mu \psi - F^\nu_{\phantom{\nu}\ta}
\Gamma^\ta_{\mu\nu} \psi \\ 
&=& \pa_\mu \psi - 2F^\nu_{\phantom{\nu}\ta}(\delta^\ta_\mu\phi_\nu +
\delta^\ta_\nu\phi_\mu\ - g_{\mu\nu}\phi^\ta) \psi \\
&=& \pa_\mu \psi + 2l\phi_\mu\psi + 2S_\mu^{\phantom{\mu}\ta}\phi_\ta
\psi, 
\end{array}
\eeq
where we have defined $S_{\mu\nu} \equiv F_{\mu\nu} - F_{\nu\mu}$.
This is exactly the covariant derivative found using nonlinear
realisation. To extend the covariant derivative for general
coordinate transformations, we have to introduce a vielbein, i.e., we
have to include the symmetric part of $F^\mu_{\phantom{\mu}\nu}$. This
is not possible using only the conformal group, since no appropriate
connection can be defined for this purpose.

We define the Riemann tensor \cite{Nakahara:2003th,Weinberg:1972}
\beq
R_{\mu\ta\nu}^{\phantom{\mu\ta\nu}\rho} = \pa_\mu\Gamma^\rho_{\ta\nu}
- \pa_\ta\Gamma^\rho_{\mu\nu} +
\Gamma^\si_{\ta\nu}\Gamma^\rho_{\mu\si} -
\Gamma^\si_{\mu\nu}\Gamma^\rho_{\ta\si}
\eeq
the Ricci tensor
\beq
R_{\mu\nu} = R_{\mu\ta\nu}^{\phantom{\mu\ta\nu}\ta}
\eeq
and the curvature scalar
\beq
R = g^{\mu\nu}R_{\mu\nu}.
\eeq
We can decompose the Ricci tensor into three linearly independent rank
two tensors
\beq
\begin{array}{rcl}
R_{(\mu\nu)} &=& \frac{1}{2}(R_{\mu\nu} + R_{\nu\mu}) \\ 
&=& 2\{ \pa_\nu\phi_\mu + \pa_\mu\phi_\nu - 4\phi_\mu\phi_\nu +
4g_{\mu\nu}(\phi_\ta\phi^\ta + \frac{1}{4}\pa_\ta\phi^\ta) \} \\
R_{[\mu\nu]} &=& \frac{1}{2}(R_{\mu\nu} - R_{\nu\mu}) = 2\{
\pa_\nu\phi_\mu - \pa_\mu\phi_\nu \} \\
g_{\mu\nu}R &=& 12g_{\mu\nu}(2\phi_\ta\phi^\ta + \pa_\ta\phi^\ta).
\end{array}
\eeq
For a non-Riemannian geometry, i.e., $\pa_\mu\si \neq 2\phi_\mu$, the
antisymmetric part of the Ricci tensor will not vanish. We can combine
these tensors to form covariant objects, e.g.,
\beq
\begin{array}{rcl}
\D_\mu \phi_\nu &=& \frac{1}{8}\left(2R_{(\mu\nu)} -
  \frac{1}{3}g_{\mu\nu}R - R_{[\mu\nu]}\right) \\
&=& \pa_\mu\phi_\nu + g_{\mu\nu}\phi_\ta \phi^\ta - 2\phi_\mu\phi_\nu
\end{array}
\eeq
which is the covariant derivative of $\phi_\mu$.

In conclusion, we get the same covariant derivatives of both preferred
fields and more general fields of degree $l$, despite if we use
nonlinear realisation techniques or explicitly define a metric tensor
and a connection based on the preferred fields.

\subsection{Simultaneous realisation}

Having now found the derivatives covariant under the G$_{11}$ group
and the conformal group separately, we want to construct quantities
covariant under both these groups simultaneously
\cite{West:2000ga,Borisov:1975bn}. The role the preferred field
$\sigma$ played in the conformal realisation makes it natural to
identify it with the diagonal elements of the metric
\beq
m_\mu^{\phantom{\mu}a} = \bar{m}_\mu^{\phantom{\mu}a} +
\si\delta_\mu^a, \hspace{1 cm} \bar{m}_\mu^{\phantom{\mu}\mu} = 0,
\eeq
where $m_\mu^{\phantom{\mu}a}$ is the Goldstone field defining the
vielbein from the G$_{11}$ realisation. The fields
$\bar{m}_\mu^{\phantom{\mu}a}$, $A_{c_1c_2c_3}$ and $A_{c_1\ldots
  c_6}$ transform under conformal transformations as their indices
suggest.

The covariant derivative of $A_{c_1c_2c_3}$ is according to
\eqnref{covd_A3} 
\beq
\begin{array}{rcl}
\tilde{D}_a A_{c_1c_2c_3} &=&
\bein^{\mu}_{\phantom{\mu}a}\{\pa_\mu A_{c_1c_2c_3} +
3(\bein^{-1}\pa_\mu\bein)_{[c_1}^{\phantom{[c_1}b}A_{|b|c_2c_3]}\} \\
&=& \bar{\bein}^{\mu}_{\phantom{\mu}a}\{ e^{-\si}\pa_\mu
A_{c_1c_2c_3} + 3e^{-\si}[
(\bar{\bein}^{-1}\pa_\mu\bar{\bein})_{[c_1}^{\phantom{[c_1}b}A_{|b|c_2c_3]}
\\ 
& & + \pa_\mu\si \delta_{[c_1}^{b}A_{|b|c_2c_3]} ] \},
\end{array}
\eeq
where $\bar{\bein}_{\mu}^{\phantom{\mu}a} =
(e^{\bar{m}})_{\mu}^{\phantom{\mu}a} =
(e^m)_{\mu}^{\phantom{\mu}a}e^{-\si}$. Knowing the covariant
derivative of an arbitrary field under conformal transformations
\eqnref{conf_covd} and assuming $\pa_\mu\si = 2\phi_\mu$, we find
\beq
\eqnlab{G11_conf}
\begin{array}{rcl}
\tilde{D}_a A_{c_1c_2c_3} &=&
\bar{\bein}^{\mu}_{\phantom{\mu}a}\{ \D_\mu A_{c_1c_2c_3} + 3e^{-\si}[
-\eta_{\mu[c_1}\pa^b\si A_{|b|c_2c_3]} \\
& & + \pa_{[c_1}\si A_{|\mu|c_2c_3]} + 
(\bar{\bein}^{-1}\pa_\mu\bar{\bein})_{[c_1}^{\phantom{[c_1}b}A_{|b|c_2c_3]}
+ \pa_\mu\si A_{c_1c_2c_3} ]\},
\end{array}
\eeq
where $A_{c_1c_2c_3}$ has the degree $l = 0$. Let us consider the
above equation at order $(\bar{m})^0$
\beq
\begin{array}{rcl}
\tilde{D}_a A_{c_1c_2c_3} &\approx& \D_a A_{c_1c_2c_3} + 3e^{-\si}[
-\eta_{a[c_1}\pa^b\si A_{|b|c_2c_3]} + \pa_{[c_1}\si A_{|a|c_2c_3]} \\
& & + \pa_a\si A_{c_1c_2c_3} ].
\end{array}
\eeq
The contributions from the last three terms vanishes only if we
completely antisymmetrise in the indices $a$, $c_1$, $c_2$ and $c_3$
\beq
\begin{array}{rcl}
\tilde{D}_{[a} A_{c_1c_2c_3]} &=& \D_{[a} A_{c_1c_2c_3]} + 3e^{-\si}[
-\eta_{[ac_1}\pa^b\si A_{|b|c_2c_3]} \\ 
& & + \pa_{[c_1}\si A_{ac_2c_3]} + \pa_{[a}\si A_{c_1c_2c_3]} ] \\
&=& \D_{[a} A_{c_1c_2c_3]}.
\end{array}
\eeq
The quantity simultaneously covariant under both the conformal group
and the G$_{11}$ group is
\beq
\begin{array}{rcl}
\tilde{F}_{c_1c_2c_3c_4} &\equiv& 4\tilde{D}_{[c_1} A_{c_2c_3c_4]} \\
&=& 4\{ \bein^{\mu}_{\phantom{\mu}[c_1} \pa_{|\mu|}
A_{c_2c_3c_4]} + 3\bein^{\mu}_{\phantom{\mu}[c_1} 
(\bein^{-1}\pa_{|\mu|}\bein)_{c_2}^{\phantom{c_2}b}A_{|b|c_3c_4]} \}
\end{array}
\eeq
A similar analysis for the field $A_{c_1\ldots c_6}$ yields a field
strength with seven indices
\beq
\begin{array}{rcl}
\tilde{F}_{c_1\ldots c_7} &\equiv& 7\tilde{D}_{[c_1} A_{c_2\ldots c_7]} \\
&=& 7\{ \bein^{\mu}_{\phantom{\mu}[c_1} \pa_{|\mu|}
A_{c_2\ldots c_7]} + 6\bein^{\mu}_{\phantom{\mu}[c_1} 
(\bein^{-1}\pa_{|\mu|}\bein)_{c_2}^{\phantom{c_2}b}A_{|b|c_3\ldots
  c_7]} \\
& & + 5\tilde{F}_{[c_1c_2c_3c_4}A_{c_5c_6c_7]} \}.
\end{array}
\eeq

For a matter field $\psi$ of degree $l=0$ the covariant derivatives
under the conformal and the G$_{11}$ group are
\beq
\begin{array}{rcll}
\tilde{D}_a \psi &=& \bein^\mu_{\phantom{\mu}a}\pa_\mu \psi +
\frac{1}{2} \om_{ab}^{\phantom{ab}c} S^b_{\phantom{b}c} \psi &
\textrm{the~G$_{11}$~group} \\
\D_\mu \psi &=& e^{-\si}(\pa_\mu +
S_\mu^{\phantom{\mu}\rho}\pa_\rho\si) \psi &
\textrm{the~conformal~group},
\end{array}
\eeq
respectively. Using $\bar{\bein}^{\mu}_{\phantom{\mu}a} =
\bein^{\mu}_{\phantom{\mu}a} e^\si$ we find
\beq
\eqnlab{sim_covd}
\begin{array}{rcl}
\tilde{D}_a \psi &=& \bar{\bein}^\mu_{\phantom{\mu}a}e^{-\si}
\pa_\mu \psi + \frac{1}{2}
\bar{\bein}^{\mu}_{\phantom{\mu}a}e^{-\si} \om_{\mu
  b}^{\phantom{\mu b}c} S^b_{\phantom{b}c} \psi \\
&=& \bar{\bein}^\mu_{\phantom{\mu}a}\{ \D_\mu -
e^{-\si}S_\mu^{\phantom{\mu}\rho}\pa_\rho\si + \frac{1}{2}
e^{-\si} \om_{\mu b}^{\phantom{\mu b}c} S^b_{\phantom{b}c} \} \psi,
\end{array}
\eeq
requiring the expression to be covariant also under the conformal
group will yield a constraint, putting the part in \Eqnref{sim_covd}
which is not covariant under the conformal group to zero. The
consistent constraint is
\beq
\eqnlab{connection_constraint}
\Omega_{a[bc]} - \Omega_{b(ac)} + \Omega_{c(ab)} = 0,
\eeq
which gives the relation between the spin connection and the vielbein
\eqnref{connect_vielbein} if we insert the definition $\Omega_{abc}
\equiv \bein^{\mu}_{\phantom{\mu}a} \bein^{\rho}_{\phantom{\rho}b}
\pa_\mu \bein_{\rho c} - \omega_{abc}$. We can also rewrite
$\Omega_{abc}$ as
\beq
\begin{array}{rcl}
\Omega_{abc} &=& \bar{\bein}^{\mu}_{\phantom{\mu}a}
  \bar{\bein}^{\rho}_{\phantom{\rho}b} \D_\mu \bar{\bein}_{\rho c} +
  e^{-\si}\bar{\bein}^{\mu}_{\phantom{\mu}a}
    \bar{\bein}^{\rho}_{\phantom{\rho}b}[ -\eta_{\mu\rho}\pa^d \si
    \bar{\bein}_{dc} - \eta_{\mu c}\pa^d\si \bar{\bein}_{\rho d} \\
& & + \pa_\rho \si \bar{\bein}_{\mu c} + \pa_c\si \bar{\bein}_{\rho\mu}]
    + \eta_{bc}\pa_a\si - \omega_{abc},
\end{array}
\eeq
where $\D_\mu \bar{\bein}_{\rho c}$ denotes the covariant derivative
of $\bar{\bein}_{\rho c}$ under the conformal group. Putting this into
\Eqnref{connection_constraint} gives the relation
\beq
\omega_{abc} - 2\eta_{a[b}\pa_{c]}\si =
\bar{\bein}^{\mu}_{\phantom{\mu}a}
\bar{\bein}^{\rho}_{\phantom{\rho}[b} \D_{|\mu} \bar{\bein}_{\rho| c]}
- \bar{\bein}^{\mu}_{\phantom{\mu}b}
\bar{\bein}^{\rho}_{\phantom{\rho}(a} \D_{|\mu} \bar{\bein}_{\rho| c)}
+ \bar{\bein}^{\mu}_{\phantom{\mu}c}
\bar{\bein}^{\rho}_{\phantom{\rho}(a} \D_{|\mu} \bar{\bein}_{\rho| b)},
\eeq
which we can use to write
\beq
\tilde{D}_a \psi = \bar{\bein}^\mu_{\phantom{\mu}a}\D_\mu \psi +
\frac{1}{2} \left\{ \bar{\bein}^{\mu}_{\phantom{\mu}a}
  \bar{\bein}^{\rho}_{\phantom{\rho}b} \D_{\mu}
  \bar{\bein}_{\rho}^{\phantom{\rho}c} -
  \bar{\bein}^{\mu}_{\phantom{\mu}b} 
  \bar{\bein}^{\rho c} \D_{\mu} \bar{\bein}_{\rho a} +
  \bar{\bein}^{\mu c} \bar{\bein}^{\rho}_{\phantom{\rho}a} \D_{\mu}
  \bar{\bein}_{\rho b} \right\} S^b_{\phantom{b}c} \psi.
\eeq
This shows that $\tilde{D}_a \psi$ is simultaneously covariant under
both the G$_{11}$ and the conformal group if the constraint
\eqnref{connection_constraint} is satisfied.

To identify with general relativity, we define the covariant
derivative of a vector $A_\mu = \bein_\mu^{\phantom{\mu}a}A_a$ as
\beq
\begin{array}{rcl}
D_\mu A_\nu &\equiv& \bein_\mu^{\phantom{\mu}a}
\bein_\nu^{\phantom{\nu}b} \tilde{D}_a A_b \\
&=& \bein_\mu^{\phantom{\mu}a} \bein_\nu^{\phantom{\nu}b} \left\{
\bein^\ta_{\phantom{\ta}a}\pa_\ta A_b +
\frac{1}{2}\omega_{ac}^{\phantom{ac}d}
(S^c_{\phantom{c}d})_b^{\phantom{b}f}A_f\right\} \\
&=& \pa_\mu A_\nu - (\bein^\rho_{\phantom{\rho}a}\pa_\mu
\bein_\nu^{\phantom{\nu}a} - \bein_\mu^{\phantom{\mu}a}
\bein_\nu^{\phantom{\nu}b} \bein^\rho_{\phantom{\rho}c}
\omega_{ab}^{\phantom{ab}c})A_\rho \\
&=& \pa_\mu A_\nu - \Gamma_{\mu\nu}^{\rho}A_\rho,
\end{array}
\eeq
where $(S^c_{\phantom{c}d})_b^{\phantom{b}f} =
(\eta^c_{\phantom{c}b}\eta_d^{\phantom{d}f} - \eta^{cf}\eta_{db})$ for
vectors. Using the relation between the spin connection and the
vielbein \eqnref{connect_vielbein}, we can show that
$\Gamma_{\mu\nu}^{\rho}$ is just the Christoffel symbol
\beq
\Gamma_{\mu\nu}^{\rho} = \bein^\rho_{\phantom{\rho}a}\pa_\mu
\bein_\nu^{\phantom{\nu}a} - \bein_\mu^{\phantom{\mu}a}
\bein_\nu^{\phantom{\nu}b} \bein^\rho_{\phantom{\rho}c}
\omega_{ab}^{\phantom{ab}c} = \frac{1}{2}g^{\rho\ta}(\pa_\mu
g_{\nu\ta} + \pa_\nu g_{\mu\ta} - \pa_\ta g_{\mu\nu})
\eeq
with $g_{\mu\nu} = \bein_\mu^{\phantom{\mu}a}
\bein_\nu^{\phantom{\nu}b} \eta_{ab}$ being the metric tensor.

The invariant equations of motion can now be formed out of the field
strengths, $\tilde{F}_{c_1c_2c_3c_4}$ and $\tilde{F}_{c_1\ldots c_7}$,
and the Riemann tensor
\beq
\begin{array}{rcl}
R_b^{\phantom{b}c} &=& d\om_b^{\phantom{b}c} + \om_b^{\phantom{b}d}
\om_d^{\phantom{d}c} = \frac{1}{2}dx^\mu \we dx^\nu R_{\mu\nu
  b}^{\phantom{\mu\nu b}c} \hspace{1 cm} \Rightarrow \\
R_{\mu\nu b}^{\phantom{\mu\nu b}c} &=& \pa_\mu\om_{\nu
  b}^{\phantom{\nu b}c} - \pa_\nu\om_{\mu b}^{\phantom{\mu b}c} +
\om_{\mu b}^{\phantom{\mu b}d} \om_{\nu d}^{\phantom{\nu d}c} - 
\om_{\nu b}^{\phantom{\nu b}d} \om_{\mu d}^{\phantom{\mu d}c}.
\end{array}
\eeq
The spin connection will enter only through these quantities. The
equations of motion have also to be covariant under local Lorentz
transformations. The only possible non-trivial equations are then
\beq
\tilde{F}_{c_1c_2c_3c_4} = \frac{1}{7!}
\varepsilon_{c_1\ldots c_{11}}\tilde{F}^{c_5\ldots c_{11}} 
\eeq
and
\beq
\begin{array}{rcl}
& & R_{\mu\nu b}^{\phantom{\mu\nu b}c} \bein_c^{\phantom{c}\nu}
\bein_a^{\phantom{a}\mu} - \frac{1}{2} \eta_{ab}R_{\mu\nu
  d}^{\phantom{\mu\nu d}c} \bein_c^{\phantom{c}\nu} \bein^{d\mu} = \\
&=& \frac{C}{4} \left( \tilde{F}_{ac_1c_2c_3}
  \tilde{F}_{b}^{\phantom{b}c_1c_2c_3} - \frac{1}{6}\eta_{ab}
  \tilde{F}_{c_1c_2c_3c_4} \tilde{F}^{c_1c_2c_3c_4} \right),
\end{array}
\eeq
where $C$ is a numerical constant. The first equation is precisely
\Eqnref{dualA}, while the second equation is the Einstein's equation
of gravity.

\subsection{Gauge symmetry}

The reducing matrix in the G$_{11}$ realisation 
\beq
g = e^{x^f P_f}e^{m_a^{\phantom{a}b}K^a_{\phantom{a}b}} \expo
\left(\frac{1}{3!}A_{c_1c_2c_3}R^{c_1c_2c_3} +
  \frac{1}{6!}A_{c_1\ldots c_6}R^{c_1\ldots c_6} \right), 
\eeq
transforms as $g \rightarrow g_0 g$ under a pure global
transformation. If we take
\beq
g_0 = \expo \left(\frac{1}{3!}c_{\mu_1\mu_2\mu_3}
  \delta_{a_1a_2a_3}^{\mu_1\mu_2\mu_3} R^{a_1a_2a_3} +
  \frac{1}{6!}c_{\mu_1\ldots\mu_6} \delta_{a_1\ldots a_6}^{\mu_1\ldots
    \mu_6}R^{a_1\ldots a_6} \right) 
\eeq
then
\beq
\begin{array}{rcl}
g \rightarrow g_0 g &=& e^{x^{f}P_{f}}
e^{m_a^{\phantom{a}b}K^a_{\phantom{a}b}} \expo\left\{ \right.
  \frac{1}{3!}(A_{a_1a_2a_3} + c_{a_1a_2a_3})R^{a_1a_2a_3} \\
& & + \frac{1}{6!}(A_{a_1\ldots a_6} + c_{a_1\ldots a_6} +
  20c_{[a_1a_2a_3}A_{a_4a_5a_6]})R^{a_1\ldots a_6} \left. \right\}.
\end{array}
\eeq
The variations of the Goldstone fields are
\beq
\begin{array}{rcl}
\delta h_a^{\phantom{a}b} &=& 0 \\
\delta A_{a_1a_2a_3} &=& c_{a_1a_2a_3} \\
\delta A_{a_1\ldots a_6} &=& c_{a_1\ldots a_6} +
20c_{[a_1a_2a_3}A_{a_4a_5a_6]}, \\
\end{array}
\eeq
where $c_{\mu_1\mu_2\mu_3} = \bein_{\mu_1}^{\phantom{\mu_1}a_1}
\bein_{\mu_2}^{\phantom{\mu_2}a_2} \bein_{\mu_3}^{\phantom{\mu_3}a_3}
c_{a_1a_2a_3}$. Using \Eqnref{SCT_infin} with $l=-p$, the fields with
curved indices will transform under special conformal transformations
according to
\beq
\begin{array}{rcl}
\delta A_{\mu_1\ldots\mu_p} &=& (2x^\nu\beta_\nu x^\mu\pa_\mu - x^2
\beta^\mu\pa_\mu) A_{\mu_1\ldots\mu_p} + 2p\beta^\mu x_\mu
A_{\mu_1\ldots\mu_p} \\
& & + 2p(\beta_{[\mu_1}x^{\rho}A_{|\rho|\mu_2\ldots\mu_p]} - x_{[\mu_1}
\beta^{\rho}A_{|\rho|\mu_2\ldots\mu_p]}),
\end{array}
\eeq
where $p = \{3,6\}$ and $A_{\mu_1\ldots\mu_p} \equiv
(e^{\bar{m}})_{\mu_1}^{\phantom{\mu_1}a_1} \ldots
(e^{\bar{m}})_{\mu_p}^{\phantom{\mu_p}a_p} e^{p\si} A_{a_1\ldots
  a_p}$.

We assume now that $\delta A_{\mu_1\ldots\mu_p} =
c_{\mu_1\ldots\mu_p}$. Defining $\Lambda^{(1)}_{\mu_2\ldots\mu_p} =
x^{\rho}c_{\rho\mu_2\ldots\mu_p}$, the transformations can be written
as 
\beq
\begin{array}{rcl}
\delta_c A_{\mu_1\ldots\mu_p} &=& c_{\mu_1\ldots\mu_p} =
\pa_{[\mu_1}\Lambda^{(1)}_{\mu_2\ldots\mu_p]} = 
\pa_{[\mu_1}x^{\rho}c_{|\rho|\mu_2\ldots\mu_p]} \\
\delta_\beta A_{\mu_1\ldots\mu_p} &=& (2x^\nu\beta_\nu x^\mu\pa_\mu -
x^2 \beta^\mu\pa_\mu) A_{\mu_1\ldots\mu_p} +
2p(\beta_{[\mu_1}x^{\rho}A_{|\rho|\mu_2\ldots\mu_p]} \\
& & - x_{[\mu_1} \beta^{\rho}A_{|\rho|\mu_2\ldots\mu_p]}) +
2p\beta^\mu x_\mu A_{\mu_1\ldots\mu_p}.
\end{array}
\eeq
Taking the commutation relation between these we get
\beq
\begin{array}{rcl}
\com{\delta_c,\delta_\beta}A_{\mu_1\ldots\mu_p} &=&
\delta_c(\delta_\beta A_{\mu_1\ldots\mu_p}) - \delta_\beta(\delta_c
A_{\mu_1\ldots\mu_p}) \\ 
&=& \delta_c\{ (2x^\nu\beta_\nu x^\mu\pa_\mu -
x^2 \beta^\mu\pa_\mu) A_{\mu_1\ldots\mu_p} +
2p(\beta_{[\mu_1}x^{\rho}A_{|\rho|\mu_2\ldots\mu_p]} \\
& & - x_{[\mu_1} \beta^{\rho}A_{|\rho|\mu_2\ldots\mu_p]}) +
2p\beta^\mu x_\mu A_{\mu_1\ldots\mu_p}\} - \delta_\beta
c_{\mu_1\ldots\mu_p} \\
&=& 2p(\beta_{[\mu_1}x^{\rho}c_{|\rho|\mu_2\ldots\mu_p]} - x_{[\mu_1}
\beta^{\rho}c_{|\rho|\mu_2\ldots\mu_p]} + \beta^\mu x_\mu
c_{\mu_1\ldots\mu_p}) \\
&=& \pa_{[\mu_1}\Lambda^{(2)}_{\mu_2\ldots\mu_p]} \equiv
\delta_{\Lambda^{(2)}} A_{\mu_1\ldots\mu_p}
\end{array}
\eeq
with
\beq
\Lambda^{(2)}_{\mu_2\ldots\mu_p} = (2px^\nu\beta_\nu x^\rho -
x^2\beta^\rho) c_{\rho\mu_2\ldots\mu_p} -
2(p-1)x_{[\mu_2}\beta^{\rho}x^\nu c_{|\nu\rho|\mu_3\ldots\mu_p]}.
\eeq
We have in the calculations used that $\pa_\mu c_{\mu_1\ldots\mu_p} =
0$ and $\delta_\beta c_{\mu_1\ldots\mu_p} = 0$.

Generally $\delta_\beta(\delta_{\Lambda^{(r)}} A_{\mu_1\ldots\mu_p}) =
0$, which results in 
\beq
\begin{array}{rcl}
\com{\delta_{\Lambda^{(r)}},\delta_\beta}A_{\mu_1\ldots\mu_p} &=&
\delta_{\Lambda^{(r)}}(\delta_\beta A_{\mu_1\ldots\mu_p}) \\
&=& (2x^\nu\beta_\nu x^\mu\pa_\mu - x^2 \beta^\mu\pa_\mu)
\pa_{[\mu_1}\Lambda^{(r)}_{\mu_2\ldots\mu_p]} + \\
& & + 2p(\beta_{[\mu_1}x^{\rho}
\pa_{|\rho|}\Lambda^{(r)}_{\mu_2\ldots\mu_p]} - x_{[\mu_1}
\beta^{\rho} \pa_{|\rho|}\Lambda^{(r)}_{\mu_2\ldots\mu_p]}) \\
& & + 2p\beta^\mu x_\mu \pa_{[\mu_1}\Lambda^{(r)}_{\mu_2\ldots\mu_p]}
\\
&=& \pa_{[\mu_1}\Lambda^{(r+1)}_{\mu_2\ldots\mu_p]} \equiv
\delta_{\Lambda^{(r+1)}} A_{\mu_1\ldots\mu_p},
\end{array}
\eeq
where
\beq
\begin{array}{rcl}
\Lambda^{(r+1)}_{\mu_2\ldots\mu_p} &=& \{2x^\nu\beta_\nu x^\mu\pa_\mu -
x^2 \beta^\mu\pa_\mu +
2(p-1)x^\nu\beta_\nu\}\Lambda^{(r)}_{\mu_2\ldots\mu_p} \\ 
& & + 2(p-1)\{
\beta_{[\mu_2}x^{\rho}\Lambda^{(r)}_{|\rho|\mu_3\ldots\mu_p]} -
x_{[\mu_2} \beta^{\rho}\Lambda^{(r)}_{|\rho|\mu_3\ldots\mu_p]}\}
\end{array}
\eeq
is defined recursively. It is then clear that one can obtain gauge
transformations $\Lambda^{(r)}_{\mu_2\ldots\mu_p}$ of
$A_{\mu_1\ldots\mu_p}$ at arbitrary order $r$ in $x^\mu$, by taking
repeated commutators with the special conformal transformations. We
have thus shown that simultaneous realisation under the G$_{11}$ and
the conformal group leads to local gauge transformations of the
fields.

\chapter{Lorentzian Kac-Moody algebras}
\chlab{algebra}

The field content of the eleven dimensional supergravity suggests that
M-theory possesses some rank eleven symmetry algebra. There are also
mounting evidence that Kac-Moody algebras of infinite type and
generalised Kac-Moody (Borcherds) algebras might appear in the guise
of duality symmetries in string and M-theory. It is therefore
conjectured that M-theory includes a rank eleven Lorentzian Kac-Moody
symmetry denoted E$_{11}$. Some fragments of this symmetry have indeed
been found in the maximal supergravities in ten and eleven dimensions.
Before we continue with eleven dimensional supergravity, we will in
this chapter look at the general theory of Lorentzian Kac-Moody
algebras mainly following the reference \cite{Gaberdiel:2002db}. Some
special attention will then be given to the group E$_{11}$. Since all
the hyperbolic Kac-Moody algebras have a real principal so(2,1)
subalgebra, it is interesting to see the constraints obtained by
imposing the existence of an so(2,1) subalgebra on the Lorentzian
Kac-Moody algebras.

\section{General theory}

A finite dimensional simple Lie algebra \textbf{g} is defined through
its Cartan matrix \cite{Fuchs:1997jv}
\beq
\eqnlab{cartan}
A_{ij} = \frac{2}{(\alpha_i,\alpha_i)} (\alpha_i,\alpha_j),
\eeq
where $\alpha_i$ are the simple roots of the algebra, this definition
implies automatically that $A_{ii}=2$. The fundamental weights are
defined by
\beq
\eqnlab{root_weight}
\frac{2}{(\alpha_i,\alpha_i)} (\alpha_i,\la_j) = \delta_{ij},
\eeq
using which we can express the inverse of the Cartan matrix as
\beq
A^{-1}_{ij} = \frac{2}{(\alpha_i,\alpha_i)} (\la_i,\la_j).
\eeq
Since the simple roots constitute a basis for the root lattice, the
highest root of a finite dimensional Lie algebra can be written as
\beq
\theta = \displaystyle\sum_{i=1}^{r}n_i\alpha_i,
\eeq
where $r$ is the rank of the algebra and $n_i$ are called Kac labels 
\beq
n_i = \frac{2}{(\alpha_i,\alpha_i)} (\theta,\la_i).
\eeq
The Coxeter number is defined as
\beq
h(\textbf{g}) \equiv 1 + \displaystyle\sum_{i=1}^{r}n_i
\eeq
with $\sum_{i=1}^{r}n_i$ being the height of the highest root. In the
same way as the simple roots form a basis for the root lattice, the
fundamental weights span the weight lattice and we define the Weyl
vector by
\beq
\rho \equiv \displaystyle\sum_{i=1}^{r}
\frac{2}{(\alpha_i,\alpha_i)}\la_i. 
\eeq

We can now generalise the definitions to a Kac-Moody algebra, which is
specified through a Cartan matrix satisfying \cite{Kac:1990gs}
\beq
\begin{array}{rcl}
A_{ii} &=& 2 \\
A_{ij} &=& \textrm{negative integers or zero, } i \neq j \\
A_{ij} = 0 &\Rightarrow& A_{ji} = 0.
\end{array}
\eeq
In addition to these criteria, we will here consider only symmetric
generalised Cartan matrices, i.e., $A_{ij}=A_{ji}$. The assumption
that $A_{ij}$ is symmetric means that we have chosen all the simple
roots to be space-like with $(\alpha_i,\alpha_i)=2$. If the Cartan
matrix is positive definite, the algebra is finite dimensional. If the
Cartan matrix is positive semi-definite, the algebra is of affine
type. Moreover, we call the Kac-Moody algebra Lorentzian, if the
Cartan matrix is non-singular and possesses precisely one negative
eigenvalue. Among the Lorentzian algebras, the hyperbolic ones are
characterised by the property that the entries of the associated
inverse Cartan matrix have a definite sign
\cite{Ruuska:1991ne,Nicolai:2001ir}, $A_{ij}^{-1} \le 0$. In the case
of finite dimensional or affine algebras, the algebra is referred as
simply-laced if the Cartan matrix is symmetric, and the only allowed
values for the off-diagonal elements are $0$ and $(-1)$. All the roots
are then of the same length, and the simple roots will span an even
lattice $\La_R(A)$. 

The generators of the Kac-Moody algebra can be divided into the
generators of the Cartan subalgebra $H_i$ and the generators of the
positive roots $E_\alpha$ and the negative roots $F_\alpha$. The roots
are defined as eigenvectors of the Cartan generators under the adjoint
action. All the root generators can be obtained from those
corresponding to the simple roots ($E_i,F_i$) using the Lie brackets.
The generators are taken to obey the Chevalley-Serre relations
\beq
\eqnlab{CSrelations}
\begin{array}{rcl}
\com{H_i,H_j} &=& 0 \\
\com{H_i,E_j} &=& A_{ij}E_j \\
\com{H_i,F_j} &=& -A_{ij}F_j \\
\com{E_i,F_j} &=& \delta_{ij}H_i \\
(\textrm{ad}_{X_i})^{1-A_{ij}} X_j &=& 0 \hspace{0.5 cm} \textrm{ for }
i\neq j,
\end{array}
\eeq
where $X_i$ is either $E_i$ or $F_i$, and $(\textrm{ad}_{X_i})$ denotes
the adjoint representation of the generator $X_i$. The Chevalley-Serre
relations are invariant under the Cartan involution
\beq
E_\alpha \rightarrow -F_\alpha, \hspace{0.5 cm} F_\alpha \rightarrow
-E_\alpha, \hspace{0.5 cm} H_\alpha \rightarrow -H_\alpha,
\eeq
with $\alpha$ being any positive root. The generators $(E_\alpha -
F_\alpha)$ will then be even under the Cartan involution and form the
maximal compact subalgebra of the original Kac-Moody algebra. The
remaining generators $(E_\alpha + F_\alpha)$ and $H_\alpha$ are odd
under the involution.

The information contained in the Cartan matrix can also be presented
as an unoriented graph with $r$ nodes called a Dynkin diagram. Each
simple root is represented by a node, and the number of links between
the nodes $i$ and $j$ equals $\max\{|A_{ij}|,|A_{ji}|\}$. The Dynkin
diagram of a Lorentzian Kac-Moody algebra is a connected diagram
possessing at least one node whose deletion yields a diagram whose
connected components are of finite type except for at most one of
affine type \cite{Gaberdiel:2002db}. The Dynkin diagram of a
hyperbolic Kac-Moody algebra, on the other hand, is a connected
diagram such that deletion of any one node leaves a (possibly mutually
disconnected) set of connected Dynkin diagrams, each of which is of
finite type except for at most one of affine type.

Let $C$ denote the overall Dynkin diagram, yielding the reduced
diagram $C_R$ after having deleted one node. The node deleted is
called the central node, and the reduced diagram can possibly contain
mutually disconnected diagrams $C_1, C_2, \ldots, C_n$, see
\Figref{GeneralDynkin}.
\begin{figure}
  \begin{center}
    \includegraphics[scale=0.5]{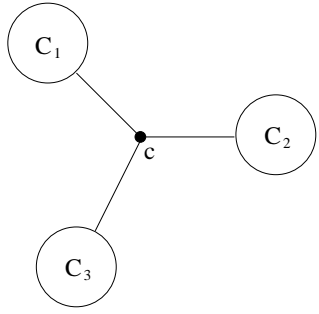}
    \caption{\textit{The Dynkin diagram $C$ of a Kac-Moody
        algebra. After having deleted the central node, the reduced
        diagram will consist of three disconnected diagrams.}}
    \figlab{GeneralDynkin}
  \end{center}
\end{figure}
The Cartan matrix of $C_R$ is obtained from the Cartan matrix of $C$
by deleting the row and column corresponding to the central node. The
number of links between the central node $c$ and node $i$ is denoted
by
\beq
\eta_i = -A_{ci},
\eeq
where $A_{ij}$ is the Cartan matrix of $C$ and is symmetric by
assumption. Since the Cartan matrix is non-singular, we can write the
fundamental weights as
\beq
\la_i = \displaystyle\sum_{j=1}^{r} (A^{-1})_{ij}\alpha_j
\eeq
satisfying \Eqnref{root_weight}. For hyperbolic algebras all the
fundamental weights lie inside the same light-cone, since $A^{-1}_{ij}
= \frac{2}{(\alpha_i,\alpha_i)} (\la_i,\la_j)$ are all negative or
zero. For Lorentzian algebras which are not of hyperbolic type, some
of the fundamental weights have to be space-like. We can now form a
weight lattice $\Lambda_w(A)$, which is dual to the root lattice,
using the fundamental weights. It is interesting to form the quotient
\beq
\eqnlab{conjugacy_class}
\Lambda_w(A)/\Lambda_R(A) = Z(A),
\eeq
where $Z(A)$ is a finite abelian group containing $|Z(A)|$ elements
and relates to the Cartan matrix of $C$ via
\beq
\eqnlab{detA_conjclass}
\det A = \pm |Z(A)|.
\eeq
We can also define the Weyl vector $\rho$
\beq
\rho = \displaystyle\sum_{i=1}^{r}\frac{2}{(\alpha_i,\alpha_i)} \la_i
= \displaystyle\sum_{i,j=1}^{r}(A^{-1})_{ij}
\frac{2\alpha_j}{(\alpha_j,\alpha_j)} 
\eeq
satisfying $(\rho,\alpha_i)=1$. Whether all the coefficients
$\sum_{i=1}^{r}(A^{-1})_{ij}$ have the same sign or not is crucial for
the existence of a principal so(2,1) subalgebra. In the finite case
the coefficients $\sum_{i=1}^{r}(A^{-1})_{ij}$ are all positive, while
they are all negative in the hyperbolic case. In the Lorentzian case
mixed signs are possible, but at least one of them is negative. More
about real principal so(2,1) subalgebras can be found in
Appendix~\chref{principal}.

\section{The root system}

We will now construct the simple roots of a rank $r$ Dynkin diagram
$C$ of Lorentzian type, by using the simple roots of the corresponding
reduced Dynkin diagram $C_R$.

\subsection{The reduced algebra is finite}

Let us first assume that the reduced diagram corresponds to a finite
dimensional Lie algebra. Since $C_R$ is of finite type, the simple
roots $\alpha_1, \alpha_2, \ldots, \alpha_{r-1}$ are linearly
independent and span a $(r-1)$-dimensional Euclidean space. We let
these roots also be the simple roots of the overall diagram $C$, and
then add the central node to the set of simple roots. The central node
can be taken as
\beq
\alpha_c = -\nu + x, \hspace{0.5 cm} \textrm{with } \nu =
\displaystyle\sum_{i=1}^{r-1} \eta_i\la_i =
-\displaystyle\sum_{i=1}^{r-1} A_{ci}\la_i
\eeq
satisfying the constraints $(\alpha_c,\alpha_i)=A_{ci}$. We have used
$\la_i$ to denote the fundamental weights of the reduced diagram
$C_R$, while $x$ is a direction orthogonal to the root space of $C_R$.
Demanding the central node to be space-like leads to
\beq
2 = A_{cc} = (\alpha_c,\alpha_c) = \nu^2 + x^2,
\eeq
which determines the sign of $x^2$. The root space of $C$ is Euclidean
or Lorentzian depends on whether $x^2$ is positive or negative, this
is because we have assumed the root space of $C_R$ to be Euclidean. If
$x^2$ vanishes, the root space will have a positive semi-definite
metric, and thus corresponds to an affine Lie algebra.

We use $l_c, l_1, l_2, \ldots, l_{r-1}$ to denote the fundamental
weights of $C$, and similarly let $\la_1, \la_2, \ldots, \la_{r-1}$
denote the fundamental weights of $C_R$. These two sets are related to
each other by
\beq
l_i = \la_i + \frac{(\nu,\la_i)}{x^2} x, \hspace{0.5 cm} l_c =
\frac{1}{x^2} x
\eeq
providing that $x^2$ is non-vanishing, i.e., the overall diagram is not
of affine type. The overall Weyl vector is then given by
\beq
R = \displaystyle\sum_{i=c}^{r-1} l_i = \rho + \frac{[1 +
  (\nu,\rho)]}{x^2} x,
\eeq
where $\rho = \sum_{i=1}^{r-1}\la_i$ is the Weyl vector for the
reduced diagram $C_R$, noting that $(\alpha_i,\alpha_i)=2$ for
symmetric Cartan matrices. Taking the scalar product of the Weyl
vector with the fundamental weight $l_c$
\beq
(R,l_c) = \frac{[1 + (\nu,\rho)]}{x^2},
\eeq
we see that at least one of the coefficients in the expansion of the
Weyl vector $R$ in terms of simple roots is negative if $x^2$ is
chosen negative, since $(\nu,\rho)$ is positive when the reduced
diagram is of finite type.

\subsection{The reduced algebra is finite or affine}

Assume now the connected components of the reduced diagram are allowed
to be of both finite and affine type. Since the Cartan matrix $B$ for
the reduced diagram $C_R$ is obtained from the Cartan matrix $A$ for
the overall diagram $C$, we can relate the determinants of these two
matrices by
\beq
\eqnlab{A_subA}
\det A = x^2 \det B = (2-\nu^2)\det B.
\eeq
This can be shown by using the definition $A_{ij} =
\frac{2}{(\alpha_i,\alpha_i)}(\alpha_i,\alpha_j)$, writing out the
simple roots in an orthogonal basis and observing that $x$ is
orthogonal to all the simple roots of $C_R$. Noting $(B^{-1})_{ij} =
\frac{2}{(\alpha_i,\alpha_i)}(\la_i,\la_j)$ we find
\beq
\eqnlab{detA_calc}
\begin{array}{rcl}
\det A &=& \left( 2 - \displaystyle\sum_{i,j=1}^{r-1} \eta_i
  (B^{-1})_{ij} \eta_j \right) \det B \\
&=& 2\det B - \displaystyle\sum_{i,j=1}^{r-1} \eta_i (\textrm{adj }
  B)_{ij} \eta_j,
\end{array}
\eeq
where the adjugate matrix $(\textrm{adj }B)_{ij} = \det B
(B^{-1})_{ij}$ is the matrix of cofactors for $B$. If $B$ is singular
we just drop the first term on the right hand side. When $C_R$ is
disconnected with the connected components $C_1, \ldots, C_n$, the
Cartan matrix $B$ can be written in a block diagonal form
\beq
B = \textrm{diag}(B_1,B_2,\ldots,B_n),
\eeq
where $B_\beta$ is the Cartan matrix of the $\beta$:th component
$C_\beta$. Adopting the notation $\Delta_\beta = \det B_\beta$,
\Eqnref{detA_calc} becomes
\beq
\eqnlab{detA_reduced}
\det A = \Delta_1\ldots\Delta_n \left( 2 -
  \displaystyle\sum_{\beta=1}^{n} \left\{
    \displaystyle\sum_{i,j\in{C_\beta}} \frac{\eta_i (\textrm{adj
        }B_\beta)_{ij} \eta_j}{\Delta_\beta} \right\} \right),
\eeq
allowing evaluation of determinants of Cartan matrices iteratively.
For consistency we put $\Delta_n = 1$ and $\textrm{adj}B_n = 0$ if the
diagram is empty. Also, if $p$ of the connected components of $C_R$
correspond to affine algebras, then $\det A$ will have a $(p-1)$-fold
zero. Unless $p=1$ the overall algebra is neither affine nor
Lorentzian, since it will have one negative and $(p-1)$ zero
eigenvalues.

Using \Eqnref{detA_reduced} on the Dynkin diagram for the exceptional
algebras e$_N$ and choosing the tip of the shortest leg to be the
central node, one finds
\beq
\det A(\textrm{e}_N) = 9 - N.
\eeq
This indicates that e$_N$ is Lorentzian for $N\ge 10$.

We construct now the simple roots of $C$ by using the simple roots of
each connected components of $C_R$. If $C_\beta$ is of finite type,
the simple roots $\alpha_i$ belonging to this diagram are linearly
independent, whereas in the case of an affine diagram the simple roots
are linearly dependent satisfying
\beq
\displaystyle\sum_{i\in{C_\beta}} n_i\alpha_i = 0.
\eeq
The positive integers $n_i$ are the Coxeter labels for the affine
diagram $C_\beta$. The simple roots of the overall diagram are given
by
\beq
\begin{array}{rcl}
a_i &=& \alpha_i + \eta_i k = \alpha_i - A_{ci}k, \hspace{0.5 cm}
i\in{C_R} \\
a_c &=& -(k + \bar{k}),
\end{array}
\eeq
where the vectors $k$ and $\bar{k}$ span the even self-dual Lorentzian
lattice $\Pi_{1;1}$ satisfying $k^2 = \bar{k}^2 = 0$ and $(k,\bar{k})
= 1$, see Appendix~\chref{lattice}.

Let us concentrate on the Lorentzian Lie algebras, i.e., algebras
which when deleting the central node yields only one affine component
and the rest of finite type. Let also the affine component of the
reduced diagram be $C_1$, this means that $\Delta_1 = \det B_1 = 0$
and \Eqnref{detA_reduced} becomes
\beq
\eqnlab{detA_affine_calc}
\det A = -\Delta_2\ldots\Delta_n \displaystyle\sum_{i,j\in{C_1}}
\eta_i (\textrm{adj }B_1)_{ij} \eta_j.
\eeq
Note that by doing so the dependence on $\eta_i = -A_{ci}$ for
$i\not\in{C_1}$ has disappeared. Since $C_1$ is a connected diagram
corresponding to a simply-laced affine algebra $\g^{(1)}$, its
adjugate matrix can be written as
\beq
\eqnlab{adjB_affine}
(\textrm{adj }B_1)_{ij} = |Z(A(\g))| n_i n_j,
\eeq
where $\g^{(1)}$ is the affine extension of $\g$. \Eqnref{adjB_affine}
is motivated by the fact that $\sum_j (B_1)_{ij} (\textrm{adj
}B_1)_{jk} = \delta_{ik} \det B_1 = 0$ and $\sum_j B_{ij} n_j = 0$ for
an affine algebra. The specific form of \Eqnref{adjB_affine} comes
about due to that $(\textrm{adj }B_1)_{ij}$ is symmetric, and
$(\textrm{adj }B_1)_{00} = \det A(\g) = |Z(A(\g))|$ according to
\Eqnref{detA_conjclass}. Inserting \Eqnref{adjB_affine} into
\eqnref{detA_affine_calc} gives
\beq
\eqnlab{detA_affine}
\det A = -\Delta_2\ldots\Delta_n |Z(A(\g))| \left(
  \displaystyle\sum_{i\in{C_1}} n_i \eta_i \right)^2,
\eeq
and thus $\det A$ is explicitly negative.

It is of particular interest to study root lattices that are even,
integral and self-dual Lorentzian. All the Cartan matrices considered
so far guarantee an even, integral Lorentzian lattice. For the
self-duality, the determinant of the Cartan matrix has to equal
$(-1)$.  From \Eqnref{detA_affine} we see that imposing this leads to
the condition that all the connected parts $C_2, \ldots, C_n$ have to
give rise to self-dual lattices, i.e., $\Delta_i = 1$ for $i =
2,\ldots,n$, while $C_1$ must be the affine extension of an algebra
with self-dual lattice. Since the only finite dimensional algebra
having self-dual lattice is e$_8$, each diagram $C_i$ is corresponding
to an e$_8$ algebra except for $C_1$, which corresponds to the affine
algebra e$_9$. Finally the sum $\sum_{i\in{C_1}} n_i\eta_i$ must equal
unity also, meaning that all the $\eta_i$ in $C_1$ must vanish except
the one belonging to the node with Coxeter label equals one, i.e., the
affine node. These unique Lorentzian self-dual lattices are denoted by
$\Pi_{8n+1;1}$ and have the dimensions $(8n+2)$, see
\Figref{SelfDual}. When $n\ge2$ there exist many Cartan matrices
corresponding to the same lattice, since we are still free to choose
the numbers $\eta_i, i\not\in{C_1}$.
\begin{figure}
  \begin{center}
    \includegraphics[scale=0.5]{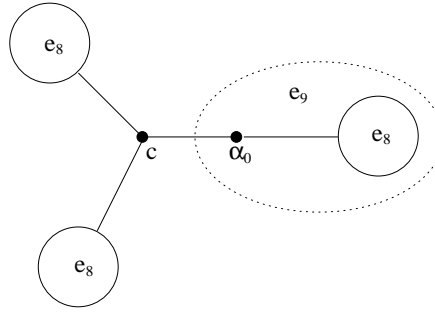}
    \caption{\textit{The Dynkin diagram of a Kac-Moody algebra with
        self-dual Lorentzian root lattice. The root lattice in this
        example will be $\Pi_{25;1}$.}}
    \figlab{SelfDual}
  \end{center}
\end{figure}

To simplify notations we include all the diagrams of finite type in
$C_2$, which now is allowed to be disconnected. The fundamental
weights of the overall diagram $C$ are
\beq
\begin{array}{rcl}
l_\beta &=& \la_\beta, \hspace{0.5 cm} \beta\in{C_2} \\
l_c &=& -k \\
l_i &=& \la_i - \frac{n_i}{\eta}(k - \bar{k} + \nu), \hspace{0.5 cm}
i\in{C_1},
\end{array}
\eeq
where the vectors $k$ and $\bar{k}$ again span the lattice
$\Pi_{1;1}$, and $\la_\beta$ are the fundamental weights of $C_2$. The
algebra corresponding to $C_1$ is the untwisted affine extension of
$\g$, for which $\la_i$ are the fundamental weights. We also define
\beq
\eqnlab{nu_eta}
\begin{array}{rcl}
\nu &=& \nu(\g) + \nu(C_2) \equiv \displaystyle\sum_{i=1}^{r(\g)}
\eta_i\la_i + \displaystyle\sum_{\beta\in{C_2}} \eta_\beta\la_\beta \\
\eta &\equiv& \displaystyle\sum_{i\in{C_1}} n_i\eta_i,
\end{array}
\eeq
with $r(\g)$ being the rank of $\g$. The Weyl vector will be 
\beq
R(C) = \rho(C_2) + \rho(\g) - \frac{h(\g)}{\eta}(k-\bar{k}+\nu) - k,
\eeq
with $\rho(C_2)$ and $\rho(\g)$ being the Weyl vectors of $C_2$ and
$\g$ respectively, and the Coxeter number is given by
\beq
h(\g) = \displaystyle\sum_{j=0}^{r(\g)}n_j.
\eeq

\subsection{Real principal so(2,1) subalgebra}

Taking the scalar product 
\beq
(R,l_c) = -\frac{h(\g)}{\eta} < 0,
\eeq
we see that the only three dimensional principal subalgebra allowed is
so(2,1), but for the algebra to really contain a real principal
so(2,1) subalgebra the Weyl vector squared
\beq
\eqnlab{Weyl_squared}
R^2 = [\rho(C_2) - \frac{h(\g)}{\eta}\nu(C_2)]^2 + [\rho(\g) -
\frac{h(\g)}{\eta}\nu(\g)]^2 - \frac{2h(\g)[h(\g)+\eta]}{\eta^2}
\eeq
must be negative \cite{Gaberdiel:2002db,Nicolai:2001ir}.

Let us first consider the case where the central node is linked to the
affine node by a single link, i.e., $\eta_i = \delta_{i0}$ for
$i\in{C_1}$. \Eqnref{Weyl_squared} is then simplified to 
\beq
R^2 = \rho(\g)^2 - 2h(\g)[h(\g)+1] + [\rho(C_2) - h(\g)\nu(C_2)]^2, 
\eeq
since $\eta=1$ and $\nu(\g)=0$. Using the Freudenthal-de Vries strange
formula \cite{Fuchs:1997jv} on the simply-laced finite algebra $\g$
\beq
\rho(\g)^2 = \frac{h(\g)[h(\g)+1]r(\g)}{12},
\eeq
we finally arrive at
\beq
R^2 = \frac{h(\g)[h(g)+1][r(\g)-24]}{12} + [\rho(C_2)-h(\g)\nu(C_2)]^2.
\eeq
A minimal condition for the existence of principal so(2,1) subalgebra
is thus given by that the rank of $\g$ has to be less than 24,
allowing only a finite number of possibilities for $\g$. If $C_2$ is
empty, that is if $C$ is an over-extension of $\g$, then the condition
$R^2<0$ reduces to $r(\g)<24$.

Consider now the remaining configurations where the central node is
linked by a single link to the node $\star$ in $C_1$ such that
$n_{\star}\ge2$, i.e., $\eta_i = \delta_{i\star}$ for
$i,\star\in{C_1}$. Note that all the nodes with $n_i = 1$ are related
by the symmetry of the affine Dynkin diagram, hence these nodes are
equivalent. The only simply-laced semi-simple finite algebras
containing Coxeter labels greater than one are e$_6$, e$_7$, e$_8$ and
d$_r$, so $C_1$ has to be the affine extension of one of these
algebras. Calculations with e$_6$, e$_7$ and e$_8$ show that the sum
of the last two terms in \Eqnref{Weyl_squared} is negative, thus the
sign of $R^2$ is depending on the choice of $C_2$. Similarly, one
finds for d$_r$ that the sum of the last two terms in
\Eqnref{Weyl_squared} is negative only if $r<26$. The conclusion is
then that whenever $C_1$ is linked to the central node by a single
link, imposing the overall algebra to contain a principal so(2,1)
subalgebra, i.e., $R^2<0$, the number of possibilities to choose $C_1$
is finite.

Having found the restrictions on $C_1$ for the overall algebra to
contain a principal so(2,1) subalgebra, we then try to find
corresponding upper bounds for the rank of $C_2$ assuming the central
node is linked to the node $\star$ in $C_1$ via a single link.

First we observe the condition $R^2<0$ can be written as
\beq
\eqnlab{R2_neg}
[\rho(C_2)-h_0\nu(C_2)]^2 \le M_0(\g),
\eeq
where $h_0 \equiv \frac{h(\g)}{\eta}$ and $M_0(\g) \equiv
\frac{2h(\g)[h(\g)+\eta]}{\eta^2} - [\rho(\g)-h_0\nu(\g)]^2$ only
depend on $\g$. It can be verified explicitly that $h_0 =
\frac{1}{n_\star}\sum_{j=0}^{r(\g)}n_j \ge 2$ for all the simply-laced
Lie algebras $\g$. If $C_2$ is not connected, we can split the left
hand side of \Eqnref{R2_neg} into a sum over the connected simple
components, where for each component the factor
$[\rho(C_2^i)-h_0\nu(C_2^i)]^2$ is strictly positive. It is then clear
that the number of simple components in $C_2$ is finite for a given
$\g$. What remains now is to show that the rank of each simple
component must be bounded, and we do this for the special cases where
$C_2$ is a$_r$ or d$_r$. In the rest of this section we use the
notations $\rho = \rho(C_2)$ and $\nu = \nu(C_2)$.

\subsubsection{The case of a$_r$}

For the Lie algebras a$_r$, the vector
$\nu=\displaystyle\sum_{\beta\in{\textrm{a}_r}}\eta_\beta \la_\beta$
can be written in an orthogonal basis $\{e_i\}$ as
\beq
\nu = \displaystyle\sum_{i=1}^{r+1}l_i e_i,
\eeq
where the coefficients obey $l_{i+1}-l_i = -\eta_i$. Writing
$(\rho-h_0\nu)$ in the same basis we get
\beq
(\rho-h_0\nu) = \displaystyle\sum_{i=1}^{r+1}\left( \frac{r+2-2i}{2} -
  h_0l_i \right)e_i = \displaystyle\sum_{i=1}^{r+1}m_i e_i,
\eeq
since $\rho = \displaystyle\sum_{i=1}^{r+1}\frac{r+2-2i}{2}e_i$ for
a$_r$ \cite{Fuchs:1997jv}. Observing that $m_{i+1}-m_i = h_0\eta_i-1$,
with $h_0\ge2$ and $\eta_i\in{\mathbb N_0}$ we find
$|m_{i+2}|\ge|m_i|\ge\frac{1}{2}$ by induction. This can then be used
to give
\beq
(\rho-h_0\nu)^2 = \displaystyle\sum_{i=1}^{r+1}|m_i|^2 \ge
\frac{r}{8}, 
\eeq
putting which into \Eqnref{R2_neg} yields an upper bound
$\frac{r}{8}\le M_0(\g)$ for the rank of allowed a$_r$.

\subsubsection{The case of d$_r$}

Assume first that $\nu$ is not a spinor weight. By analogy with a$_r$,
we use $\nu = \displaystyle\sum_{i=1}^{r}l_i e_i$ and $\rho =
\displaystyle\sum_{i=1}^{r}(r-i)e_i$ for d$_r$ \cite{Fuchs:1997jv} to
obtain
\beq
(\rho-h_0\nu) = \displaystyle\sum_{i=1}^{r}(r-i-h_0l_i)e_i.
\eeq
Since all the coefficients of $e_i$ are integers and at most
$\frac{r}{h_0}+1$ of them vanish, we get
\beq
(\rho-h_0\nu)^2 = \displaystyle\sum_{i=1}^{r}(r-i-h_0l_i)^2 \ge
\displaystyle\sum_{k=1}^{r-(\frac{r}{h_0}+1)}1 = r(1-\frac{1}{h_0})-1.
\eeq
If $\nu$ is a spinor weight, each $l_i$ is then half-odd-integer. The
calculations will remain the same, but with $h_0$ replaced by $h_0/2$
when $h_0$ is even. The rank of $C_2$ is thus again bounded. 

For the case where the central node is linked to the affine node of
$C_1$ only, and is connected to one node of each simple component of
$C_2$ by a single link, one can show that the rank of the overall
algebra with principal so(2,1) subalgebra can never exceed 42,
regardless the choices of $C_1$ and $C_2$ \cite{Gaberdiel:2002db}.

\section{Very extended Lie algebras}

In this section we will look at the special class of Lorentzian
Kac-Moody algebras, obtained by joining an affine Kac-Moody algebra
$\g^{(1)}$ via the affine node to a central node that links in turn to
the single finite dimensional Lie algebra a$_1$. The resulting algebra
is called very extended, see \Figref{VeryExtended}.
\begin{figure}
  \begin{center}
    \includegraphics[scale=0.5]{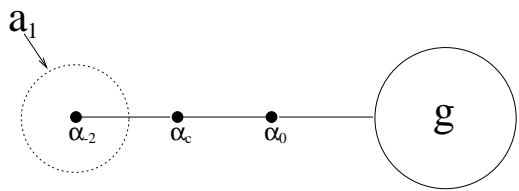}
    \caption{\textit{The Dynkin diagram of a very extended Kac-Moody
        algebra.}} 
    \figlab{VeryExtended}
  \end{center}
\end{figure}

We start with a finite dimensional semi-simple Lie algebra $\g$ of
rank $r$, and from now on we do not require that $\g$ has a symmetric
Cartan matrix. The simple roots of $\g$ are denoted $\alpha_i$,
$i=1,\ldots,r$. These are normalised such that $\theta^2=2$, where
$\theta$ denotes the highest root. The lattice spanned by the simple
roots is called $\La_g$ and the corresponding Cartan matrix is defined
as in \Eqnref{cartan}. The only exception to our analysis is when $\g$
is a$_1$, and we do not consider that case. The very extension of $\g$
is obtained by adding three new simple roots.

\benu
\item Add the simple root 
\beq
\alpha_0 = k - \theta, \hspace{0.5 cm} (\alpha_0,\alpha_0) = 2,
\eeq
where $k\in{\Pi_{1;1}}$. The corresponding Lie algebra is just the
affine extension of $\g$ and we denote it by $\g_0$. The root lattice
of $\g_0$ is a part of $\La_g\oplus\Pi_{1;1}$, more precisely those
vectors $x$ in this lattice which are orthogonal to $k$, i.e.,
$(x,k)=0$ for all $x\in{\La_{g_0}}$. The Cartan matrix is 
\beq
\mathbf{A_{g_0}} = \left( \begin{array}{cc}
\mathbf{A_g} & \begin{array}{c} q_1 \\ \vdots \\ q_r \end{array} \\
\begin{array}{ccc} q'_1 & \ldots & q'_r \end{array} & 2
\end{array} \right),
\eeq
where we have introduced $q'_i \equiv (\alpha_0,\alpha_i)$ and $q_i
\equiv \frac{2}{(\alpha_i,\alpha_i)}(\alpha_i,\alpha_0)$. Since $\g_0$
is an affine Lie algebra, the determinant of its Cartan matrix
vanishes, i.e., $\det A_{g_0} = 0$.

\item Add yet another simple root
\beq
\alpha_{-1} = -(k + \bar{k}), \hspace{0.5 cm}
(\alpha_{-1},\alpha_{-1}) = 2. 
\eeq
The resulting algebra is an over-extended Lie algebra denoted by
$\g_{-1}$. The root lattice $\La_{g_{-1}}$ is now the whole $\La_{g}
\oplus \Pi_{1;1}$, indicating that the algebra is Lorentzian. The
other scalar products involving the new simple root $\alpha_{-1}$ are
$(\alpha_{-1},\alpha_{0}) = -1$ and $(\alpha_{-1},\alpha_i) = 0$,
using which we find the Cartan matrix
\beq
\mathbf{A_{g_{-1}}} = \left( \begin{array}{ccc}
\mathbf{A_g} & \begin{array}{c} q_1 \\ \vdots \\ q_r \end{array} &
\begin{array}{c} 0 \\ \vdots \\ 0 \end{array} \\
\begin{array}{ccc} q'_1 & \ldots & q'_r \end{array} & 2 & -1 \\
\begin{array}{ccc} 0 & \ldots & 0 \end{array} & -1 & 2
\end{array} \right).
\eeq
The determinant of the Cartan matrix is then
\beq
\eqnlab{detA_overextension}
\det A_{g_{-1}} = 2\det A_{g_{0}} - \det A_g = -\det A_g,
\eeq
verifying that $\g_{-1}$ is Lorentzian.

\item The even further enlarged algebra $\g_{-2}$ is obtained by
  adding the simple root
\beq
\alpha_{-2} = k - (l + \bar{l}), \hspace{0.5 cm}
(\alpha_{-2},\alpha_{-2}) = 2,
\eeq 
where $l$ and $\bar{l}$ span a new $\Pi_{1;1}$ lattice. The whole root
lattice consists of vectors in $\La_g \oplus \Pi_{1;1} \oplus
\Pi_{1;1} = \La_{g_{-1}} \oplus \Pi_{1;1}$ which are orthogonal to the
time-like vector $s = l-\bar{l} = (1,1)$. The Cartan matrix of
$\g_{-2}$ has the form
\beq
\mathbf{A_{g_{-2}}} = \left( \begin{array}{cccc}
\mathbf{A_g} & \begin{array}{c} q_1 \\ \vdots \\ q_r \end{array} &
\begin{array}{c} 0 \\ \vdots \\ 0 \end{array} &
\begin{array}{c} 0 \\ \vdots \\ 0 \end{array} \\
\begin{array}{ccc} q'_1 & \ldots & q'_r \end{array} & 2 & -1 & 0 \\
\begin{array}{ccc} 0 & \ldots & 0 \end{array} & -1 & 2 & -1 \\
\begin{array}{ccc} 0 & \ldots & 0 \end{array} & 0 & -1 & 2
\end{array} \right),
\eeq
due to that $(\alpha_{-2},\alpha_{-1}) = -1$ and
$(\alpha_{-2},\alpha_{0}) = (\alpha_{-2},\alpha_{i}) = 0$. Clearly this
is again a Lorentzian Kac-Moody algebra
\beq
\eqnlab{detA_veryextension}
\det A_{g_{-2}} = 2\det A_{g_{-1}} - \det A_{g_0} = -2\det A_g.
\eeq
The same value for the determinant is obtained using
\Eqnref{detA_affine}.
\eenu

The fundamental weights of the over-extended algebra are
\beq
\eqnlab{weights_over}
\begin{array}{rcl}
\la_i &=& \la^f_i - (\la^f_i,\theta)(k-\bar{k}), \hspace{0.5 cm}
i=1,\ldots,r \\
\la_0 &=& -(k-\bar{k}) \\
\la_{-1} &=& -k,
\end{array}
\eeq
with $\la^f_i$ being the fundamental weights of $\g$, whereas the
fundamental weights of the very extended algebra are
\beq
\eqnlab{weights_very}
\begin{array}{rcl}
\la_i &=& \la^f_i -
(\la^f_i,\theta)[k-\bar{k}-\frac{1}{2}(l+\bar{l})], \hspace{0.5 cm}
i=1,\ldots,r \\ 
\la_0 &=& -[k-\bar{k}-\frac{1}{2}(l+\bar{l})] \\
\la_{-1} &=& -k \\
\la_{-2} &=& -\frac{1}{2}(l+\bar{l}).
\end{array}
\eeq
These together with the corresponding simple roots satisfy
\Eqnref{root_weight} for the respective algebra. For the
over-extended algebra $\g_{-1}$ the Weyl vector is given by
\beq
\rho = \rho^f + h\bar{k} - (h+1)k,
\eeq
where $\rho^f$ is the Weyl vector and $h = 1+\sum_{i=1}^{r}n_i$ is the
Coxeter number of $\g$, respectively. Similarly,
\beq
\rho = \rho^f + h\bar{k} - (h+1)k - \frac{1}{2}(1-h)(l+\bar{l})
\eeq
is the Weyl vector of the very extended Kac-Moody algebra.

The weight lattice spanned by the fundamental weights is just the
lattice $\La_w$ dual to the root lattice $\La_R$, $\La_w =
\La_R^{*}$. Generally we have 
\beq
(\La_1 \oplus \La_2)^{*} = \La_1^{*} \oplus \La_2^{*}
\eeq
for two arbitrary lattices, and this is sufficient for us to find the
weight lattices of these extended algebras. The weight lattice of
$\g_{-1}$ is 
\beq
\La_{g_{-1}}^{*} = \La_g^{*} \oplus \Pi_{1;1}^{*} = \La_g^{*} \oplus
\Pi_{1;1},
\eeq
the last equality is due to the self-duality of $\Pi_{1;1}$. Inserting
this into \Eqnref{conjugacy_class} we find
\beq
Z_{g_{-1}} = \frac{\La_{g_{-1}}^{*}}{\La_{g_{-1}}} =
\frac{\La_g^{*}\oplus\Pi_{1;1}}{\La_g\oplus\Pi_{1;1}} =
\frac{\La_g^{*}}{\La_g} = Z_g
\eeq
in consistency with the relation between the determinants in
\eqnref{detA_overextension}. The weight lattice of $\g_{-2}$, on the
other hand, is given by
\beq
\La_{g_{-2}}^{*} = \La_g^{*} \oplus \Pi_{1;1} \oplus
\{(r,-r):2r\in{\mathbb Z}\},
\eeq
where the convention in Appendix~\chref{lattice} has been used. Since
the root lattice of $\g_{-2}$ is $\La_{g_{-2}} = \La_g \oplus
\Pi_{1;1} \oplus \{(t,-t):t\in{\mathbb Z}\}$, we obtain
\beq
Z_{g_{-2}} = \frac{\La_{g_{-2}}^{*}}{\La_{g_{-2}}} =
\frac{\La_g^{*}\oplus\Pi_{1;1}\oplus\{(r,-r):2r\in{\mathbb Z}\}}
{\La_g\oplus\Pi_{1;1}\oplus\{(t,-t):t\in{\mathbb Z}\}} = Z_g \times
\mathbb Z_2. 
\eeq
Again, the result is consistent with the determinant relation in
\Eqnref{detA_veryextension}. 

\subsection{Self-dual lattices}

Let us study the even self-dual lattices a little more in detail.  We
start with a finite dimensional semi-simple rank $r$ Lie algebra $\g$,
for which the root lattice $\La_g$ is an even self-dual lattice of
dimension $r$ or a sublattice of a such lattice. Even self-dual
Euclidean lattices exist only in dimensions $D = 8n$, $n\in{\mathbb
  N}$.

\subsubsection{The root lattice of e$_{11}$}

In eight dimensions, i.e., for $n=1$, the only lattice fulfilling these
criteria is the root lattice of e$_8$. The set of simple roots for
e$_8$ written in an orthogonal basis is
\beq
\begin{array}{rcl}
\alpha_1 &=& (0,0,0,0,0,1,-1,0) \\
\alpha_2 &=& (0,0,0,0,1,-1,0,0) \\
\alpha_3 &=& (0,0,0,1,-1,0,0,0) \\
\alpha_4 &=& (0,0,1,-1,0,0,0,0) \\
\alpha_5 &=& (0,1,-1,0,0,0,0,0) \\
\alpha_6 &=& (-1,-1,0,0,0,0,0,0) \\
\alpha_7 &=& (\frac{1}{2},\frac{1}{2},\frac{1}{2},\frac{1}{2},
\frac{1}{2},\frac{1}{2},\frac{1}{2},\frac{1}{2}) \\ 
\alpha_8 &=& (1,-1,0,0,0,0,0,0),
\end{array}
\eeq
combining these with the Coxeter numbers as coefficients yields the
highest root
\beq
\theta = (0,0,0,0,0,0,-1,1).
\eeq
We denote the corresponding affine, over-extended and very extended
algebras by e$_9$, e$_{10}$ and e$_{11}$, respectively. The affine
root is 
\beq
\alpha_0 = k-\theta = (-\theta;(1,0)),
\eeq
whereas the over-extended root is given by
\beq
\alpha_{-1} = -(k+\bar{k}) = (\mathbf{0};(-1,1)).
\eeq
These together with the simple roots of e$_8$ span the root lattice of
e$_{10}$, the resulting lattice $\La_{e_8} \oplus \Pi_{1;1}$ is even,
self-dual and Lorentzian. Such lattices occur only in dimensions
$D=8n+2$, $n\in{\mathbb N}$, and are unique in each dimension, hence
we denote them by $\Pi_{8n+1;1}$. It is then obvious that the root
lattice of e$_{10}$ is $\Pi_{9;1} = \La_{e_8} \oplus \Pi_{1;1}$.
Adding the very extended root
\beq
\alpha_{-2} = k-(l+\bar{l}) = (\mathbf{0};(1,0);(-1,1))
\eeq
extends the span of the simple roots to being part of the lattice
\beq
\Pi_{10;2} = \Pi_{9;1} \oplus \Pi_{1;1} = \La_{e_8} \oplus \Pi_{1;1}
\oplus \Pi_{1;1}, 
\eeq
which is the unique even self-dual lattice with signature
$(10,2)$. The Dynkin diagram of the algebra e$_{11}$ can be seen in
\Figref{DynkinE11}. 
\begin{figure}
  \begin{center}
    \includegraphics[scale=0.6]{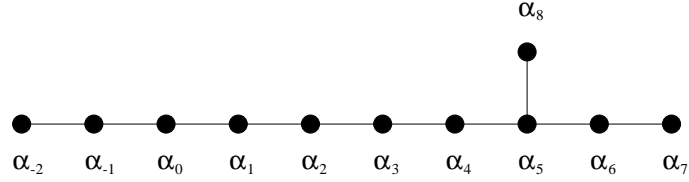}
    \caption{\textit{The Dynkin diagram of the Kac-Moody algebra
        e$_{11}$.}} 
    \figlab{DynkinE11}
  \end{center}
\end{figure}
To find the number of conjugacy classes for the e$_{11}$ weight
lattice we use \Eqnref{conjugacy_class} 
\beq
Z_{e_{11}} = \frac{\La_{e_{11}}^{*}}{\La_{e_{11}}} = \mathbb Z_2.
\eeq
The e$_{11}$ algebra is of great importance, since it is conjectured
to be a symmetry of M-theory.

\subsubsection{The root lattice of k$_{27}$}

Another algebra important in string theory is the very extension of
the algebra d$_{24}$. The root lattice of d$_{24}$ is by itself not
self-dual, but it is a sublattice of one of the Niemeier lattices
\cite{Conway:1998}.  Niemeier lattices are the even self-dual
Euclidean lattices in dimension 24, and there are totally 24 of them.
The simple roots of d$_{24}$ in an orthogonal basis are
\beq
\begin{array}{rcl}
\alpha_i &=& (\underbrace{0,\ldots,0}_{i-1},1,-1,
\underbrace{0,\ldots,0}_{23-i}), \hspace{0.5 cm} i = 1,\ldots,23 \\
\alpha_{24} &=& (\underbrace{0,\ldots,0}_{22},1,1).
\end{array}
\eeq
As mentioned before, the root lattice of d$_{24}$ is not self-dual and 
\beq
Z_{d_{24}} = \frac{\La_{d_{24}}^{*}}{\La_{d_{24}}} = \mathbb Z_2
\times \mathbb Z_2,
\eeq
which is consistent with the fact that $\det A_{d_{24}} = 4$. We can
construct a 24-dimensional self-dual lattice by adding the point
\beq
g = \frac{1}{2}(\underbrace{1,\ldots,1}_{24}) = \la_{24} \hspace{0.5
  cm} \in{\La_{d_{24}}^{*}}
\eeq
to the lattice $\La_{d_{24}}$, the resulting lattice is then
$\La_{d_{24}}^N = \La_{d_{24}}^{*}/\mathbb Z_2$. This new point
satisfies $(g,g)=6$ and $2g\in{\La_{d_{24}}}$.

Following the general procedure we add an affine and an over-extended
root
\beq
\begin{array}{rcl}
\alpha_0 &=& k-\theta = ((-1,-1,\underbrace{0,\ldots,0}_{22});(1,0))
\\ 
\alpha_{-1} &=& -(k+\bar{k}) = (\mathbf{0};(-1,1))
\end{array}
\eeq
to the collection of simple roots for d$_{24}$, respectively. The root
lattice of the corresponding algebra k$_{26}$ is thus part of the even
self-dual lattice $\Pi_{25;1} = \La_{d_{24}}^N \oplus \Pi_{1;1} =
\La_{k_{26}}^{*}/\mathbb Z_2$ in 26 dimensions. Adding also the very
extended root
\beq
\alpha_{-2} = k-(l+\bar{l}) = (\mathbf{0};(1,0);(-1,1)),
\eeq
we obtain the algebra k$_{27}$ with its root lattice being contained
in 
\beq
\La_{d_{24}} \oplus \Pi_{1;1} \oplus \Pi_{1;1}
\eeq
and 
\beq
Z_{k_{27}} = \frac{\La_{k_{27}}^{*}}{\La_{k_{27}}} =
\frac{\La_{d_{24}}^{*}}{\La_{d_{24}}} \times \mathbb Z_2 = \mathbb Z_2
\times \mathbb Z_2 \times \mathbb Z_2.
\eeq
The algebra k$_{27}$, with a Dynkin diagram as in \Figref{DynkinK27},
has been thought to be a symmetry of the 26-dimensional closed bosonic
string.
\begin{figure}
  \begin{center}
    \includegraphics[scale=0.33]{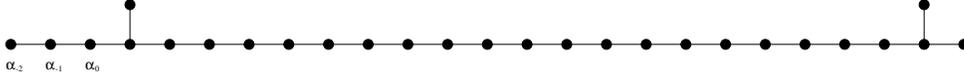}
    \caption{\textit{The Dynkin diagram of the Kac-Moody algebra
        k$_{27}$.}} 
    \figlab{DynkinK27}
  \end{center}
\end{figure}

\subsubsection{The root lattice of m$_{19}$}

Let us also look at the very extension of the rank 16 algebra
d$_{16}$, the procedure is almost identical to the case of
k$_{27}$. The root lattice of d$_{16}$ satisfy again
\beq
Z_{d_{16}} = \frac{\La_{d_{16}}^{*}}{\La_{d_{16}}} = \mathbb Z_2
\times \mathbb Z_2,
\eeq
and it can be made self-dual by adding the point
\beq
g = \frac{1}{2}(\underbrace{1,\ldots,1}_{16}), \hspace{0.5 cm} (g,g) =
4.
\eeq
The resulting even self-dual lattice is $\La_{d_{16}}^s =
\La_{d_{16}}^{*}/\mathbb Z_2$. 

The root lattice of the over-extended algebra m$_{18}$ is a sublattice
of the 18-dimensional even self-dual Lorentzian lattice $\Pi_{17;1} =
\La_{d_{16}}^s \oplus \Pi_{1;1} = \La_{m_{18}}^{*}/\mathbb Z_2$.
Finally, adding the very extended root
\beq
\alpha_{-2} = (\mathbf{0};(1,0);(-1,1))
\eeq
yields an algebra denoted by m$_{19}$, whose root lattice is part of 
\beq
\La_{d_{16}} \oplus \Pi_{1;1} \oplus \Pi_{1;1}.
\eeq

It is remarkable that construction of the very extended Lorentzian
Kac-Moody algebras based on Euclidean self-dual lattices leads to
algebras with fundamental importance in physics. The obtained
algebras, e$_{11}$ and k$_{27}$, are conjectured to be symmetries of
M-theory and the 26-dimensional bosonic string, respectively. The
symmetries of these two theories are thus deeply related to the unique
even self-dual Lorentzian lattices in ten and 26 dimensions,
respectively. Studying the Dynkin diagrams of e$_{11}$ and k$_{27}$,
\Figsref{DynkinE11} and \figref{DynkinK27}, we conclude that e$_{11}$
is a subalgebra of k$_{27}$, indicating that the whole M-theory is
possibly described by the bosonic string in 26 dimensions. Also, it
seems natural that there should exist an 18-dimensional string with
m$_{19}$ as its symmetry.

\subsection{Principal so(2,1) subalgebra}

The condition for the existence of principal so(2,1) subalgebra is 
\beq
\eqnlab{principal_cond_KacMoody}
\displaystyle\sum_a A^{-1}_{ab} = \displaystyle\sum_a
\frac{2}{(\alpha_a,\alpha_a)} (l_a,l_b) \le 0,
\eeq
where $a$ and $b = -2,-1,0,1,\ldots,r$. Putting in the fundamental
weights for the over-extended Kac-Moody algebras given in
\Eqnref{weights_over} we get
\beq
\begin{array}{rcl}
\eqnlab{principal_over}
\displaystyle\sum_a A^{-1}_{a,-1} &=& -h \\
\displaystyle\sum_a A^{-1}_{a,0} &=& -(2h+1) \\
\displaystyle\sum_a A^{-1}_{a,j} &=& -\frac{1}{2}n_j
(\alpha_j,\alpha_j)(2h+1) + \displaystyle\sum_i (A^f)^{-1}_{ij},
\end{array}
\eeq
where $n_j$, $h = 1+\sum_{i=1}^r n_i$, $(A^f)^{-1}_{ij}$ are the
Coxeter labels, the Coxeter number and the inverse Cartan matrix of
the finite dimensional Lie algebra $\g$, respectively. The
corresponding expressions for the very extended Kac-Moody algebras are
\beq
\begin{array}{rcl}
\eqnlab{principal_very}
\displaystyle\sum_a A^{-1}_{a,-2} &=& -\frac{1}{2}(h-1) \\
\displaystyle\sum_a A^{-1}_{a,-1} &=& -h \\
\displaystyle\sum_a A^{-1}_{a,0} &=& -\frac{3}{2}(h+1) \\
\displaystyle\sum_a A^{-1}_{a,j} &=& -\frac{3}{4}n_j
(\alpha_j,\alpha_j)(h+1) + \displaystyle\sum_i (A^f)^{-1}_{ij}.
\end{array}
\eeq
The only sums not satisfying the condition
\eqnref{principal_cond_KacMoody} automatically are $\sum_a
A^{-1}_{a,j}$, where information about the specific finite dimensional
Lie algebra $\g$ is required.

As an illustrative example we take $\g$ to be a$_r$, which satisfy
$(\alpha_j,\alpha_j) = 2$, $n_j = 1$ and $h = r+1$, see reference
\cite{Fuchs:1997jv}. Combining the fundamental weights $\la_i =
\sum_{j=1}^i e_j - \frac{i}{r+1} \sum_{j=1}^{r+1} e_j$ gives
\beq
\begin{array}{rcl}
\displaystyle\sum_i (A^f)^{-1}_{ij} &=& \displaystyle\sum_i
\frac{2}{(\alpha_i,\alpha_i)} (\la_i,\la_j) =
\displaystyle\sum_{i=1}^j i\frac{r+1-j}{r+1} +
\displaystyle\sum_{i=j+1}^r j\frac{r+1-i}{r+1} \\ 
&=& \frac{j}{2}(r+1-j),
\end{array}
\eeq
which in turn can be used in \Eqnref{principal_very} to yield
\beq
\displaystyle\sum_a A^{-1}_{a,j} = -\frac{3}{2} (r+2) +
\frac{j}{2}(r+1-j)
\eeq
for the very extension of a$_r$. Maximising the above equation
\beq
(\displaystyle\sum_a A^{-1}_{a,j})_{\mathrm{max}} = \left\{
\begin{array}{rcll}
\frac{1}{8}(r^2-10r-23) &,& j = \frac{r+1}{2} & \textrm{when $r$ is
  odd} \\ 
\frac{1}{8}(r^2-10r-24) &,& j = \frac{r}{2} & \textrm{when $r$ is
  even}
\end{array} \right.
\eeq
shows that $\sum_a A^{-1}_{a,j}$ is negative only if $r\le 12$. Hence,
the rank of a$_r$ cannot exceed 12 for the corresponding very
extension to contain a principal so(2,1) subalgebra.

Performing similar calculations for all the simple finite dimensional
Lie algebras, we find the constraints summarised in
\Tabref{subalgebra}. 
\begin{table}[!h]
  \begin{center}
    \begin{tabular}{l|ll}
      $\g$      & Over-extension         & Very extension \\ \hline 
      a$_r$     & $r \le 16$             & $r \le 12$ \\ 
      d$_r$     & $r \le 16$             & $r \le 12$ \\ 
      e$_r$     & $r = 6,7,8$            & $r = 6,7,8$ \\ 
      b$_r$     & $r \le 15$             & $r \le 11$ \\ 
      c$_r$     & $r \le 8$              & $r \le 6$ \\ 
      f$_4$     & yes                    & yes \\ 
      g$_2$     & yes                    & yes \\
     \end{tabular}
    \caption{\textit{The algebras with principal so(2,1)
      subalgebras.}} 
  \tablab{subalgebra}
  \end{center}
\end{table}
Note that e$_{11}$ contains a principal so(2,1) subalgebra, while
neither k$_{27}$ nor m$_{19}$ does since these are the very extensions
of the Lie algebras d$_{24}$ and d$_{16}$, respectively.

\chapter{Symmetries in M-theory}
\chlab{symmetry}

When eleven dimensional supergravity is dimensionally reduced, the
obtained scalars have coset space symmetries (\Tabref{coset}). This
coset construction for the the scalar sectors has been extended to
include also the gauge fields for the maximal supergravity theories
\cite{Cremmer:1998ct,Cremmer:1998px}. For space-time dimensions larger
than three, the coset space E$_n$/F$_n$ is based on the normal real
forms of the finite dimensional semi-simple Lie algebras from the
E-series, with the subgroup invariant under the Cartan involution
taken as the local subgroup F$_n$. These subgroups are precisely the
maximal compact subgroups of the non-compact groups E$_n$. There is
also some evidence \cite{Julia:1985,Nicolai:1987jk} that dimensional
reduction of eleven dimensional supergravity to two dimensions and one
dimension will yield the affine extension of e$_8$ (called e$_9$) and
the hyperbolic algebra e$_{10}$, respectively. Continuing in the same
spirit, it has now been conjectured that the eleven dimensional
supergravity itself has a coset symmetry based on the very extended
Kac-Moody algebra e$_{11}$.  The linearly realised local symmetry is
assumed to be the subalgebra invariant under the Cartan involution.
The generators in this subalgebra have the form $(E_\alpha -
F_\alpha)$ for an arbitrary positive root $\alpha$, $E_\alpha$ and
$F_\alpha$ are the positive and negative root generators,
respectively. It then follows that $E_\alpha \sim F_\alpha$, and the
coset representatives can all be taken to belong to the Borel
subalgebra of e$_{11}$. Note that whenever we use the word algebra in
this chapter, we actually mean the normal real form of the complex
algebra, except for the local subalgebras where we mean their compact
real forms.

It was shown in \Chref{nonlinear} that the field equations of the
bosonic sector of eleven dimensional supergravity arise from
simultaneous nonlinear realisation under the conformal group and the
group G$_{11}$. In this coset formulation all the bosonic fields of
the theory, i.e., the graviton and the rank three gauge field are
treated as Goldstone bosons. For the final symmetry to be the
Kac-Moody algebra e$_{11}$, we have to enlarge the algebra g$_{11}$
defined in \Eqsref{IGL11} and \eqnref{G11}. There are two ways to go
\cite{West:2001as}. The first one is to add more generators to
g$_{11}$, which also belong to the local subalgebra. The Cartan form
will now transform under a larger local subgroup, see
\Eqnref{Cartan_transf}. Since the number of the preferred fields
remains the same, most of the calculations are unchanged. The only
constraint for the local subalgebra is that the field equations
obtained must be invariant under the corresponding larger local
subgroup. The second method is to perform the nonlinear realisation
with a larger algebra than g$_{11}$ and some suitable local
subalgebra, the difference now is that we do not demand the field
content to remain the same. Instead, we have to show that this gives
an alternative description of eleven dimensional supergravity, with
the same on-shell degrees of freedom. To find the total symmetry, we
let the theory in both cases be simultaneously nonlinearly realised
under the conformal group. The hope is that the resulting theory
contains a coset symmetry G/H, with G being a group corresponding to a
Kac-Moody algebra and H its subgroup invariant under the Cartan
involution. It seems also reasonable that the symmetries E$_n$ and
F$_n$ found in the dimensional reductions of eleven dimensional
supergravity are subgroups of G and H, respectively. Again, the
natural candidate for G is E$_{11}$.

\section{The Borel subalgebras}

The starting point for finding the final global symmetry algebra g is
the algebra g$_{11}$ from \Chref{nonlinear}, whose generators can be
decomposed as 
\beq
\begin{array}{rcl}
\textrm{g}_{11}^{+} &=& \left\{K^a_{\phantom{a}b}~(a<b),a = 1,\ldots,11;
  \hspace{0.2 cm} R^{c_1c_2c_3};\hspace{0.2 cm} R^{c_1\ldots
  c_6}\right\} \\

\textrm{g}_{11}^{0} &=& \left\{H_a\equiv K^a_{\phantom{a}a} -
K^{a+1}_{\phantom{a+1}a+1}, a = 1,\ldots,10;\hspace{0.2 cm}  D =
  \sum_{a=1}^{11} K^a_{\phantom{a}a}\right\}
\end{array}
\eeq

g$_{11}^{-} = K^a_{\phantom{a}b}$ ($a>b$) and $P_a$ with $a =
1,\ldots,11$. The role the generators $P_a$ play is quite different
from the other ones and it should be noted that there are no preferred
fields associated with them. Let us thus for now concentrate on the
generators $K^a_{\phantom{a}b}$, $R^{c_1c_2c_3}$ and $R^{c_1\ldots
  c_6}$. The basic requirement on the final Kac-Moody algebra g is
that it contains all the generators in g$_{11}$, except $P_a$ which
will be treated separately. We demand the set of positive root
generators of g to contain g$_{11}^{+}$. Also, the Cartan subalgebra
of g must contain g$_{11}^{0}$. For notational simplicity we introduce
g$_{11}^{0+} = \textrm{g}_{11}^0 \oplus \textrm{g}_{11}^{+}$. Note
that the generators $K^a_{\phantom{a}b}$ form a gl(11) $\cong$ A$_{10}
\oplus$ u(1) algebra, with $H_a$, $K^a_{\phantom{a}b}$ ($a<b$) and
$K^a_{\phantom{a}b}$ ($a>b$) being the Cartan generators, positive and
negative root generators of A$_{10}$, respectively\footnote{In this
  chapter all the classical simple Lie algebras are denoted by capital
  letters, e.g. A$_r$.}. Since the final algebra g is assumed to
include the finite dimensional exceptional algebras, we analyse this
aspect a little more in detail.

\subsection{The Borel subalgebra of e$_7$}

Consider the simple Lie algebra e$_7$. Restricting the indices to only
take the values $\{5,\ldots,11\}$, the positive root generators of
e$_7$ can be identified as
\beq
\hat{K}^i_{\phantom{i}j} = K^i_{\phantom{i}j} -
\frac{1}{7}\delta^i_j \displaystyle\sum_{l=5}^{11}
K^l_{\phantom{l}l} \hspace{0.5 cm} (i<j), \hspace{0.5 cm}
S_i = \frac{1}{6!}\varepsilon_{ii_1\ldots i_6}R^{i_1\ldots i_6}
\eeq
and $R^{i_1i_2i_3}$. Similarly, the Cartan generators of e$_7$ can be
found as
\beq
H_i = K^i_{\phantom{i}i} - K^{i+1}_{\phantom{i+1}i+1} \hspace{0.5 cm}
\textrm{and} \hspace{0.5 cm} \hat{D} = \displaystyle\sum_{l=5}^{11}
K^l_{\phantom{l}l}.
\eeq
These generators together with $\hat{K}^i_{\phantom{i}j} =
K^i_{\phantom{i}j}$ ($i>j$) obey the commutation relations
\beq
\begin{array}{rcl}
\com{\hat{K}^i_{\phantom{i}j},\hat{K}^k_{\phantom{k}l}} &=& \delta^k_j
\hat{K}^i_{\phantom{i}l} - \delta^i_l \hat{K}^k_{\phantom{k}j} \\
\com{\hat{K}^i_{\phantom{i}j},R^{k_1k_2k_3}} &=& 3\delta^{[k_1}_j
R^{|i|k_2k_3]} - \frac{3}{7}\delta^i_j R^{k_1k_2k_3} \\
\com{\hat{K}^i_{\phantom{i}j},S_k} &=& -\delta^i_k S_j + \frac{1}{7}
\delta^i_jS_k \\
\com{R^{i_1i_2i_3},R^{i_4i_5i_6}} &=& 2\varepsilon^{i_1\ldots i_6j} S_j
  \\
\com{R^{i_1i_2i_3},S_k} &=& 0 \\
\com{\hat{D},\hat{K}^i_{\phantom{i}j}} &=& 0 \\
\com{\hat{D},R^{i_1i_2i_3}} &=& 3R^{i_1i_2i_3} \\
\com{\hat{D},S_i} &=& 6S_i \\
\com{S_i,S_j} &=& 0,
\end{array}
\eeq
which are precisely the commutation relations of the Borel subalgebra
of e$_7$ written with respect to its A$_6$ subalgebra
\cite{West:2001as}. Note that the generators
$\hat{K}^i_{\phantom{i}j}$ ($i>j$) are the negative root generators of
A$_6$ and thus not included in the Borel subalgebra of e$_7$. This
shows that the Borel subalgebra of e$_7$ is also a symmetry of the
eleven dimensional supergravity, just as the A$_6$ subalgebra of
e$_7$. The whole e$_7$ can be found as a symmetry if we enlarge the
set of negative roots generators $\{\hat{K}^i_{\phantom{i}j},(i>j)\}$
to also include the generators $S^i$ and $R_{i_1i_2i_3}$. If these
generators are also contained in the local subalgebra, the equations
of motion will then remain the same. The whole e$_7$ is generated by
the simple root generators
\beq
E_i = K^i_{\phantom{i}i+1}, i = 5,\ldots,10; \hspace{0.3 cm} E_{11} =
R^{9,10,11} 
\eeq
and the Cartan generators
\beq
H_i = K^i_{\phantom{i}i} - K^{i+1}_{\phantom{i+1}i+1}, i =
5,\ldots,10; \hspace{0.3 cm} H_{11} = K^9_{\phantom{9}9} +
K^{10}_{\phantom{10}10} + K^{11}_{\phantom{11}11} -
\frac{1}{3}\hat{D}.
\eeq
The new local subalgebra is A$_7$.

\subsection{The Borel subalgebra of e$_8$}

Having found the Borel subalgebra of e$_7$ as a symmetry algebra of
eleven dimensional supergravity in the previous subsection, we also
suggested how to incorporate the whole e$_7$ as a symmetry. Perform
now a similar analysis for the algebra e$_8$. Introduce the generators
\beq
\eqnlab{e8_gen}
\begin{array}{rl}
H_i = K^i_{\phantom{i}i} - K^{i+1}_{\phantom{i+1}i+1}; & \hat{D} =
\displaystyle\sum_{l=4}^{11} K^l_{\phantom{l}l} \\ 
\hat{K}^i_{\phantom{i}j} = K^i_{\phantom{i}j} -
\frac{1}{8}\delta^i_j \displaystyle\sum_{l=4}^{11}
K^l_{\phantom{l}l} \hspace{0.5 cm} (i<j); &
S_{k_1k_2} = \frac{1}{6!}\varepsilon_{k_1k_2i_1\ldots i_6}
R^{i_1\ldots i_6} \\
\hat{K}^i_{\phantom{i}j} = K^i_{\phantom{i}j} \hspace{0.5 cm} (i>j); &
R^{i_1i_2i_3},
\end{array}
\eeq
where the indices take the values $\{4,\ldots,11\}$. Using the
commutation relations of g$_{11}$ we find the Lie brackets
\beq
\eqnlab{e8LB}
\begin{array}{rcl}
\com{\hat{K}^i_{\phantom{i}j},\hat{K}^k_{\phantom{k}l}} &=& \delta^k_j
\hat{K}^i_{\phantom{i}l} - \delta^i_l \hat{K}^k_{\phantom{k}j} \\
\com{\hat{K}^i_{\phantom{i}j},R^{k_1k_2k_3}} &=& 3\delta^{[k_1}_j
R^{|i|k_2k_3]} - \frac{3}{8}\delta^i_j R^{k_1k_2k_3} \\
\com{\hat{K}^i_{\phantom{i}j},S_{k_1k_2}} &=& -2\delta^i_{[k_1}
S_{|j|k_2]} + \frac{2}{8} \delta^i_j S_{k_1k_2} \\ 
\com{R^{i_1i_2i_3},R^{i_4i_5i_6}} &=& \varepsilon^{i_1\ldots i_6jk}
S_{jk} \\
\com{\hat{D},\hat{K}^i_{\phantom{i}j}} &=& 0 \\
\com{\hat{D},R^{i_1i_2i_3}} &=& 3R^{i_1i_2i_3} \\
\com{\hat{D},S_{ij}} &=& 6S_{ij} \\
\com{S_{i_1i_2},S_{j_1j_2}} &=& 0 \\
\com{S_{i_1i_2},R^{j_1j_2j_3}} &=& 0.
\end{array}
\eeq
Comparing \Eqnref{e8LB} with the Lie brackets of e$_8$ written with
respect to its regular A$_7$ subalgebra, we can identify
$\hat{K}^i_{\phantom{i}j}$ with the generators of A$_7$. $\hat{D}$
spans the eighth dimension in the Cartan subalgebra for e$_8$,
$R^{i_1i_2i_3}$ and $S_{i_1i_2}$ are positive root generators
\cite{West:2001as}. The adjoint representation of e$_8$ decomposes as
$248 = 1 + 63 + (56+\bar{28}+\bar{8}) + (\bar{56}+28+8)$ with
\beq
\begin{array}{rcl}
1: & \hat{D} & \textrm{Cartan subalgebra} \\
63: & \hat{K}^i_{\phantom{i}j} & \textrm{Regular A$_7$ subalgebra} \\
 & & \textrm{Cartan, positive and negative root generators} \\
(56+\bar{28}+\bar{8}): & & \textrm{Positive root generators} \\
 & & R^{i_1i_2i_3} \hspace{0.2 cm} (56) \hspace{0.2 cm} \textrm{and}
 \hspace{0.2 cm} S_{i_ii_2} \hspace{0.2 cm} (\bar{28}) \\
(\bar{56}+28+8): & & \textrm{Negative root generators}.
\end{array}
\eeq

As can be seen, there is a missing $\bar{8}$ of the Borel subalgebra
of e$_8$. The way to solve this is to assume these missing generators
form an ideal, which is trivially realised. The full non-trivial
realisation of the Borel subalgebra can be obtained by adding the
generators $R^{c_1\ldots c_8,d}$ to the algebra g$_{11}$. These new
generators are antisymmetric in the indices $c_1,\ldots,c_8$. The
commutation relations of g$_{11}$ modify to
\beq
\eqnlab{G11_modified}
\begin{array}{rcl}
\com{K^a_{\phantom{a}b},P_c} &=& -\delta^a_c P_b; \hspace{0.2 cm} 
\com{P_a,P_b} = 0 \\
\com{K^a_{\phantom{a}b},K^c_{\phantom{c}d}} &=& \delta^c_b
K^a_{\phantom{a}d} - \delta^a_d K^c_{\phantom{c}b} \\
\com{K^a_{\phantom{a}b},R^{c_1c_2c_3}} &=&
3\delta^{[c_1}_bR^{|a|c_2c_3]} = \delta^{c_1}_bR^{ac_2c_3} +
\delta^{c_2}_bR^{ac_3c_1} + \delta^{c_3}_bR^{ac_1c_2} \\
\com{K^a_{\phantom{a}b},R^{c_1\dots c_6}} &=&
6\delta^{[c_1}_bR^{|a|c_2 \dots c_6]} \\
\com{R^{c_1c_2c_3},R^{c_4c_5c_6}} &=& 2R^{c_1\ldots c_6} \\
\com{R^{c_1\ldots c_6},R^{d_1\ldots d_6}} &=& 0 \\
\com{R^{c_1\ldots c_6},R^{d_1d_2d_3}} &=& 3R^{c_1\ldots
  c_6[d_1d_2,d_3]} \\
\com{R^{c_1\ldots c_8,a},R^{d_1d_2d_3}} &=& 0; \hspace{0.2 cm}
\com{R^{c_1\ldots c_8,a},R^{d_1\ldots d_6}} = 0; \hspace{0.2 cm}
\com{R^{c_1\ldots c_8,a},R^{d_1\ldots d_8,b}} = 0 \\
\com{K^a_{\phantom{a}b},R^{c_1\ldots c_8,d}} &=&
8\delta^{[c_1}_bR^{|a|c_2 \dots c_8],d} + \delta^d_b R^{c_1\ldots
  c_8,a},
\end{array}
\eeq
with $a,b,\ldots = 1,\ldots,11$.  Compared to \Eqnref{G11}, the
commutator $\com{R^{c_1\ldots c_6},R^{d_1d_2d_3}}$ is now
non-vanishing. To satisfy the Jacobi identities, the new generators
must in addition obey
\beq
\eqnlab{9gen_constr}
R^{[c_1\ldots c_8,d]} = 0.
\eeq

Restricting the indices to the values $\{4,\ldots,11\}$ yields the
generators in \Eqnref{e8_gen} as well as $R^{i_1\ldots i_8,j} =
\varepsilon^{i_1\ldots i_8}S^j$. The corresponding Lie brackets are
given by the first eight rows in \Eqnref{e8LB} as well as the
relations
\beq
\begin{array}{rcl}
\com{\hat{K}^i_{\phantom{i}j},S^k} = \delta^k_j S^i - \frac{1}{8}
\delta^i_jS^k &;& \com{S_{i_1i_2},R^{j_1j_2j_3}} =
3\delta^{[j_1j_2}_{i_1i_2}S^{j_3]} \\
\com{R^{i_1i_2i_3},S^j} = 0 &;& \com{S_{i_1i_2},S^j} = 0 \\
\com{\hat{D},S^i} &=& 9S^i.
\end{array}
\eeq
Excluding the generators $\hat{K}^i_{\phantom{i}j}$ ($i>j$), these
define precisely the Borel subalgebra of e$_8$.

\subsection{The Borel subalgebra of e$_{11}$}

Suppose the final symmetry of M-theory is a Kac-Moody algebra g. Since
the coset formulation of the bosonic sector of eleven dimensional
supergravity uses only the g$_{11}$ algebra and the conformal algebra,
one should find the Cartan and simple root generators of the searched
for Kac-Moody algebra inside g$_{11}$. The generators needed to
generate the whole g$_{11}^{+}$ are
\beq
\eqnlab{e11_simple}
E_a = K^a_{\phantom{a}a+1}, a=1,\ldots,10; \hspace{0.2 cm} E_{11} =
R^{9,10,11}, 
\eeq
where $E_a$, $a=1,\ldots,10$ are also the simple root generators of
the finite dimensional algebra A$_{10}$. Obviously, the generators in
\eqnref{e11_simple} are natural choices for the simple root generators
of the final Kac-Moody algebra. Similarly, the set g$_{11}^0$ can be
identified with the Cartan subalgebra of g, but the basis choice
\beq
\eqnlab{e11_Cartan}
H_a = K^a_{\phantom{a}a} - K^{a+1}_{\phantom{a+1}a+1}, a=1,\ldots,10;
\hspace{0.2 cm} H_{11} = K^9_{\phantom{9}9} + K^{10}_{\phantom{10}10}
+ K^{11}_{\phantom{11}11} - \frac{1}{3}D,
\eeq
with $D = \sum_{a=1}^{11}K^a_{\phantom{a}a}$, makes the embeddings of
e$_7$ and e$_8$ more explicit. The generators in \eqnref{e11_simple}
and \eqnref{e11_Cartan} form the Chevalley-Serre basis of g.  Applying
the Chevalley-Serre relations, we then find the Cartan matrix of the
very extended Kac-Moody algebra e$_{11}$. The Dynkin diagram of
e$_{11}$ shows that it indeed contains e$_7$, e$_8$ and A$_{10}$ as
subalgebras. In fact, e$_{11}$ also contains all the algebras e$_n$
listed in \Tabref{coset}. Using the Lie brackets of g$_{11}$ it is
straightforward to verify that the Chevalley-Serre relations
\eqnref{CSrelations} involving the positive root generators are
satisfied. However, we have not yet found any symmetries in eleven
dimensional supergravity that correspond to the negative root
generators, which must exist and obey the remaining Chevalley-Serre
relations for the theory to be invariant under a Kac-Moody algebra.
Moreover, it is insufficient with only the g$_{11}$ commutation
relations to generate the whole Borel subalgebra of g, because some of
the positive root generators form a trivially realised ideal in eleven
dimensional supergravity. We have already encounted this problem when
trying to identify the Borel subalgebra of e$_8$. Later we will show
that this leads to an alternative description of eleven dimensional
supergravity.

The local subalgebra h in eleven dimensions must include all the local
subalgebras f$_n$ obtained by dimensional reduction in \Tabref{coset}.
In addition, it must contain so(10,1), i.e., the local subalgebra we
used in \Chref{nonlinear}. The proposed subalgebra, which is the one
invariant under the Cartan involution, fulfills all the criteria. The
real form chosen for h should be non-compact, since it must contain
the non-compact so(10,1) as a subalgebra. The non-zero components in
the Killing form are \cite{Fuchs:1997jv}
\beq
\left\{ \begin{array}{rcl}
\kappa(H_a,H_b) &=& I_{\mathrm{adj}} A^{ab} \\
\kappa(E_\alpha,F_\beta) &\sim& \delta_{\alpha,\beta},
\end{array}
\right.
\eeq
where $A^{ab}$ is the Cartan matrix of e$_{11}$.

Taking e$_{11}$ and its subalgebra invariant under the Cartan
involution to be the final global symmetry algebra and the local
subalgebra, respectively, will yield an infinite number of preferred
fields. This is due to the fact that e$_{11}$ is infinite dimensional.
The problem may be solved by the inverse Higgs mechanism
\cite{Ivanov:1975zq}, where the Goldstone fields superfluous for the
content of the theory can be eliminated by expressing them in terms of
other preferred fields. The remaining Goldstone bosons will then
correspond to the generators in \eqnref{G11_modified}.

The addition of $R^{c_1\ldots c_8,b}$ gives some new insight on the
relation between eleven dimensional and IIA supergravities. Separating
off the eleventh index we define
\beq
\begin{array}{c}
\tilde{K}^a_{\phantom{a}b} = K^a_{\phantom{a}b}; \hspace{0.2 cm}
\tilde{R}^a = K^a_{\phantom{a}11}; \hspace{0.2 cm} \tilde{R}^{a_1a_2}
= R^{a_1a_211}; \hspace{0.2 cm} \tilde{R}^{a_1a_2a_3} = R^{a_1a_2a_3}
\\
\tilde{R}^{a_1\ldots a_5} = R^{a_1\ldots a_5 11}; \hspace{0.2 cm}
\tilde{R}^{a_1\ldots a_6} = -R^{a_1\ldots a_6}; \hspace{0.2 cm}
\tilde{R}^{a_1\ldots a_7} = \frac{1}{2}R^{a_1\ldots a_7 11,11} \\
\tilde{R}^{a_1\ldots a_8} = \frac{3}{8}R^{a_1\ldots a_8,11};
\hspace{0.2 cm} \tilde{R} = \frac{1}{12}\left(
  -\displaystyle\sum_{a=1}^{10} K^a_{\phantom{a}a} +
  8K^{11}_{\phantom{11}11} \right),
\end{array}
\eeq
with the indices $a,\ldots = 1,\ldots,10$. The algebra made of these
generators is exactly the one, on which nonlinear realisation of IIA
supergravity is based
\cite{West:2000ga,West:2001as,Schnakenburg:2002xx}. This indicates
that the final symmetry algebra of IIA is also e$_{11}$. Observe that
we can make the identification $R^{a_1\ldots a_8,11} =
-8R^{11a_1\ldots a_7,a_8}$ due to \Eqnref{9gen_constr}.

\section{Levels and representations}

To understand the Kac-Moody algebra e$_{11}$ a little better, we do
some systematic analysis. Consider the Lorentzian Kac-Moody algebras
from \Chref{algebra}, i.e., algebras whose Dynkin diagram $C$
possesses at least one node $\alpha_c$ such that deleting it leads to
finite dimensional semi-simple Lie algebras and at most one affine
algebra. For simplicity we assume that the remaining Lie algebra is an
irreducible finite dimensional algebra with the corresponding reduced
Dynkin diagram $C_R$. The simple roots of $C_R$ are denoted by
$\alpha_i$, $i=1,\ldots,r-1$, while the central node is defined as
\beq
\alpha_c = -\nu + x, \hspace{0.5 cm} \nu =
-\displaystyle\sum_{i=1}^{r-1}A_{ci}\la_i, 
\eeq
where $x$ is a vector in a space orthogonal to $\alpha_i$, and $\la_i$
are the fundamental weights of $C_R$. We will also assume that
$(\alpha_c,\alpha_c) = 2$, and hence $x^2=2-\nu^2$. Any positive root
can be written as
\beq
\alpha = l\alpha_c + \displaystyle\sum_{i=1}^{r-1}m_i\alpha_i = lx -
\Lambda, 
\eeq
where 
\beq
\Lambda = -l\displaystyle\sum_{i=1}^{r-1}A_{ci}\la_i -
\displaystyle\sum_{i,j=1}^{r-1} m_i A^f_{ij} \la_j
\eeq
and $A^f_{jk}$ is the Cartan matrix for $C_R$. The integers $l$ and
$m_i$ are non-negative. Moreover, $l$ is called the level and counts
the number of times the root $\alpha_c$ occurs. For a fixed $l$, the
root space of $C$ can be described by its representation content with
respect to $C_R$. This can be done due to the fact that $\Lambda$ is a
positive linear combination of the fundamental weights of $C_R$.

An irreducible representation is completely specified by its highest
weight. Thus, at a certain level, whether or not a representation of
$C_R$ is contained in $C$ depends on the existence of positive
integers $p_j$ satisfying
\beq
p_j \equiv (\Lambda,\alpha_j^\vee) = -lA_{cj} -
\displaystyle\sum_{k=1}^{r-1} m_k A^f_{kj} \ge 0,
\eeq
with $l$ and $m_k$ being non-negative integers. We can then write the
constraint for the allowed $C_R$ representations in $C$ as
\beq
\eqnlab{highest_weight}
\displaystyle\sum_{j=1}^{r-1}(A^f)^{-1}_{jk} p_j = -l
\displaystyle\sum_{j=1}^{r-1}A_{cj}(A^f)^{-1}_{jk} - m_k,
\eeq
where $p_j$ are the components of the highest weight in the Dynkin
basis that uniquely defines a representation of $C_R$. Note that the
left hand side and the first term on the right hand side are both
non-negative, while the numbers $m_k$ take the values
$m_k=0,1,2,\ldots$ \cite{Kac:1990gs}. Assuming that we are only
dealing with symmetric Cartan matrices, the length of the roots are
bounded by $\alpha^2 = 2,1,0,\ldots$. Generally we have
\beq
\eqnlab{root_squared}
\alpha^2 = l^2x^2 + \displaystyle\sum_{i,j=1}^{r-1}p_i p_j
(\la_i,\la_j) = l^2\frac{\det{A_C}}{\det{A_{C_R}}} +
\displaystyle\sum_{i,j=1}^{r-1}p_i (A^f)^{-1}_{ij} p_j =
2,1,0,\ldots
\eeq
where \Eqnref{A_subA} has been used. For fixed $l$ the first term in
\eqnref{root_squared} is also fixed. It then automatically follows
that in the root space of $C$, we can never take a positive root to a
negative one by acting with raising and lowering generators of $C_R$.
Any representation of $C_R$ must be contained entirely in the positive
root space of $C$, but with a copy also in the negative root space due
to the symmetry.

The \Eqsref{highest_weight} and \eqnref{root_squared} are all we need
to find potential highest weight vectors of representations of the
reduced subalgebra $C_R$, among the vectors in the positive root
lattice of $C$ which have length squared $2,1,0,\ldots$. For a certain
$C_R$-representation to actually belong to the root generators of $C$,
the Serre relations and the Jacobi identities must be satisfied. Those
vectors leading to generators which do not satisfy these additional
requirements belong in fact to the weight lattice of $C$. Furthermore,
the root multiplicity of the Kac-Moody algebras is an unsolved issue,
except for a few special cases.

\subsection{Low level representations of e$_{11}$}

Using the concept of levels, we can systematically study the
representation content of e$_{11}$. More specifically, we can express
all the root generators of e$_{11}$ in terms of representations of
A$_{10}$. The Dynkin diagram of e$_{11}$ is given in
\Figref{DynkinE11}. We consider the node above the horizontal line as
the preferred node $\alpha_c = \alpha_{11}$, deleting which yields the
Lie algebra A$_{10}$. The nodes on the horizontal line are denoted by
$\alpha_i$, $i=1,\ldots,10$ from left to right. We use $D$ when
referring to the dimension of the very extended Kac-Moody algebra, in
this case $D=11$. Let us now level by level go through the positive
root lattice.

The level zero contains the adjoint representation of A$_{10}$ and the
generator $D$. The inverse Cartan matrix of A$_{D-1}$ is
\cite{Fuchs:1997jv,Gaberdiel:2002db} 
\beq
(A^f)^{-1}_{jk} = \left\{ \begin{array}{rcl}
\frac{j(r+1-k)}{r+1} = \frac{j(D-k)}{D} &,& j\le k \\
\frac{k(r+1-j)}{r+1} = \frac{k(D-j)}{D} &,& j\ge k
\end{array} \right.,
\eeq
inserting which into \Eqnref{highest_weight} with $D=11$ will give
\beq
\eqnlab{e11_high_w}
\displaystyle\sum_{j\le k}j(11-k)p_j + \displaystyle\sum_{j>k} k(11-j)
p_j = -11n_k + l\left\{ \begin{array}{ccl}
3k &,& k \le 8 \\
8(11-k) &,& k=9,10 \\
\end{array}\right.,
\eeq
$n_k = 0,1,2,\ldots$. Also, \Eqnref{root_squared} becomes
\beq
\alpha^2 = -\frac{2}{11}l^2 + \frac{1}{11}\displaystyle\sum_{i\le j}
p_i i(11-j) p_j + \frac{1}{11}\displaystyle\sum_{i>j} p_i j(11-i) p_j
\le 2,1,0,\ldots,
\eeq
where $\det{A_{e_{n}}} = 9-n$ and $\det{A_{A_n}} = n+1$ have been
used. It is then straightforward to work out the A$_{10}$
representations for every fixed value of $l$. At the lowest levels
these are given in \Tabref{e11_level}.
\begin{table}[h!]
  \begin{center}
    \begin{tabular}{l|l|l}
      Level & Highest weight & A$_{10}$ representation\\ \hline 
      $l = 0$  & $(1,0,0,0,0,0,0,0,0,1)$ & $K^a_{\phantom{a}b}$
      \hspace{1 cm} (A$_{10}\oplus D$)\\ 
      $l = 1$  & $(0,0,0,0,0,0,0,1,0,0)$ & $R^{c_1c_2c_3}$ \\
      $l = 2$  & $(0,0,0,0,1,0,0,0,0,0)$ & $R^{c_1\ldots c_6}$ \\
      $l = 3$  & $(0,0,1,0,0,0,0,0,0,1)$ & $R^{c_1\ldots c_8,d}$
      ($R^{[c_1\ldots c_8,d]}=0$) \\ 
               & $(0,1,0,0,0,0,0,0,0,0)$ & $\star$ $R^{c_1\ldots c_9}$
      \\ 
      $l = 4$  & $(0,1,0,0,0,0,0,1,0,0)$ & \\
               & $(1,0,0,0,0,0,0,0,0,2)$ & \\
               & $(1,0,0,0,0,0,0,0,1,0)$ & $\star$ \\
               & $(0,0,0,0,0,0,0,0,0,1)$ & \\
    \end{tabular}
    \caption{\textit{The A$_{10}$ representations occurring at the
        lowest levels of e$_{11}$
        \cite{Nicolai:2003fw,Kleinschmidt:2003mf,Kleinschmidt:2003jf}.
        The representations marked with $\star$ do not belong to the
        root space of e$_{11}$, but they do belong to the weight
        space.}}
    \tablab{e11_level}
  \end{center}
\end{table}
Note that the highest weight of each representation is written in the
Dynkin basis. All these generators appear in the nonlinear realisation
of eleven dimensional supergravity, except $R^{c_1\ldots c_9}$ which
are omitted from the e$_{11}$ algebra since they do not fulfill the
Jacobi identities and have also vanishing outer multiplicities
\cite{Nicolai:2003fw,Kleinschmidt:2003jf}. To obtain the whole
e$_{11}$ algebra up to level three, we must also add the negative root
generators $R_{c_1c_2c_3}$, $R_{c_1\ldots c_6}$ and $R_{c_1\ldots
  c_8,d}$. The commutation relations are those given in
\eqnref{G11_modified}, omitting the first row, together with
\beq
\begin{array}{rcl}
\com{K^a_{\phantom{a}b},R_{c_1c_2c_3}} &=&
-3\delta^a_{[c_1}R_{|b|c_2c_3]} \\
\com{K^a_{\phantom{a}b},R_{c_1\ldots c_6}} &=&
-6\delta^a_{[c_1}R_{|b|c_2\ldots c_6]} \\
\com{K^a_{\phantom{a}b},R_{c_1\ldots c_8,d}} &=&
-8\delta^a_{[c_1}R_{|b|c_2\ldots c_8],d} - \delta^a_d R_{c_1\ldots
  c_8,b} \\
\com{R_{c_1c_2c_3},R_{c_4c_5c_6}} &=& 2R_{c_1\ldots c_6} \\
\com{R_{a_1\ldots a_6},R_{b_1b_2b_3}} &=& 3R_{a_1\ldots
  a_6[b_1b_2,b_3]} \\
\com{R^{a_1a_2a_3},R_{b_1b_2b_3}} &=&
36\delta^{[a_1a_2}_{[b_1b_2}K^{a_3]}_{\phantom{a_3]}b_3]} -
4\delta^{a_1a_2a_3}_{b_1b_2b_3} D \\
\com{R_{a_1a_2a_3},R^{b_1\ldots b_6}} &=& -\frac{6!}{3!}
\delta^{[b_1b_2b_3}_{a_1a_2a_3}R^{b_4b_5b_6]},
\end{array}
\eeq
where $\delta^{a_1a_2}_{b_1b_2} \equiv \delta^{[a_1}_{b_1}
\delta^{a_2]}_{b_2}$. The Cartan and simple root generators are given
in \eqnref{e11_Cartan} and \eqnref{e11_simple}, respectively.

The fundamental weights of e$_{11}$ are
\beq
\eqnlab{e11_weight}
l_i = \la_i + (\nu,\la_i)\frac{x}{x^2}, \hspace{0.3 cm} l_c =
\frac{x}{x^2}, 
\eeq
where $\la_i$ are the fundamental weights of A$_{10}$ and $\nu =
-\sum_i A_{ci}\la_i = \la_8$. The representation $R^{c_1\ldots c_9}$
we omitted corresponds to the vector $(3x-\la_2)$, which we now
recognise as a negative fundamental weight $-l_2$ of
e$_{11}$. Considering $-l_2$ as the lowest weight of an e$_{11}$
representation, we find in the root string the vector
\beq
-l_2 + \alpha_2 + \ldots + \alpha_8 + \alpha_{11} = 4x - \la_1 -
\la_9,
\eeq
this vector appears also at level four in \Tabref{e11_level}. The
A$_{10}$ representation corresponding to this vector must as well be
discarded from the root generators of e$_{11}$, since they are
included in a basic representation of e$_{11}$.

The subalgebra invariant under the Cartan involution is generated by
$(E_{\alpha}-F_{\alpha})$ and at low levels it includes the generators
\beq
\begin{array}{rcl}
J_{ab} &=& K^{c}_{\phantom{c}b}\eta_{ac} -
K^{c}_{\phantom{c}a}\eta_{bc} \\
S_{a_1a_2a_3} &=& R^{b_1b_2b_3}\eta_{b_1a_1}\eta_{b_2a_2}\eta_{b_3a_3}
- R_{a_1a_2a_3} \\
S_{a_1\ldots a_6} &=& R^{b_1\ldots
  b_6}\eta_{b_1a_1}\ldots\eta_{b_6a_6} + R_{a_1\ldots a_6} \\
S_{a_1\ldots a_8,c} &=& R^{b_1\ldots b_8,d}\eta_{b_1a_1}\ldots
\eta_{b_8a_8}\eta_{dc} - R_{a_1\ldots a_8,c}.
\end{array}
\eeq
These generators belong to the maximal compact subalgebra of the
normal real form of e$_{11}$. The generators $J_{ab}$ constitute a
representation for the Lorentz algebra, while $S_{a_1a_2a_3}$ and
$S_{a_1\ldots a_6}$ obey
\beq
\eqnlab{involution_inv}
\begin{array}{rcl}
\com{S^{a_1a_2a_3},S_{b_1b_2b_3}} &=& -36\delta^{[a_1a_2}_{[b_1b_2}
J^{a_3]}_{\phantom{a_3]}b_3]} +
2S^{a_1a_2a_3}_{\phantom{a_1a_2a_3}b_1b_2b_3} \\
\com{S_{a_1a_2a_3},S^{b_1\ldots b_6}} &=& -\frac{6!}{3!}
\delta^{[b_1b_2b_3}_{a_1a_2a_3}S^{b_4b_5b_6]} - 3S^{b_1\ldots
  b_6}_{\phantom{b_1\ldots b_6}[a_1a_2,a_3]}.
\end{array}
\eeq

\subsection{Low level representations of A$_{D-3}^{+++}$}

Let us first consider the very extension of A$_8$, see
\Figref{DynkinA8ppp}. 
\begin{figure}
  \begin{center}
    \includegraphics[scale=0.5]{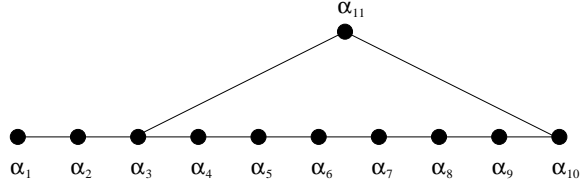}
    \caption{\textit{The Dynkin diagram of the Kac-Moody algebra
        A$_8^{+++}$.}} 
    \figlab{DynkinA8ppp}
  \end{center}
\end{figure}
We take $\alpha_{11}$ to be the preferred node, deleting which gives
the finite dimensional A$_{10}$. The subalgebra A$_{10}$ is used to
decompose the root system of A$_8^{+++}$. \Eqnref{highest_weight}
becomes 
\beq
\eqnlab{A8ppp_high_w}
\displaystyle\sum_{j\le k}j(11-k)p_j + \displaystyle\sum_{j>k} k(11-j)
p_j = -11m_k + l\left\{ \begin{array}{ccl}
9k &,& k = 1,2 \\
33-2k &,& k \ge 3
\end{array}\right.,
\eeq
where $m_k = 0,1,2,\ldots$. Using also \Eqnref{root_squared} we find
the highest weights of A$_{10}$ representations in the root lattice of
A$_8^{+++}$, the lowest levels are given in \Tabref{A8ppp_level}.
\begin{table}[h!]
  \begin{center}
    \begin{tabular}{l|l|l}
      Level & Highest weight & A$_{10}$ representation\\ \hline 
      $l = 0$  & $(1,0,0,0,0,0,0,0,0,1)$ & $K^a_{\phantom{a}b}$
      \hspace{1 cm} (A$_{10}\oplus D$)\\
      $l = 1$  & $(0,0,1,0,0,0,0,0,0,1)$ & $R^{c_1\ldots c_8,d}$
      ($R^{[c_1\ldots c_8,d]}=0$) \\
               & $(0,1,0,0,0,0,0,0,0,0)$ & $\star$ $R^{c_1\ldots c_9}$
      \\ 
      $l = 2$  & $(0,0,0,0,1,0,0,0,0,1)$ & \\
               & $(0,0,0,0,0,1,0,0,1,0)$ & \\ 
               & $(1,0,1,0,0,0,0,0,0,0)$ & \\
               & $(1,0,0,1,0,0,0,0,0,1)$ & $\star$ \\
    \end{tabular}
    \caption{\textit{The A$_{10}$ representations occurring at the
        lowest levels of A$_8^{+++}$
        \cite{Kleinschmidt:2003mf,Kleinschmidt:2003jf}.  The
        representations marked with $\star$ belong only to the weight
        space.}}
    \tablab{A8ppp_level}
  \end{center}
\end{table}
Since A$_8$ is a subalgebra of e$_8$, the very extended A$_8$ must as
well be a subalgebra of e$_{11}$. This means that all the A$_{10}$
representations occurring in the adjoint representation of A$_8^{+++}$
have to appear in the adjoint representation of e$_{11}$. Comparing
the levels $3l$ and $l$ in \eqnref{e11_high_w} and
\eqnref{A8ppp_high_w}, respectively, we find these coincide if we at
the same time make the identifications
\beq
n_k = \left\{ \begin{array}{lcl}
m_k &,& k=1,2 \\
m_k + l(k-3) &,& k=3,\ldots,8 \\
m_k + l(21-2k) &,& k=9,10.
\end{array} \right.
\eeq
It then follows that those A$_{10}$ representations found at level $l$
for A$_8^{+++}$ will also be found at level $3l$ for e$_{11}$, this is
verified by comparing \Tabsref{e11_level} and \tabref{A8ppp_level}.

We generalise now the results to A$_{D-3}^{+++}$ of arbitrary
dimension $D$, the Dynkin diagram in \Figref{DynkinA8ppp} can readily
be generalised. Omitting the preferred node, we find the finite
dimensional Lie algebra A$_{D-1}$. At level zero we have again the
generators $K^a_{\phantom{a}b}$ forming an adjoint representation for
the algebra A$_{D-1}\oplus D$.  The A$_{D-1}$ representations at all
the other levels are obtained by using \Eqsref{highest_weight} and
\eqnref{root_squared}. In particular, at level one we have
\beq
\begin{array}{rlll}
l = 1: & p_3=1, p_{D-1}=1 & R^{c_1\ldots c_{(D-3)},d} & (R^{[c_1\ldots
  c_{(D-3)},d]} = 0) \\
 & p_{2}=1 & R^{c_1\ldots c_{(D-2)}}.
\end{array}
\eeq
The solution $p_2=1$ does not belong to the root space of
A$_{D-3}^{+++}$, instead, it is the negative fundamental weight $-l_2
= x - \la_2$. Observing that the root string of $-l_2$ includes the
representation $(p_1=1, p_4=1, p_{D-1}=1)$ at level two, we also must
exclude this A$_{D-1}$ representation from the root space. This
explains the exclusion of $R^{c_1\ldots c_9}$ from the A$_8^{+++}$
algebra.

The level zero and one generators of A$_{D-3}^{+++}$ obey the Lie
brackets 
\beq
\begin{array}{rcl}
\com{K^a_{\phantom{a}b},K^c_{\phantom{c}d}} &=& \delta^c_b
K^a_{\phantom{a}d} - \delta^a_d K^c_{\phantom{c}b} \\
\com{K^a_{\phantom{a}b},R^{c_1\ldots c_{(D-3)},d}} &=&
(D-3)\delta^{[c_1}_bR^{|a|c_2 \dots c_{(D-3)}],d} + \delta^d_b
R^{c_1\ldots c_{(D-3)},a}
\end{array}
\eeq
plus those involving the negative root generators
$R_{c_1\ldots c_{(D-3)},d}$.

\section{Dual formulation of gravity}

The preferred fields in the G$_{11}$ nonlinear realisation are
$m_a^{\phantom{a}b}$, $A_{c_1c_2c_3}$ and $A_{c_1\ldots c_6}$. The
gauge fields are related through the equations of motion
\eqnref{dualA}, while $m_a^{\phantom{a}b}$ describes gravity.  Since
including the Borel subalgebra of e$_8$ requires the addition of the
generators $R^{c_1\ldots c_8,d}$, there must exist a preferred field
of the form $A_{c_1\ldots c_8,d}$. The relation between the gauge
fields $A_{c_1c_2c_3}$ and $A_{c_1\ldots c_6}$ indicates the existence
of some first order formulation of gravity using the fields
$m_a^{\phantom{a}b}$ and $A_{c_1\ldots c_8,d}$. Regarding the lower
index on $m_a^{\phantom{a}b}$ as some internal index, the field
strength is found to be $f_{a_1a_2}^{\phantom{a_1a_2}b}$, with
$f_{a_1\ldots a_9}^{\phantom{a_1\ldots a_9}b}$ being its ``dual gauge
field''. We will now construct this dual formulation of gravity
explicitly following the references \cite{West:2001as,West:2002jj}.

\subsection{The Einstein-Hilbert action}

In a $D$ dimensional space-time the Einstein-Hilbert action is given
by \cite{Weinberg:1972}
\beq
S = \int d^Dx\,eR(\bein_\mu^{\phantom{\mu}a},\omega_{\mu
  a}^{\phantom{\mu a}b}(\bein))
\eeq
with $e = \sqrt{|\det g_{\mu \nu}|}$. Defining
$C_{\mu\nu}^{\phantom{\mu\nu}a} \equiv
\pa_\mu\bein_\nu^{\phantom{\nu}a} -
\pa_\nu\bein_\mu^{\phantom{\mu}a}$, the action can be written as
\beq
\eqnlab{C_EH}
S = \int d^Dx\,e\left(C_{ca,}^{\phantom{ca,}a}
  C^{cb,}_{\phantom{cb,}b} - \frac{1}{2}C_{ab,c} C^{ac,b} -
  \frac{1}{4} C_{ab,c} C^{ab,c}\right),  
\eeq 
where \eqnref{connect_vielbein} has been used. There is an equivalent
way to write the action
\beq
\eqnlab{YC_EH}
S = \frac{1}{2}\int d^Dx\,e\left(Y^{ab,c}C_{ab,c} +
  \frac{1}{2}Y^{ab,c}Y_{ac,b} - \frac{1}{2(D-2)}\eta_{bc} Y^{ab,c}
  Y_{af,}^{\phantom{af,}f} \right)
\eeq
by introducing some field $Y^{ab,c}$. Varying \Eqnref{YC_EH}
\beq
\delta S = \frac{1}{2}\int d^Dx\,e\left\{ C_{ab,c} + Y_{ac,b} -
  \frac{1}{D-2}\eta_{bc}Y_{af,}^{\phantom{af,}f} \right\} \delta
Y^{ab,c} = 0
\eeq
yields
\beq
\eqnlab{YC_rel}
\begin{array}{rcl}
Y_{[a|c|,b]} &=& -C_{ab,c} + \frac{1}{D-2}\eta_{c[b}
Y_{a]f,}^{\phantom{a]f,}f} \hspace{2 cm} \Rightarrow \\
Y_{ab,c} &=& C_{bc,a} + C_{ca,b} - C_{ab,c} +
2(\eta_{bc}C_{af,}^{\phantom{af,}f} -
\eta_{ac}C_{bf,}^{\phantom{bf,}f}).
\end{array}
\eeq
Putting \eqnref{YC_rel} into \eqnref{YC_EH} gives \eqnref{C_EH}, and
thus shows the equivalence. Introduce the field $Y_{c_1\ldots
  c_{(D-2)},}^{\phantom{c_1\ldots c_{(D-2)},}d}$ dual to $Y^{ab,c}$
\beq
\eqnlab{Y_dual}
Y^{ab,d} \equiv \frac{1}{(D-2)!}\varepsilon^{abc_1\ldots c_{(D-2)}}
Y_{c_1\ldots c_{(D-2)},}^{\phantom{c_1\ldots c_{(D-2)},}d}.
\eeq
Note that the indices denoted by Roman letters are flat. Expressed in
terms of the dual field the action becomes
\beq
\eqnlab{Ydual_EH}
\begin{array}{rcl}
S &=& \frac{1}{2(D-2)!} \int d^Dx\,\left\{
  \varepsilon^{\mu\nu\ta_1\ldots \ta_{(D-2)}} Y_{\ta_1\ldots
    \ta_{(D-2)},}^{\phantom{\ta_1\ldots \ta_{(D-2)},}d} C_{\mu\nu,d}
  \right. \\
& & + e\left[-\frac{D-3}{2(D-2)}Y^{c_1\ldots c_{(D-2)},a} Y_{c_1\ldots
  c_{(D-2)},a} \right. \\
& & \left. \left. + \frac{D-2}{2} Y^{d_1\ldots
  d_{(D-3)}c,}_{\phantom{d_1\ldots d_{(D-3)}c,}c} Y_{d_1\ldots
  d_{(D-3)}d,}^{\phantom{d_1\ldots d_{(D-3)}d,}d} - \frac{1}{2}
  Y^{c_1\ldots c_{(D-3)}a,d}Y_{c_1\ldots c_{(D-3)}d,a} \right]
  \right\},
\end{array}
\eeq
observe that $\varepsilon^{a_1\ldots a_D} \equiv e^{-1}
\bein_{\mu_1}^{\phantom{\mu_1}a_1} \ldots
\bein_{\mu_D}^{\phantom{\mu_D}a_D} \varepsilon^{\mu_1\ldots\mu_D}$ to
emphasise the fact that $\varepsilon^{1\ldots D} = 1$ despite the type
of indices.

Changing the indices $a$ and $b$ in \Eqnref{YC_rel} to curved
space-time indices
\beq
Y_{\mu\nu,c} = C_{\nu c,\mu} + C_{c\mu,\nu} - C_{\mu\nu,c} +
2(\bein_{\nu c}C_{\mu f,}^{\phantom{\mu f,}f} -
\bein_{\mu c}C_{\nu f,}^{\phantom{\nu f,}f})
\eeq
and using the definition \eqnref{Y_dual}, we get
\beq
\eqnlab{eom_grav_Y}
\begin{array}{rl}
\frac{e^{-1}}{(D-2)!} &
 \varepsilon_{\mu\nu}^{\phantom{\mu\nu}\ta_1\ldots \ta_{(D-2)}}
 Y_{\ta_1\ldots \ta_{(D-2)},c} \\  
 & = -C_{\mu\nu,c} + C_{c \mu,\nu} - C_{c \nu,\mu} + 2(\bein_{\nu
 c}C_{\mu f,}^{\phantom{\mu f,}f} - \bein_{\mu c}C_{\nu
 f,}^{\phantom{\mu f,}f}). 
\end{array}
\eeq
On the other hand, varying the action \eqnref{Ydual_EH} with respect
to the vielbein $\bein_{\mu a}$, while regarding $Y_{\ta_1\ldots
  \ta_{(D-2)},}^{\phantom{\ta_1\ldots \ta_{(D-2)},}a}$ as some
independent field gives
\beq
\begin{array}{rcl}
\delta S &=& \frac{1}{2(D-2)!}\int d^Dx \{
2\varepsilon^{\mu\nu\ta_1\ldots \ta_{(D-2)}}\pa_\nu Y_{\ta_1\ldots
  \ta_{(D-2)},}^{\phantom{\ta_1\ldots \ta_{(D-2)},}a} \\
& & + [\textrm{terms of order }(Y_{\ta_1\ldots
  \ta_{(D-2)},}^{\phantom{\ta_1\ldots \ta_{(D-2)},}a})^2] \} \delta
\bein_{\mu a} = 0
\end{array}
\eeq
or
\beq
\varepsilon^{\mu\ta_1\ldots \ta_{(D-1)}}\pa_{\ta_1} Y_{\ta_2\ldots
  \ta_{(D-1)},}^{\phantom{\ta_2\ldots \ta_{(D-1)},}a} = [\textrm{terms
  of order }(Y_{\ta_1\ldots \ta_{(D-2)},}^{\phantom{\ta_1\ldots
    \ta_{(D-2)},}a})^2].
\eeq
To lowest order in $Y_{\ta_1\ldots \ta_{(D-2)},}^{\phantom{\ta_1\ldots
    \ta_{(D-2)},}a}$ the equations of motion for gravity become
\beq
\eqnlab{eom_grav_e}
\varepsilon^{\mu\ta_1\ldots \ta_{(D-1)}}\pa_{\ta_1} Y_{\ta_2\ldots
  \ta_{(D-1)},}^{\phantom{\ta_2\ldots \ta_{(D-1)},}a} = 0,
\eeq
leading to the conclusion
\beq
\eqnlab{lin_solution}
Y_{\ta_1\ldots \ta_{(D-2)},a} = \pa_{[\ta_1}m_{\ta_2\ldots
  \ta_{(D-2)}],a}.
\eeq
We have thus shown that there exists a dual formulation of gravity
based on the fields $\bein_\mu^{\phantom{\mu}a}$ and $m_{\ta_1\ldots
  \ta_{(D-3)}}^{\phantom{\ta_1\ldots \ta_{(D-3)},}a}$. In particular
when $D=11$, the preferred field corresponding to the generators
$R^{c_1\ldots c_8,d}$ does really belong to the field content of
eleven dimensional supergravity. One finds that the local Lorentz
transformation $\delta\bein_{\mu\nu}=\la_{\mu\nu}$ is a symmetry only
if $\delta m_{\ta_2\ldots \ta_{(D-2)},a}=
-\varepsilon_{\ta_2\ldots\ta_{(D-2)}a\rho\ka}\la^{\rho\ka}$
\cite{West:2002jj}. 

Based on these considerations, it is conjectured that pure gravity in
$D$ dimensions could be described as nonlinear realisation based on
the very extended A$_{D-3}$. This is also consistent with the
observation that A$_{D-3}^{+++}$ is a subalgebra of e$_{11}$, since
eleven dimensional supergravity contains gravity. The preferred fields
$m_a^{\phantom{a}b}$ and $m_{c_1\ldots c_{(D-3)},d}$ used in the dual
formulation of gravity correspond precisely to the generators
$K^a_{\phantom{a}b}$ and $R^{c_1\ldots c_{(D-3)},d}$ at levels zero
and one for A$_{D-3}^{+++}$. The major difference is that
$R^{[c_1\ldots c_{(D-3)},d]} = 0$, while $m_{c_1\ldots c_{(D-3)},d}$
does not have a such constraint.

\subsection{The field equations}

Instead of starting from the Einstein-Hilbert action, we now want to
derive the Einstein's equation from field equations
\cite{West:2002jj}. Define a $(D-2)$-form
\beq
Y_a \equiv \varepsilon_{ab_1\ldots b_{(D-1)}} \bein^{b_1} \we \ldots
\we \bein^{b_{(D-3)}} \we \om^{b_{(D-2)}b_{(D-1)}}
\eeq
with $\bein^a = \dxmu\bein_{\mu}^{\phantom{\mu}a}$ and $\om^{ab} =
\dxmu\om_{\mu}^{\phantom{\mu}ab}$. Written as an antisymmetric tensor
we get
\beq
\eqnlab{Y_tensor}
Y_{\rho_1\ldots\rho_{(D-2)},a} = (D-2)!\,\varepsilon_{ab_1\ldots
  b_{(D-1)}}\bein_{[\rho_i}^{\phantom{[\rho_i}b_1}\ldots
\bein_{\rho_{(D-3)}}^{\phantom{\rho_{(D-3)}}b_{(D-3)}}
\om_{\rho_{(D-2)}]}^{\phantom{\rho_{(D-2)}]}b_{(D-2)}b_{(D-1)}}.
\eeq
Define also a $(D-1)$-form
\beq
\begin{array}{rcl}
M_a &\equiv& -\varepsilon_{ab_1\ldots b_{(D-1)}}\{ (D-3)
\om^{b_1}_{\phantom{b_1}f} \we \bein^f \we \bein^{b_2} \we \ldots \we
\bein^{b_{(D-3)}} \we \om^{b_{(D-2)}b_{(D-1)}} \\
& & + (-1)^{D-3} \bein^{b_1} \we \ldots \we \bein^{b_{(D-3)}} \we
\om^{b_{(D-2)}f} \we \om_f^{\phantom{f}b_{(D-1)}} \}
\end{array}
\eeq
or as a tensor 
\beq
\eqnlab{M_tensor}
\begin{array}{rl}
M_{\rho_1\ldots \rho_{(D-1)},a} & =  -(D-1)!\varepsilon_{ab_1\ldots
  b_{(D-1)}} \times \\ 
& \{(D-3)\om_{[\rho_1\phantom{b_1}|f|}^{\phantom{[\rho_1}b_1}
  \bein_{\rho_2}^{\phantom{\rho_2}f}
  \bein_{\rho_3}^{\phantom{\rho_3}b_2} \ldots
  \bein_{\rho_{(D-2)}}^{\phantom{\rho_{(D-2)}}b_{(D-3)}}
  \om_{\rho_{(D-1)}]}^{\phantom{\rho_{(D-1)}]}b_{(D-2)}b_{(D-1)}} \\
& +(-1)^{D-3} \bein_{[\rho_1}^{\phantom{[\rho_1}b_1} \ldots
  \bein_{\rho_{(D-3)}}^{\phantom{\rho_{(D-3)}}b_{(D-3)}}
  \om_{\rho_{(D-2)}}^{\phantom{\rho_{(D-2)}}b_{(D-2)}f}
  \om_{\rho_{(D-1)}]f}^{\phantom{\rho_{(D-1)}]f}b_{(D-1)}} \}.
\end{array}
\eeq 
The equations of motion are then
\beq
\eqnlab{nonlinear_eom}
\begin{array}{rcl}
dY_a - M_a &=& 0 \hspace{1 cm} \textrm{or} \\
\varepsilon^{\mu\nu\rho_1\ldots\rho_{(D-2)}} \{ (D-1)\pa_\nu
Y_{\rho_1\ldots \rho_{(D-2)},a} - M_{\nu\rho_1\ldots \rho_{(D-2)},a}
\} &=& 0.
\end{array}
\eeq
The torsion and the curvature tensor are defined as
\beq
\begin{array}{rcl}
T^a = d\bein^a + \om^a_{\phantom{a}b}\we\bein^b &\Rightarrow&
T_{\mu\nu}^{\phantom{\mu\nu}a} =
2(\pa_{[\mu}\bein_{\nu]}^{\phantom{\nu]}a} +
\om_{[\mu\phantom{a}|b|}^{\phantom{[\mu}a}
\bein_{\nu]}^{\phantom{\nu]}b}) \\
R^a_{\phantom{a}b} = d\om^a_{\phantom{a}b} +
\om^a_{\phantom{a}c}\we\om^c_{\phantom{c}b} &\Rightarrow& 
R_{\mu\nu\phantom{a}b}^{\phantom{\mu\nu}a} =
2(\pa_{[\mu}\om_{\nu]\phantom{a}b}^{\phantom{\nu]}a} +
\om_{[\mu\phantom{a}|c|}^{\phantom{[\mu}a}
\om_{\nu]\phantom{c}b}^{\phantom{\nu]}c}), \\ 
\end{array}
\eeq
respectively, with $R_{\nu b} = \bein^{\mu}_{\phantom{\mu}a}
R_{\mu\nu\phantom{a}b}^{\phantom{\mu\nu}a}$ and $R = \bein^{\nu b}
\bein^{\mu}_{\phantom{\mu}a}
R_{\mu\nu\phantom{a}b}^{\phantom{\mu\nu}a}$. Using
\Eqsref{Y_tensor}, \eqnref{M_tensor} and \eqnref{nonlinear_eom}, we
find
\beq
\begin{array}{c}
\varepsilon_{ab_1\ldots b_{(D-1)}} \varepsilon^{\mu\nu\rho_1\ldots
  \rho_{(D-2)}} \{ \frac{D-3}{2}
  T_{\nu\rho_1}^{\phantom{\nu\rho_1}b_1}
  \bein_{\rho_2}^{\phantom{\rho_2}b_2} \ldots
  \bein_{\rho_{(D-3)}}^{\phantom{\rho_{(D-3)}}b_{(D-3)}}
  \om_{\rho_{(D-2)}}^{\phantom{\rho_{(D-2)}}b_{(D-2)}b_{(D-1)}} \\
 \\
+ (-1)^{D-3} \bein_\nu^{\phantom{\nu}b_1}
  \bein_{\rho_1}^{\phantom{\rho_1}b_2} \ldots
  \bein_{\rho_{(D-4)}}^{\phantom{\rho_{(D-4)}}b_{(D-3)}}
  \om_{\rho_{(D-3)}}^{\phantom{\rho_{(D-3)}}b_{(D-2)}f}
  \om_{\rho_{(D-2)}f}^{\phantom{\rho_{(D-2)}f}b_{(D-1)}} \\
 \\
+ \bein_{\rho_1}^{\phantom{\rho_1}b_1} \ldots
  \bein_{\rho_{(D-3)}}^{\phantom{\rho_{(D-3)}}b_{(D-3)}}
  \pa_\nu\om_{\rho_{(D-2)}}^{\phantom{\rho_{(D-2)}}b_{(D-2)}b_{(D-1)}}
  \} = 0.
\end{array}
\eeq
Noticing the sum of the last two terms as the Einstein tensor, we
finally get 
\beq
\begin{array}{c}
\varepsilon_{ab_1\ldots b_{(D-1)}} \varepsilon^{\mu\nu\rho_1\ldots
  \rho_{(D-2)}} \{ \frac{D-3}{2}
  T_{\nu\rho_1}^{\phantom{\nu\rho_1}b_1}
  \bein_{\rho_2}^{\phantom{\rho_2}b_2} \ldots
  \bein_{\rho_{(D-3)}}^{\phantom{\rho_{(D-3)}}b_{(D-3)}}
  \om_{\rho_{(D-2)}}^{\phantom{\rho_{(D-2)}}b_{(D-2)}b_{(D-1)}} \} \\
 \\
-2(-1)^{D-2}e(D-3)! (R^\mu_{\phantom{\mu}a} -
  \frac{1}{2}\bein^\mu_{\phantom{\mu}a} R) = 0.
\end{array}
\eeq
If we use the Levi-Civita connection the torsion will vanish, then the
Einstein's equation in vacuum is obtained. \Eqnref{nonlinear_eom} is
second order in space-time derivatives, to find an equation system
with only first order derivatives we introduce the fields
\beq
\bein_\mu^{\phantom{\mu}a} \equiv (e^m)_\mu^{\phantom{\mu}a}
,\hspace{0.5 cm} m_{c_1\ldots c_{(D-3)},d}, \hspace{0.5 cm}
k_{c_1\ldots c_{(D-2)},d}.
\eeq
The equation system
\beq
\eqnlab{linear_eom}
\left\{ \begin{array}{l}
Y_a - k_a = dm_a + \Omega_a^{\phantom{a}b} \we m_b \\
dk_a + \Omega_a^{\phantom{a}b} \we k_b - \Omega_a^{\phantom{a}b} \we
Y_b - M_a = 0
\end{array} \right.
\eeq
with $\Omega_a^{\phantom{a}b} = \dxmu \Omega_{\mu a}^{\phantom{\mu
    a}b}$, $m_a = dx^{\rho_1} \we \ldots \we dx^{\rho_{(D-3)}}
m_{\rho_{1}\ldots\rho_{(D-3)},a}$ and $k_a = dx^{\rho_1} \we \ldots
\we dx^{\rho_{(D-2)}} k_{\rho_{1}\ldots\rho_{(D-2)},a}$, is just what
we are looking for, since putting the first line into the second
yields
\beq
dY_a - (d\Omega_a^{\phantom{a}b} + \Omega_a^{\phantom{a}c} \we
\Omega_c^{\phantom{c}b}) \we m_b - M_a = 0
\eeq
or
\beq
\eqnlab{grav_field_eqn}
\left\{ \begin{array}{l}
dY_a - M_a = 0 \\
d\Omega_a^{\phantom{a}b} + \Omega_a^{\phantom{a}c} \we
\Omega_c^{\phantom{c}b} = 0
\end{array} \right. .
\eeq
Note that the second line in \Eqnref{grav_field_eqn} is satisfied
automatically if $\Omega_{\mu a}^{\phantom{\mu a}b}$ is defined as
$\Omega_{\mu a}^{\phantom{\mu a}b} = (\bein^{-1}\pa_\mu
\bein)_a^{\phantom{a}b}$. \Eqnref{linear_eom} in terms of tensors is
\beq
\eqnlab{gravity_eom_tensor}
\left\{ \begin{array}{l}
\varepsilon^{\mu\nu\rho_1\ldots\rho_{(D-2)}}
\{Y_{\rho_1\ldots\rho_{(D-2)},a}-k_{\rho_1\ldots\rho_{(D-2)},a} -
(D-2)\hat{D}_{\rho_1} m_{\rho_2\ldots\rho_{(D-2)},a} \} = 0 \\
\varepsilon^{\mu\nu\rho_1\ldots\rho_{(D-2)}}
\{ \hat{D}_\nu k_{\rho_1\ldots\rho_{(D-2)},a} - \Omega_{\nu
  a}^{\phantom{\nu a}b}Y_{\rho_1\ldots\rho_{(D-2)},b} -
\frac{1}{D-1}M_{\nu\rho_1\ldots\rho_{(D-2)},a} \} = 0
\end{array} \right.
\eeq
with $\hat{D}_\mu m_{\rho_1\ldots\rho_{(D-3)},a} = \pa_\mu
m_{\rho_1\ldots\rho_{(D-3)},a} + \Omega_{\mu a}^{\phantom{\mu a}b}
m_{\rho_1\ldots\rho_{(D-3)},b}$.

\subsection{Nonlinear realisation of gravity}

The reformulation of gravity contains fields with the same index
structures as the low level generators of A$_{D-3}^{+++}$, written
with respect to A$_{D-1}$ representations. We can start from the other
end and try to find covariant equations, that are globally invariant
under A$_{D-3}^{+++}$ and locally invariant under its subgroup which
is invariant under the Cartan involution. We consider only the
generators at levels zero and one, namely $K^a_{\phantom{a}b}$,
$R^{c_1\ldots c_{(D-3)},d}$ and $R_{c_1\ldots c_{(D-3)},d}$. We take
\beq
\eqnlab{gravity_affine_coset}
g = e^{x^{f}P_{f}}e^{m_a^{\phantom{a}b}K^a_{\phantom{a}b}} e^{
  \hat{m}_{c_1\ldots c_{(D-3)},d}R^{c_1\ldots c_{(D-3)},d}} 
\eeq
to be the coset representative, where $\hat{m}_{[c_1\ldots
  c_{(D-3)},d]} = 0$ is also satisfied. The Cartan forms is then 
\beq
\V \equiv g^{-1}dg - \om = \dxmu\{ \bein_{\mu}^{\phantom{\mu}a}P_a +
\Omega_{\mu a}^{\phantom{\mu a}b} K^a_{\phantom{a}b} + \tilde{D}_\mu
\hat{m}_{c_1\ldots c_{(D-3)},d}R^{c_1\ldots c_{(D-3)},d} \} - \om
\eeq
with
\beq
\begin{array}{rcl}
\bein_{\mu}^{\phantom{\mu}a} &\equiv& (e^m)_{\mu}^{\phantom{\mu}a} \\
\Omega_{\mu a}^{\phantom{\mu a}b} &\equiv& \pa_\mu m_a^{\phantom{a}b}
= (\bein^{-1}\pa_\mu\bein)_a^{\phantom{a}b} \\ 
\om &=& \dxmu \om_{\mu a}^{\phantom{\mu a}b}K^{a}_{\phantom{a}b} \\
\tilde{D}_\mu \hat{m}_{c_1\ldots c_{(D-3)},d} &\equiv& \pa_\mu
\hat{m}_{c_1\ldots c_{(D-3)},d} + (D-3)
\Omega_{\mu[c_1}^{\phantom{\mu[c_1}a} \hat{m}_{|a|c_2\ldots
  c_{(D-3)}],d} \\ 
& &+ \Omega_{\mu d}^{\phantom{\mu d}a} \hat{m}_{c_1\ldots
  c_{(D-3)},a}.
\end{array}
\eeq

The Einstein's equation \eqnref{linear_eom} written in tangent indices
is indeed built up using the fields $\bein_\mu^{\phantom{\mu}a}$,
$\Omega_{ab}^{\phantom{ab}c}$, $\om_{ab}^{\phantom{ab}c}$ and
\beq
\begin{array}{l}
\tilde{D}_{c_1} \hat{m}_{c_2\ldots c_{(D-2)},d} \\
\hspace{1 cm} = \bein_{c_1}^{\phantom{c_1}\mu_1} \ldots
\bein_{c_{(D-2)}}^{\phantom{c_{(D-2)}}\mu_{(D-2)}}(\pa_{\mu_1}
\hat{m}_{\mu_2\ldots\mu_{(D-2)},d} + \Omega_{\mu_1 d}^{\phantom{\mu_1
    d}a} \hat{m}_{\mu_2\ldots\mu_{(D-2)},a}),
\end{array}
\eeq
but the appearance of $k_{c_1\ldots c_{(D-2)},d}$ and $m_{[c_1\ldots
  c_{(D-3)},d]}$ indicate that we must include higher level generators
of A$_{D-3}^{+++}$ in the nonlinear realisation. The fact that
A$_{D-3}^{+++}$ is infinite dimensional may be solved by the inverse
Higgs mechanism, leaving only a finite number of Goldstone fields as
result \cite{West:2002jj}. The final covariant objects are obtained by
simultaneous nonlinear realisation with the conformal group. Carrying
out this procedure should give \eqnref{linear_eom} as the unique
equations invariant under both groups up to the considered level.

\subsection{Some more features}

The generalisation to include matter fields is done by introducing a
rank $p$ gauge field $A_{(p)} = \frac{1}{p!}dx^{\mu_1} \we \ldots \we
dx^{\mu_p} A_{\mu_1\ldots\mu_p}$ with field strength $F_{(p+1)} =
dA_{(p)}$ or
\beq
F_{\mu_1\ldots\mu_{p+1}} = (p+1)\pa_{[\mu_1}A_{\mu_2\ldots\mu_{p+1}]}.
\eeq
The dual field strength is given by
\beq
G_{(D-p-1)} = {*}F_{(p+1)}.
\eeq
Introduce also
\beq
\left\{ \begin{array}{l}
F_{(p)a} = \frac{1}{p!}dx^{\mu_1} \we \ldots \we dx^{\mu_p}
F_{\mu_1\ldots\mu_p,a} \\
G_{(D-p-2)a} = \frac{1}{(D-p-2)!}dx^{\mu_1} \we \ldots \we
dx^{\mu_{(D-p-2)}} G_{\mu_1\ldots\mu_{(D-p-2)},a}
\end{array} \right.,
\eeq
using which we define
\beq
\hat{M}_a = M_a + \{G_{(D-p-1)} \we F_{(p)a} - (-1)^p G_{(D-p-2)a} \we
F_{(p+1)} \}.
\eeq
Replacing $M_a$ in \Eqnref{nonlinear_eom} or \eqnref{linear_eom} by
$\hat{M}_a$ we obtain
\beq
R^\mu_{\phantom{\mu}a} - \frac{1}{2}\bein^\mu_{\phantom{\mu}a}R =
\frac{(D-p)!}{(D-3)!p!} \{ F^{\mu\rho_1\ldots\rho_p}
  F_{a\rho_1\ldots\rho_p} - \frac{1}{2(p+1)}
  \bein^\mu_{\phantom{\mu}a}F^{\rho_1\ldots\rho_{p+1}}
  F_{\rho_1\ldots\rho_{p+1}} \}
\eeq
with the connection chosen so that the torsion vanishes. The right
hand side is precisely the desired contribution from the energy
momentum tensor. It can be verified explicitly that $d\hat{M}_a=0$,
which is consistent with the fact that the equations of motion
$\hat{M}_a = dY_a$ must conserve the energy and momentum.

The equation system with the linearised Einstein's equation involves
the unconstrained field $m_{c_1\ldots c_{(D-3)},d}$. However, one can
have a dual formulation of gravity which at the linearised level is
given by \cite{Curtright:1985yk,Hull:2001iu,Bekaert:2002uh}
\beq
R_{\mu\nu ab} =
\varepsilon_{\mu\nu}^{\phantom{\mu\nu}\rho_1\ldots\rho_{(D-2)}}
(\pa_a\pa_{[\rho_1} l_{\rho_2\ldots\rho_{(D-2)}],b} -
\pa_b\pa_{[\rho_1} l_{\rho_2\ldots\rho_{(D-2)}],a}),
\eeq
where $l_{[c_1\ldots c_{(D-3)},d]}=0$. Since the equation is second
order in space-time derivatives, it has an additional gauge symmetry
related to the local Lorentz transformations discussed below
\Eqnref{lin_solution}. This gauge symmetry can then be used to gauge
$l_{[c_1\ldots c_{(D-3)},d]}$ away, and hence from this point of view
$m_{[c_1\ldots c_{(D-3)},d]}$ is not really present. Trying to treat
gravity in the same way as the gauge fields, we want to have a dual
formulation of gravity, expressed by equations with only first order
space-time derivatives. To achieve this, we need only to introduce
$A_{c_1\ldots c_6}$ for the gauge field $A_{c_1c_2c_3}$, but the
non-linearity of the Einstein's equations requires an infinite number
of fields, explaining the infinite dimension of the conjectured
A$_8^{+++}$ symmetry.

\section{Nonlinear realisation based on e$_{11}$}

We want now to perform nonlinear realisation based on the generators
from the levels zero, one and two of e$_{11}$. Since the calculations
are almost identical to those in \Chref{nonlinear}, it will be done in
less detail here.

We use the subalgebra invariant under the Cartan involution as the
local subalgebra, a representative for the coset G/H is given by
\beq
g = e^{x^{f}P_{f}}e^{m_a^{\phantom{a}b}K^a_{\phantom{a}b}}
e^{\frac{1}{3!}A_{a_1a_2a_3}R^{a_1a_2a_3}} e^{\frac{1}{6!}A_{b_1\ldots
    b_6}R^{b_1\ldots b_6}} e^{\hat{m}_{c_1\ldots c_8,d}R^{c_1\ldots
    c_8,d}}
\eeq
with $\hat{m}_{[c_1\ldots c_8,d]}=0$. Calculating the Cartan form
\beq
\begin{array}{rcl}
\V &\equiv& g^{-1}dg - \om \\
&=& \dxmu\{ \bein_{\mu}^{\phantom{\mu}a}P_a +
\Omega_{\mu a}^{\phantom{\mu a}b} K^a_{\phantom{a}b} +
\frac{1}{3!}\tilde{D}_\mu A_{c_1c_2c_3}R^{c_1c_2c_3} \\
& & + \frac{1}{6!}\tilde{D}_\mu A_{c_1\ldots c_6}R^{c_1\ldots c_6}
+ \tilde{D}_\mu \hat{m}_{c_1\ldots c_8,d}R^{c_1\ldots c_8,d} \} - \om 
\end{array}
\eeq
we find
\beq
\eqnlab{e11_fields}
\begin{array}{rcl}
\bein_{\mu}^{\phantom{\mu}a} &\equiv& (e^m)_{\mu}^{\phantom{\mu}a} \\
\Omega_{\mu a}^{\phantom{\mu a}b} &\equiv&
(\bein^{-1}\pa_\mu\bein)_a^{\phantom{a}b} \\
\tilde{D}_\mu A_{c_1c_2c_3} &\equiv& \pa_\mu A_{c_1c_2c_3} +
3(\bein^{-1}\pa_\mu\bein)_{[c_1}^{\phantom{[c_1}b}A_{|b|c_2c_3]} \\
\tilde{D}_\mu A_{c_1\ldots c_6} &\equiv& \pa_\mu A_{c_1\ldots c_6} +
6(\bein^{-1}\pa_\mu\bein)_{[c_1}^{\phantom{[c_1}b}A_{|b|c_2\ldots
  c_6]} - 20A_{[c_1c_2c_3}\tilde{D}_{|\mu|}A_{c_4c_5c_6]} \\
\tilde{D}_\mu \hat{m}_{c_1\ldots c_8,d} &\equiv& \pa_\mu
\hat{m}_{c_1\ldots c_8,d} +
8(\bein^{-1}\pa_\mu\bein)_{[c_1}^{\phantom{[c_1}b}
\hat{m}_{|b|c_2\ldots c_8],d} - \textrm{Tr} \\
& & - \frac{1}{(3!)^3} A_{[c_1c_2c_3}(\tilde{D}_{|\mu|}
A_{c_4c_5c_6})A_{c_7c_8]d} - \frac{1}{3!6!}A_{[c_1\ldots
  c_6}\tilde{D}_{|\mu|}A_{c_7c_8]d}
\end{array}
\eeq
with $(-\textrm{Tr})$ denoting a term which ensures $\tilde{D}_\mu
\hat{m}_{[c_1\ldots c_8,d]} = 0$.

Simultaneous nonlinear realisation with the conformal group yields
\beq
\begin{array}{rcl}
\tilde{F}_{c_1\ldots c_4} &\equiv& 4(\bein_{[c_1}^{\phantom{[c_1}\mu}
\pa_{|\mu|}A_{c_2c_3c_4]} + \bein_{[c_1}^{\phantom{[c_1}\mu}
(\bein^{-1}\pa_{|\mu|}\bein)_{c_2}^{\phantom{c_2}b}A_{|b|c_3c_4]} +
\ldots) \\
\tilde{F}_{c_1\ldots c_7} &\equiv& 7(\bein_{[c_1}^{\phantom{[c_1}\mu}
\pa_{|\mu|}A_{c_2\ldots c_7]} + \bein_{[c_1}^{\phantom{[c_1}\mu}
(\bein^{-1}\pa_{|\mu|}\bein)_{c_2}^{\phantom{c_2}b}A_{|b|c_3\ldots
  c_7]} + \ldots \\
& & + 5\tilde{F}_{[c_1\ldots c_4}A_{c_5c_6c_7]})
\end{array}
\eeq
as the objects covariant under the both groups. Combining these we
find the equations of motion for the gauge fields as
\beq
\eqnlab{gauge_eom}
\tilde{F}^{c_1\ldots c_4} = \frac{1}{7!}\varepsilon_{c_1\ldots c_{11}}
\tilde{F}^{c_5\ldots c_{11}}.
\eeq
The equations of motion for gravity should be given by
\Eqnref{gravity_eom_tensor}, but with $\hat{D}_\mu m_{c_1\ldots
  c_8,d}$ replaced by $\tilde{D}_\mu \hat{m}_{c_1\ldots c_8,d}$ from
\eqnref{e11_fields} and the addition of the field $m_{[c_1\ldots
  c_8,d]}$.

The generator $D \equiv \sum_{a=1}^{11} K^a_{\phantom{a}a}$ can be
used to measure the level of a generator, since
\beq
\com{D,K^a_{\phantom{a}b}} = 0 \hspace{1 cm} \textrm{and} \hspace{1
  cm} \com{D,R^{9,10,11}} = R^{9,10,11}
\eeq
with $R^{9,10,11}$ corresponding to the eleventh simple root. Acting
on the coset representative with $e^{rD}$ gives
\beq
\begin{array}{rcl}
g \rightarrow g' = e^{rD}g &=&  e^{e^{-r}x^{f}P_{f}}
e^{(m_a^{\phantom{a}b}+r\delta_a^b)K^a_{\phantom{a}b}} 
e^{\frac{1}{3!}A_{a_1a_2a_3}R^{a_1a_2a_3}} \\
& & \times e^{\frac{1}{6!}A_{b_1\ldots b_6}R^{b_1\ldots b_6}}
e^{\hat{m}_{c_1\ldots c_8,d}R^{c_1\ldots c_8,d}}
\end{array}
\eeq
i.e.,
\beq
\left\{ \begin{array}{l}
x^\mu \rightarrow x'^\mu = e^{-r}x^\mu \\
m_a^{\phantom{a}b} \rightarrow {m'}_a^{\phantom{a}b} =
m_a^{\phantom{a}b} + r\delta_a^b
\end{array} \right..
\eeq
Thus, $D$ corresponds to a symmetry which leaves the equations of
motion \eqnref{gravity_eom_tensor} and \eqnref{gauge_eom} invariant.

\section{The translation generators}

The translation generators in g$_{11}$ carry one lower index and
transform under A$_{10}$, but we did not find any such generators in
e$_{11}$. The role of the translation generators is to introduce
space-time into the theory. Due to its transformation properties,
$P_a$ is equivalent to a tensor with ten upper antisymmetric indices,
and corresponds to the A$_{10}$ representation with $\la_1$ as the
highest weight. Consider thus the possibility that the translation
generators are contained in a representation of e$_{11}$. From
\Eqnref{e11_weight} we see that the e$_{11}$ representation with $-l_1
= \frac{3}{2}x - \la_1$ at level $l=\frac{3}{2}$ as the lowest weight
contains the A$_{10}$ representation $\Lambda = \la_1$. Calculating
the root string to this lowest weight $-l_1$, we find the first weight
vector where the eleventh root $\alpha_{11}$ enters to be
\beq
 - l_1 + \alpha_1 + \ldots + \alpha_8 + \alpha_{11} = \frac{5}{2}x -
 \la_9. 
\eeq
A new representation for A$_{10}$ as a tensor with two upper
antisymmetric indices then appears. Continuing in this way, we can
find all the A$_{10}$ representations contained in the e$_{11}$
representation $-l_1$, at the lowest levels these are given in
\Tabref{e11_weight}.
\begin{table}[h!]
  \begin{center}
    \begin{tabular}{l|l|l}
      Level & Highest weight & A$_{10}$ representation\\ \hline 
      $l = 3/2$ & $(1,0,0,0,0,0,0,0,0,0)$ & $P^{a_1\ldots a_{10}}$ \\  
      $l = 5/2$ & $(0,0,0,0,0,0,0,0,1,0)$ & $W^{a_1a_2}$ \\
      $l = 7/2$ & $(0,0,0,0,0,1,0,0,0,0)$ & $W^{a_1\ldots a_5}$ \\  
      $l = 9/2$ & $(0,0,0,1,0,0,0,0,0,1)$ & $W^{a_1\ldots a_7,b}$ \\  
                & $(0,0,1,0,0,0,0,0,0,0)$ & $W^{a_1\ldots a_8}$
    \end{tabular}
    \caption{\textit{The A$_{10}$ representations occurred at the
        lowest levels of the e$_{11}$ representation with $-l_1$ as
        the lowest weight \cite{West:2003fc}.}}
    \tablab{e11_weight}
  \end{center}
\end{table}

A linear representation $U(A)$ of a Lie algebra g must be a
homomorphism, i.e., $U(A_1A_2)=U(A_1)U(A_2)$ for all $A_1,A_2\in$ g.
Let $|\chi_s\rangle$ denote the states spanning a module for g, and
define the matrices
\beq
U(A)|\chi_s\rangle = (c(A))_s^{\phantom{s}t}|\chi_t\rangle.
\eeq
The matrices $c(A)$ are not the matrix representations of g since
$c(A_1A_2) = c(A_2)c(A_1)$, instead, we can use these to define $d(A)
\equiv c(A^{\ddagger})$ which preserve the algebra multiplication. The
operation $A^{\ddagger}$ is defined as $A^{\ddagger} \equiv
I^CI^I(A)$, where $I^C$ is a Cartan involution, while $I^I(A)=-A$ and
inverts the order of the operators at the same time. The matrices
$d(A)$ form a representation since
\beq
d(A_1A_2) = c(I^CI^I(A_1A_2)) = c(I^C(-A_1))c(I^C(-A_2)) =
d(A_1)d(A_2).
\eeq
We can associate a generator $X_s$ to each state $|\chi_s\rangle$ in
the module, obeying the commutation relations
\beq
\eqnlab{gX_Lie}
\com{X_s,X_t} = 0 \hspace{1 cm}\textrm{and}\hspace{1 cm} \com{X_s,A} =
d(A)_s^{\phantom{s}r}X_r, \hspace{0.2 cm} A \in \textrm{g}.
\eeq
Thus, $\{X_s\}$ and g form a semi-direct product algebra
\cite{West:2003fc}. The Jacobi identities can be checked explicitly.

Consider now the semi-direct product of e$_{11}$ and its $-l_1$
representation denoted by e$_{11}\oplus_s$L$_1$. The lowest level
generators of L$_1$ are given in \Tabref{e11_weight}. Instead of the
generators $P^{a_1\ldots a_{10}}$, we use the translation generators
\beq
P_a = \frac{1}{10!}\varepsilon_{ab_1\ldots b_{10}}P^{b_1\ldots
  b_{10}}.
\eeq
Following \Eqnref{gX_Lie} we have
\beq
\eqnlab{e11_L1_a}
\begin{array}{rcl}
\com{K^a_{\phantom{a}b},P_c} &=& -\delta^a_cP_b +
\frac{1}{2}\delta^a_bP_c \\
\com{R^{a_1a_2a_3},P_b} &=& 3\delta^{[a_1}_bW^{a_2a_3]} \\
\com{R^{a_1a_2a_3},W^{b_1b_2}} &=& W^{a_1a_2a_3b_1b_2} \\
\com{R^{a_1a_2a_3},W^{b_1\ldots b_5}} &=& W^{b_1\ldots
  b_5[a_1a_2,a_3]}. 
\end{array}
\eeq
Using the Jacobi identities and \Eqnref{e11_L1_a} we also find
\beq
\eqnlab{e11_L1_b}
\begin{array}{rcl}
\com{R^{a_1\ldots a_6},P_b} &=& -3\delta^{[a_1}_bW^{a_2\ldots a_6]} \\
\com{R_{a_1a_2a_3},P_b} &=& 0 \\
\com{R_{a_1a_2a_3},W^{b_1b_2}} &=& 12\delta^{b_1b_2}_{[a_1a_2}P_{a_3]}
\\ 
\com{R_{a_1a_2a_3},W^{b_1\ldots b_5}} &=&
120\delta^{[b_1b_2b_3}_{a_1a_2a_3}W^{b_4b_5]}
\end{array}
\eeq
Adding the vectors from the $-l_1$ representation to the root lattice
will extend the root lattice to the full weight lattice of e$_{11}$.

\chapter{Conclusions}
\chlab{conclusion}

In this thesis work some symmetry aspects of M-theory have been
studied. Here the relevant symmetry groups of a theory are the ones
used when the theory in question is reformulated as nonlinear
realisation. The technique of nonlinear realisations is used to
describe the low energy dynamics of a theory, in which a symmetry is
spontaneously broken, but at high energies the same symmetry is
expected to be linearly realised. In this way we found that the
bosonic part of eleven dimensional supergravity was globally invariant
under both the conformal group and an enlarged affine group called
G$_{11}$. In the calculations, the Lorentz group was taken as the only
local symmetry group. We then argued that M-theory itself should
contain a coset symmetry, with the very extended group E$_{11}$ and
its Cartan involution invariant subgroup being the global and the
local symmetry group, respectively. At least we showed that the
Lorentzian group E$_{11}$ satisfies some necessary conditions, e.g.,
it contains all the symmetry groups occurring in the dimensionally
reduced supergravity theories. Also, the generators of G$_{11}$ can be
identified with the lowest level generators of E$_{11}$. The Goldstone
bosons introduced in the nonlinear realisation will correspond to the
gravity degrees of freedom and the gauge fields. If the infinite
dimensional e$_{11}$ is the symmetry algebra, new symmetries that mix
gravity and the tensor gauge fields non-trivially would appear at
higher levels. During the identification of the Goldstone bosons we
had to reformulate the theory of gravity to include a field, which is
dual to the vielbein. It was then also conjectured that pure gravity
in $D$ dimensions should have a hidden Kac-Moody symmetry, namely the
very extended version of A$_{D-3}$ usually denoted as A$_{D-3}^{+++}$.
The way the space-time generators and the conformal algebra relate to
e$_{11}$ nonlinear realisation is still not very well understood.

The existence of fermions in the eleven dimensional supergravity
implies an osp(1$|$64) symmetry, which is the only algebra containing
both the supersymmetry group and the conformal group. This algebra
contains gl(32) as a subalgebra, which rotates the spinor index on the
supercharges. Being a brane rotating symmetry \cite{Barwald:1999is},
sl(32) is the most natural generalisation of the Lorentz algebra when
branes are present. Consequently, sl(32) must also be a symmetry of
M-theory and somehow be related to e$_{11}$. Indeed, the commutation
relations in \eqnref{involution_inv} are part of the commutation
relations defining the sl(32) automorphism algebra \cite{West:2003fc}.
The rank two and rank five antisymmetric tensors appeared in the
e$_{11}$ representation with the lowest weight $-l_1$
(\Tabref{e11_weight}) can be identified with the central charges of
the eleven dimensional supersymmetry algebra. Thus, sl(32) is
contained in the Cartan involution invariant subalgebra of e$_{11}$,
not as a subalgebra, but as a truncation and hence to get the entire
osp(1$|$64) the $-l_1$ representation must also be added. Nonlinear
realisation of e$_{11}\oplus_s$L$_1$ should be able to describe point
particles and branes on an equal footing.

In the familiar nonlinear realisations used in particle physics,
space-time is introduced in an ad hoc way and the reducing matrix does
not depend on space-time. As a result, a group transformation does not
change the space-time dependence of any solution. The nonlinear
realisation considered here involves space-time in a more complicated
manner, and thus one can expect that the possible group
transformations that one can carry out on any solution are much more
extensive. In this thesis we have interpreted the tensors found in the
$-l_1$ representation of e$_{11}$ as space-time translations, central
charges of the supersymmetry algebra as well as an infinite number of
other states. However, there is an alternative approach to the
translation generators. The theory is still thought to be a nonlinear
realisation of e$_{11}$, but the fields are now assumed to depend on
an auxiliary parameter and all the space-time dependence was assumed
to arise in the same way from the equations of motion of the nonlinear
realisation \cite{Damour:2002cu,Englert:2003py}. At level seven there
is an A$_{10}$ representation which also may be identified with the
translation generators, and hence providing another way to introduce
space-time into the theory. It was then conjectured that the adjoint
representation of all G$^{+++}$ contains a generator, which belongs to
the correct A representation to be identified as space-time
translations \cite{Englert:2003py}. But it was soon found that this is
not true for e.g., D$^{+++}_{n-3}$ for odd $n$ and A$^{+++}_{n-3}$ for
even $n$. Thus this mechanism cannot be applied to IIB theory and the
bosonic string D$^{+++}_{24}$ \cite{Kleinschmidt:2003jf}. Furthermore,
even if a generator with the correct A representation properties does
occur, it is usually not unique. For e$_{11}$ such a generator occurs
at levels $(7+11s)$ for any non-negative integer $s$. The question is
then which of these is the real space-time translation generator.
Moreover, one finds that some of these operators occur with outer
multiplicity greater than one, i.e., even at a given level there is
often more than one choice.  Even though the level seven candidate
does have the correct A properties, it has the problem being
non-commutative. Our usual picture of space-time translations as
commuting quantities must then change drastically. Having these
considerations in mind, it is at the present stage more plausible to
let the translation generators come from a basic representation of
e$_{11}$.

In \Chref{algebra} we have studied the simple roots of the Lorentzian
Kac-Moody algebras, but for a general Kac-Moody algebra not very much
is known. The real roots have single multiplicity, while no closed
expression exists for the imaginary roots, only the recursive Peterson
and Freudenthal formulas \cite{Kac:1990gs} can be applied. Calculating
the outer multiplicities for the vectors in \Tabref{e11_level}, one
finds that the vectors we omitted from the root space of e$_{11}$ were
the only ones having non-positive multiplicities as they should
\cite{Nicolai:2003fw,Kleinschmidt:2003mf,Kleinschmidt:2003jf}. The
real roots are those related to the simple roots by a Weyl
transformation, and the imaginary roots are Weyl conjugate to elements
in the fundamental Weyl chamber which have connected support on the
Dynkin diagram \cite{Kac:1990gs,Kleinschmidt:2003mf}. It is also known
that the Weyl transformations are related to U-duality
\cite{Kleinschmidt:2003mf}. The representation theory of a Kac-Moody
algebra is somewhat different from the one of a finite dimensional
algebra. Consider the level decomposition of a Kac-Moody algebra in
\Chref{symmetry}, at positive levels highest weight representations
are obtained, whereas at negative levels lowest weight representations
are obtained. The adjoint representation of a very extended algebra is
neither a highest nor a lowest weight representation, instead, it can
be obtained by taking tensor products of the highest and lowest weight
representations \cite{Kleinschmidt:2003jf}.

Given the Dynkin diagram of e$_{11}$, starting from the very extended
node one can embed an A$_9$ sub-Dynkin diagram in two ways. These are
the ten-dimensional type IIA and IIB superstring theories
\cite{Schnakenburg:2002xx,Schnakenburg:2001ya}. The A$_9$ subalgebra
describes the gravity sector of the theory. Generally, different ways
to embed a gravity subalgebra, i.e., an A$_{D-1}$ subalgebra with $D$
denoting the dimension of the theory obtained, inside the Dynkin
diagram correspond to generalised T-dualities
\cite{Kleinschmidt:2003mf}.

During the calculations we have used the subalgebra invariant under
the standard Cartan involution as the local symmetry, but when
considering the vielbein defining gravity line sl(n) one finds that
the corresponding local subalgebra is so(n) and not so(n-1,1).To get a
space-time with Lorentzian signature a Wick rotation is thus needed
for the actual real form of the local subalgebra of e$_{11}$
\cite{Englert:2003py,Keurentjes:2004bv}. Instead, one can define a
general involution \cite{West:2004st}
\beq
E_a \rightarrow \eta_a F_a, \hspace{0.5 cm} F_a \rightarrow \eta_a
E_a, \hspace{0.5 cm} H_a \rightarrow -H_a,
\eeq
where $\eta_a = \pm 1$ and $a$ is numbering the simple roots. For any
combination of signs $\eta_a$, the generators invariant under the
corresponding involution will constitute a real form of the local
subalgebra. Different real forms of the local subalgebra will give
rise to different space-time signatures, and all the possibilities are
connected under the Weyl group of e$_{11}$.

A theory containing e$_{11}$ as symmetry should contain an infinite
number of BPS solutions. Each of these solutions must have a field in
the theory as its source, and the objects occurring in the $-l_1$
representation are the corresponding charges. The fields of the
nonlinear realisation are contained in the group elements of E$_{11}$
when modded out by the action of the local subgroup. Consequently,
given any solution we can write down the corresponding E$_{11}$ group
element. One finds that the group elements corresponding to the common
half BPS solutions of eleven dimensional supergravity have the form
\cite{West:2004st,Cook:2004er}
\beq
g = \exp{\left(-\frac{1}{(\beta,\beta)}\beta\cdot H \ln{N}\right)}
\exp{\left\{(1-N)E_{\beta}\right\}},
\eeq
where $N$ is a harmonic function, $\beta$ is the e$_{11}$ root
corresponding to the generator $E_{\beta}$ and $H$ are the Cartan
generators. This elegant form of the group elements for the usual half
BPS branes leads to the idea that E$_{11}$ can be useful when
classifying solutions with a given amount of supersymmetry.

Except the eleven dimensional e$_{11}$ there are other very extended
Kac-Moody algebras with great importance, e.g., the conjectured
gravity symmetry A$^{+++}_{D-3}$ in $D$ dimensions. It turned out that
the low level content of the adjoint representation of a general very
extended algebra G$^{+++}$ predicted a field content, which under
nonlinearly realisation was in agreement with the oxidation theory
associated with algebra G \cite{Kleinschmidt:2003mf,Englert:2003zs}.
Another interesting example is given by k$_{27}$, which is thought to
be the symmetry of the bosonic string \cite{West:2000ga,West:2001as}.
The local subalgebra is again taken to be the subalgebra invariant
under the Cartan involution.  Nonlinear realisation based on the
lowest level generators yields the right field equations. One
interesting property of k$_{27}$ is that its Dynkin diagram contains
e$_{11}$, indicating that the closed bosonic string contains the
superstrings in ten dimensions \cite{Casher:1985ra} and maybe even
M-theory itself. The closed bosonic string on a torus is invariant
under the fake monster Lie algebra \cite{Moore:1993zc,West:1995my},
one can then ask how the fake monster Lie algebra is related to
k$_{27}$.

We have pointed out some of the physical aspects of
e$_{11}\oplus_s$L$_1$, but maybe the semi-direct product is just part
of some larger algebra.  When adding also the $-l_2$ representation to
obtain e$_{11}\oplus_s$L$_1\oplus_s$L$_2$, the translation generators
no longer need to commute. Perhaps the very extended Kac-Moody
algebras are embedded inside the so called Borcherds algebras.
M-theory is itself just one step on the way and the very extended
algebras play an important role in fixing the structure of that
theory, similar to the supersymmetry. The final algebraic structure
may include Borcherds' fake monster algebra \cite{Borcherds:1990},
which is the vertex algebra on the unique even self-dual Lorentzian
lattice in 26 dimensions.

%
%

\appendix
\pagestyle{plain}

\chapter{Differential forms}
\chlab{pform}

In this appendix we will review the basics of the differential forms
following the references
\cite{Nakahara:2003th,Flink:2001,Ambjornsson:1998}.

We define a $p$-form on an $n$-dimensional manifold $M$ as
\beq
\eqnlab{def_pform}
t \equiv \frac{1}{p!} dx^{\mu_1} \we dx^{\mu_2} \we \ldots \we
dx^{\mu_p} \, t_{\mu_1\mu_2\ldots\mu_p}, 
\eeq
where $t_{\mu_1\mu_2\ldots\mu_p}$ is a totally antisymmetric
tensor. The wedge operator on the coordinate differentials satisfies
the properties 
\beqa
\eqnlab{wedge_asso}
(dx^\lambda \we dx^\mu) \we dx^\nu &=& dx^\lambda \we (dx^\mu \we
dx^\nu) \\
\eqnlab{wedge_anti}
dx^\mu \we dx^\nu &=& - dx^\nu \we dx^\mu.
\eeqa
We see that when putting the index $\mu$ equal to $\nu$,
\Eqnref{wedge_anti} vanishes identically. This results in that there
cannot exit any forms with $p>n$ in a $n$-dimensional manifold. The
antisymmetricity of the wedge operator is made obvious through the
identification
\beq
\int dx^{\mu_1} \we \ldots \we dx^{\mu_n} = \int d^{n}x \,
\varepsilon^{\mu_1 \ldots \mu_n},
\eeq
where $\varepsilon^{\mu_1 \ldots \mu_n}$ is the total antisymmetric
tensor density with $\varepsilon^{1 2 \ldots n}=+1$.  The number of
independent components of a $p$-form is ${n \choose p}$.

The exterior product, or the wedge product, between a $p$-form and a
$q$-form is a $(p+q)$-form and it is defined as
\beq
\eqnlab{wedge_prod}
t \we w \equiv \frac{1}{p!q!} dx^{\mu_1} \we\ldots\we dx^{\mu_p} \we
dx^{\mu_{p+1}} \we\ldots\we dx^{\mu_{p+q}} \, t_{\mu_1\ldots\mu_p}
w_{\mu_{p+1}\ldots\mu_{p+q}}.
\eeq
This exterior product is linear
\beq
(a t_1 + b t_2)\we w = a t_1\we w + b t_2\we w,
\eeq
associative
\beq
t\we(w\we u)=(t\we w)\we u
\eeq
and satisfy the relation
\beq
t \we w = (-1)^{pq} \, w \we t.
\eeq

We define an exterior derivative
\beq
d \equiv \dxmu \pa_\mu,
\eeq
which acts on a $p$-form to give a $(p+1)$-form according to
\beq
\begin{array}{rcl}
\eqnlab{ext_der}
dt \equiv d\we t &=& \frac{1}{p!} dx^{\mu_1}\we dx^{\mu_2}\we\ldots\we
dx^{\mu_{p+1}} \, \pa_{\mu_1}t_{\mu_2\ldots\mu_{p+1}} \\ 
&=& \frac{1}{p!} dx^{\mu_1}\we dx^{\mu_2}\we\ldots\we dx^{\mu_{p+1}}
\, \pa_{[\mu_1}t_{\mu_2\ldots\mu_{p+1}]}\,. 
\end{array}
\eeq
Moreover, the exterior derivative satisfies
\beq
d(t\we w)=dt\we w+ (-1)^p\,t\we dw,
\eeq
and the Poincar\'e lemma
\beq
d(dt) = \frac{1}{p!}\dxmu \we \dxnu \we dx^{\mu_1} \we\ldots\we
dx^{\mu_p} \, \pa_{[\mu}\pand t_{\mu_1\ldots\mu_p]} \equiv 0.
\eeq

The so called Hodge dual of a $p$-form is defined as
\beq
\eqnlab{hodge}
(* t) \equiv
\frac{1}{p!(n-p)!}\frac{1}{\sqrt{|g|}} dx^{\mu_{p+1}} \we\ldots\we
dx^{\mu_n} \, \varepsilon_{\mu_{p+1}\ldots\mu_n\nu_1\ldots\nu_p}
t^{\nu_1\ldots\nu_p},
\eeq
where $g = \det g_{\mu\nu}$. In terms of antisymmetric tensors
\Eqnref{hodge} becomes
\beq
(*t)_{\mu_1\ldots\mu_{n-p}} =
\frac{1}{p!}\frac{1}{\sqrt{|g|}}\,\varepsilon_{\mu_1\ldots\mu_{n-p}
  \nu_1\ldots\nu_p} t^{\nu_1\ldots\nu_p}. 
\eeq
The Levi-Civita epsilon has the property
\beq
\varepsilon^{\rho_1\ldots\rho_p\mu_{p+1}\ldots\mu_n}
\varepsilon_{\rho_1\ldots\rho_p\nu_{p+1}\ldots\nu_n} =
\pm |g| p!(n-p)!\,\delta^{\mu_{p+1}}_{[\nu_{p+1}}\ldots\,
\delta^{\mu_n}_{\nu_n]}, 
\eeq
where the $-(+)$-sign is chosen whenever the signature is Lorentzian
(Riemannian). The Levi-Civita epsilon with lower indices is defined as
$\varepsilon_{\mu_1\ldots\mu_n} \equiv g_{\mu_1\nu_1} \ldots
g_{\mu_n\nu_n} \varepsilon^{\nu_1\ldots\nu_n}$. Using the definition
\eqnref{hodge} we can also deduce that
\beq
*(*t) = \left\{ \begin{array}{ll} 
 (-1)^{1+p(n-p)}\,t & \textrm{Lorentzian~space} \\
 (-1)^{p(n-p)}\,t & \textrm{Riemannian~space}
\end{array} \right..
\eeq

Having introduced the Hodge dual we can define the adjoint of the
exterior derivative, which takes a $p$-form to a $(p-1)$-form
according to
\beq
d^\da t \equiv \left\{ \begin{array}{ll} 
 (-1)^{np+n} *d* t & \textrm{Lorentzian~space} \\
 (-1)^{np+n+1} *d* t& \textrm{Riemannian~space}
\end{array} \right..
\eeq
The Laplacian is then defined by
\beq
\Delta \equiv (d+d^\da)^2 = d\cdot d^\da + d^\da \cdot d,
\eeq
taking a $p$-form to another $p$-form. According to the Hodge
decomposition theorem an arbitrary $p$-form $t_p$ can always be
written as 
\beq
t_p = d\alpha_{p-1} + d^\da\beta_{p+1} + \gamma_p,
\eeq
where $\alpha_{p-1}$ is a $(p-1)$-form, $\beta_{p+1}$ is a
$(p+1)$-form and $\gamma_p$ is a $p$-form obeying $\Delta \gamma_p =
0$. The Hodge dual is also used to define a symmetric inner product of
two $p$-forms in a $n$-dimensional manifold $M$
\beq
\eqnlab{inner_prod}
(t,w) \equiv \int_M t\we *w.
\eeq

Generally, the integral of some $p$-form $t$ over a $p$-dimensional
submanifold $\Sigma \subset M$ is
\beq
\int_\Sigma t = \int_\Sigma
d\sigma^{\alpha_1}\we\ldots\we d\sigma^{\alpha_p} \,
t^*_{\alpha_1\ldots\alpha_p},
\eeq
where 
\beq
\eqnlab{pullback}
t^*=\frac{1}{p!} d\sigma^{\alpha_1}\we\dots\we
  d\sigma^{\alpha_p} \frac{\pa x^{\mu_1}}{\pa
  \sigma^{\alpha_1}}\ldots\frac{\pa x^{\mu_p}}{\pa
  \sigma^{\alpha_p}}t_{\mu_1\ldots\mu_p}
\eeq
is the pull back of $t$ on $\Sigma$. The coordinates $\sigma^\alpha$
are used to parametrise the submanifold $\Sigma$. After having mapped
injectively between the coordinates $x$ and $\si$, we find the
integral to be
\beq
\eqnlab{diff_int}
\int_\Sigma t = \int_\Sigma
  d\sigma^1 \ldots d\sigma^p \, \frac{\pa x^{\mu_1}}{\pa
  \sigma^1}\ldots\frac{\pa x^{\mu_p}}{\pa \sigma^p}t_{\mu_1\ldots\mu_p}.
\eeq
As the end of this short review we state the Stokes theorem
\beq
\int_\Sigma dt = \int_{\pa \Sigma} t,
\eeq
where $\pa \Sigma$ is the boundary of some manifold $\Sigma$.

\chapter{The conformal group}
\chlab{conformal}

A conformal transformation of the coordinates is by definition an
invertible coordinate mapping, which leaves the metric tensor
invariant up to a scale
\beq 
\eqnlab{confcond}
g'_{\mu\nu}(x') = e^{\La(x)}g_{\mu\nu}(x).
\eeq
The space-time coordinates in $d$ dimensions are denoted by 
\beq
x^{\mu}, \hspace{1 cm} \mu = 0,1,\ldots,d-1
\eeq
where the zeroth index is referring to the time direction. We will
throughout this appendix use the signature
$\eta_{\mu\nu}=\textrm{diag}(1,-1,-1,\ldots,-1)$.

The conformal transformations do not affect the angle between two
arbitrary curves crossing each other at some point, explaining the
use of the word conformal. This property can easily be seen by
considering the angle $\theta$ between two vectors $u$ and $v$
\beq
\cos\theta = \frac{u \cdot v}{\sqrt{(u \cdot u)(v \cdot v)}}.
\eeq
Each scalar product involves a metric, thus under a conformal
transformation the scale factor $e^{\La(x)}$ on the numerator and the
denominator will cancel out.

To find the conformal generators we examine the infinitesimal
coordinate transformation $x^{\mu} \rightarrow
x'^{\mu}=x^{\mu}+\epsilon^{\mu}(x)$. The metric will then transform as 
\beq
g_{\mu\nu}(x) \rightarrow g'_{\mu\nu}(x')
= g_{\mu\nu}(x) - (\pa_\mu\epsilon_\nu + \pa_\nu\epsilon_\mu)
\eeq
to first order in $\epsilon$. For conformal transformations the
condition \eqnref{confcond} has to be satisfied, which results in 
\beq
\eqnlab{metricdiff}
\pa_\mu\epsilon_\nu + \pa_\nu\epsilon_\mu = (1 - e^{\La(x)})g_{\mu\nu}(x).
\eeq
Taking the trace on both sides, i.e., contracting with $g^{\mu\nu}$ we
find 
\beq
d(1 - e^{\La(x)}) = 2\pa_\mu\epsilon^\mu \equiv 2\pa \cdot \epsilon
\eeq
where $d$ denotes the dimension. Inserting this back into
\Eqnref{metricdiff} we get 
\beq
\eqnlab{ediff}
\pa_\mu\epsilon_\nu + \pa_\nu\epsilon_\mu = \frac{2}{d}g_{\mu\nu}\pa\cdot
\epsilon,
\eeq
to which we apply a derivative $\pa^{\nu}$ and obtain
\beq
\eqnlab{confeqn}
\pa_\mu(\pa \cdot \epsilon) + \pa^\nu\pa_\nu \epsilon_\mu =
\frac{2}{d}\pa_\mu (\pa \cdot \epsilon).
\eeq
To solve the differential equation \eqnref{confeqn} we first act on it
with $\pa^\mu$ yielding
\beq
\eqnlab{confeqneasy}
\left(2 - \frac{2}{d}\right)\pa^\mu \pa_\mu(\pa \cdot \epsilon) = 0.
\eeq
If we instead act on \Eqnref{confeqn} with $\pa_\nu$, we get
\beq
\pa^\si\pa_\si(\pa_\mu\epsilon_\nu + \pa_\nu\epsilon_\mu) + \left(2 -
  \frac{4}{d}\right)\pa_\mu\pa_\nu(\pa \cdot \epsilon) = 0,
\eeq
where we also have symmetrised over $\mu$ and $\nu$. After making use of
\Eqsref{ediff} and \eqnref{confeqneasy} we arrive at
\beq
\left(2 - \frac{4}{d}\right)\pa_\mu \pa_\nu(\pa \cdot \epsilon) = 0.
\eeq
The general solution of this equation is (for $d \neq 2$)
\cite{DiFrancesco:1997nk,Arvidsson:2001} 
\beq
\epsilon_\mu = a_\mu + b_{\mu\nu}x^\nu + c_{\mu\nu\rho}x^\nu x^\rho.
\eeq
Putting this solution into the original differential equation
\eqnref{confeqn}, we find
\beq
\eqnlab{infconf}
\epsilon^\mu = \alpha^\mu + \la x^\mu +
\omega^\mu_{\phantom{\mu}\nu}x^\nu + b^\mu x^2 - 2x^\mu b \cdot x,
\eeq
with $\omega_{\mu\nu}$ being an antisymmetric tensor. From
\Eqnref{infconf} we can read off the finite conformal transformations
on the coordinates, these are
\benu
\item translations
\beq
x^\mu \rightarrow x'^\mu = x^\mu + \alpha^\mu
\eeq

\item Lorentz transformations (spatial rotations and Lorentz boosts)
\beq
x^\mu \rightarrow x'^\mu = \omega^\mu_{\phantom{\mu}\nu}x^\nu
\eeq

\item dilations
\beq
x^\mu \rightarrow x'^\mu = \la x^\mu
\eeq

\item special conformal transformations
\beq
\eqnlab{coordSCT}
x^\mu \rightarrow x'^\mu = \frac{x^\mu+b^\mu x^2}{1+2b \cdot x+b^2x^2}.
\eeq
\eenu
The specific form of the special conformal transformations is motivated
by expanding \Eqnref{coordSCT} and keeping the lowest order terms only
\beq
\frac{x^\mu+b^\mu x^2}{1+2b \cdot x+b^2x^2} \approx (x^\mu + b^\mu
x^2)(1 - 2b \cdot x) \approx x^\mu + b^\mu x^2 - 2x^\mu b \cdot x,
\eeq
which are precisely the last two terms in \Eqnref{infconf}.

We define the generators $G_\alpha$ of the conformal group by
\beq
x \rightarrow x' = x + \epsilon^\alpha G_\alpha x,
\eeq
where $\epsilon^\alpha$ is an infinitesimal tensor. Comparing with
\Eqnref{infconf} we find the generators of the conformal algebra to be
\beq
\begin{array}{rcl}
\eqnlab{conf_gen}
P_\mu &=& \pa_\mu \\
M_{\mu\nu} &=& x_\mu \pa_\nu - x_\nu \pa_\mu \\
D &=& x^\nu\pa_\nu \\
K_\mu &=& 2x_\mu x^\nu\pa_\nu - x^2\pa_\mu,
\end{array}
\eeq
corresponding to translations, Lorentz transformations, dilations and
special conformal transformations, respectively. The commutation
relations of the conformal algebra will thus be
\beq
\eqnlab{conf_alg}
\begin{array}{rcl}
\com{D,P_\mu} &=& - P_\mu  \\
\com{D,M_{\mu\nu}} &=& 0\\
\com{D,K_\mu} &=& K_\mu \\
\com{K_\mu,K_\nu} &=& 0 \\
\com{K_\mu,P_\nu} &=& -2(\eta_{\mu\nu}D + M_{\mu\nu}) \\
\com{K_\rho,M_{\mu\nu}} &=& (\eta_{\rho\mu}K_\nu -
\eta_{\rho\nu}K_{\mu}) \\
\com{P_\mu,P_\nu} &=& 0 \\
\com{P_\rho,M_{\mu\nu}} &=& (\eta_{\rho\mu}P_\nu -
\eta_{\rho\nu}P_{\mu}) \\
\com{M_{\mu\nu},M_{\rho\sigma}} &=& (\eta_{\mu\si}M_{\nu\rho} +
\eta_{\nu\rho}M_{\mu\si} - \eta_{\mu\rho}M_{\nu\sigma} -
\eta_{\nu\sigma}M_{\mu\rho}).
\end{array}
\eeq
One can also show that there is an isomorphism between the conformal
group in $(d-1,1)$ Minkowski dimensions and the Lorentz group SO(d,2).

The conformal generators given in \eqnref{conf_gen} were above applied
directly on the coordinates, and to obtain the action of the conformal
generators on a field we have to take the intrinsic degrees of freedom
into account. We define the generators $\Sigma_{\mu\nu}$ (Lorentz
transformations), $\bar{\Delta}$ (dilations) and $\ka_\mu$ (special
conformal transformations) of the stability subgroup of $x=0$. These
generators have the property that they leave the point $x=0$
invariant. The commutation relations of the stability subgroup are
\beq
\eqnlab{stability_alg}
\begin{array}{rcl}
\com{\bar{\Delta},\Sigma_{\mu\nu}} &=& 0\\
\com{\bar{\Delta},\ka_\mu} &=& \ka_\mu \\
\com{\ka_\mu,\ka_\nu} &=& 0 \\
\com{\ka_\rho,\Sigma_{\mu\nu}} &=& (\eta_{\rho\mu}\ka_\nu -
\eta_{\rho\nu}\ka_{\mu}) \\
\com{\Sigma_{\mu\nu},\Sigma_{\rho\sigma}} &=&
(\eta_{\mu\si}\Sigma_{\nu\rho} + \eta_{\nu\rho}\Sigma_{\mu\si} -
\eta_{\mu\rho}\Sigma_{\nu\sigma} - \eta_{\nu\sigma}\Sigma_{\mu\rho})
\end{array}
\eeq
in analogy with \Eqnref{conf_alg}. The generators of the stability
subgroup will also commute with $P_\mu$, $M_{\mu\nu}$, $D$, $K_\mu$
and the coordinates $x^\mu$.

For an arbitrary field $\phi_a(x)$, the basis is chosen such that 
\beq
\eqnlab{transl_field}
P_\mu \phi_a(x) = \frac{\pa}{\pa x^\mu}\phi_a(x).
\eeq
Note that the translation generators do not act on the indices of
$\phi_a(x)$. The actions of the other generators at $x=0$ are given by
\beq
\eqnlab{stab_field}
\begin{array}{rcl}
M_{\mu\nu}\phi_a(0) &=& \Sigma_{\mu\nu}\phi_a(0) \\
D\phi_a(0) &=& \bar{\Delta}\phi_a(0) \\
K_{\mu}\phi_a(0) &=& \ka_{\mu}\phi_a(0).
\end{array}
\eeq
Using \Eqsref{transl_field} and \eqnref{stab_field} together with the
fact that
\beq
\phi_a(x) = \expo(x^{\mu}P_{\mu})\phi_a(0)
\eeq
we can find the actions of the conformal generators on arbitrary fields
\beq
\begin{array}{rcl}
P_\mu \phi(x) &=& \pa_\mu \phi(x) \\
M_{\mu\nu} \phi(x) &=& \left[(x_\mu \pa_\nu - x_\nu \pa_\mu) +
  \Sigma_{\mu\nu} \right] \phi(x) \\
D \phi(x) &=& \left[x^\nu\pa_\nu + \bar{\Delta} \right] \phi(x) \\
K_\mu \phi(x) &=& \left[(2x_\mu x_\nu\pa_\nu - x^2\pa_\mu) + 2x^\nu
  \left(\eta_{\mu\nu}\bar{\Delta} + \Sigma_{\mu\nu}\right) +
  \ka_\mu\right] \phi(x)
\end{array}
\eeq
The commutation relations in \Eqnref{conf_alg} are still satisfied.

\chapter{General coordinate transformations}
\chlab{closure}

An infinitesimal general coordinate transformation is given by
\beq
x^{\mu} \rightarrow x'^{\mu}=x^{\mu}+\epsilon^{\mu}(x),
\eeq
where $\epsilon^{\mu}$ is an arbitrary infinitesimal parameter. We can
expand $\epsilon^{\mu}$ as an infinite series in powers of $x$, and
the general form of the infinite number of generators is
\beq
\eqnlab{coord_gen}
T^{(n-1)\mu_1\ldots\mu_n}_{\phantom{(n-1)\mu_1\ldots\mu_n}\nu} =
x^{\mu_1} \ldots x^{\mu_n}\frac{\pa}{\pa x^\nu}.
\eeq
The index $(n-1)$ indicates the length dimension of the generator,
i.e., the number of $x^{\mu_i}$ minus the number of derivatives
present.

We will show that all the generators in this infinite dimensional
algebra can be obtained by commuting the generators of two finite
dimensional algebras, namely the conformal algebra and the affine
algebra. This observation was first made by the authors of
\cite{Ogievetsky:1973ik} and \cite{Ivanov:1985nu}.

The generators of the \textbf{conformal} group are $P_\mu$,
$M_{\mu\nu}$, $D$ and $K_\mu$. The realisation of these generators
using the coordinates can be found in \Eqnref{conf_gen}, while the
commutation relations of the conformal algebra are given in
\Eqnref{conf_alg}. On the other hand, the generators of the
\textbf{affine} group are $P_\mu$ and $R_{\mu\nu}$
\beq
\eqnlab{aff_gen}
\begin{array}{rcl}
P_{\mu} &=& \pa_{\mu} = T^{(-1)}_{\phantom{(-1)}\mu}\\
R_{\mu\nu} &=& x_\mu \pa_\nu = T^{(0)}_{\phantom{(0)}\mu\nu}.
\end{array}
\eeq
Note that a metric has been implicitly introduced to lower the indices
in \Eqnref{coord_gen}. The Lie brackets of the affine algebra will
then be
\beq
\eqnlab{aff_alg}
\begin{array}{rcl}
\com{P_\mu,P_\nu} &=& 0 \\
\com{P_\rho,R_{\mu\nu}} &=& \eta_{\rho\mu}P_\nu \\
\com{R_{\mu\nu},R_{\rho\sigma}} &=& (\eta_{\nu\rho}R_{\mu\si} -
\eta_{\mu\si}R_{\rho\nu}).
\end{array}
\eeq

Consider now the closure between the conformal and the affine group.
Taking the commutator between $K_\rho$ and $R_{\mu\nu}$ gives
\beq
\begin{array}{rcl}
\eqnlab{clocom}
\com{K_\rho,R_{\mu\nu}} &=& \com{(2x_\rho x^\sigma\pa_\sigma -
  x^2\pa_\rho),x_\mu \pa_\nu} \\
&=& 2x_\mu x_\nu \pa_\rho - 2x_\mu x^\si \eta_{\rho\nu} \pa_\si -
  x^\si x_\si \eta_{\mu\rho} \pa_\nu,
\end{array}
\eeq
which contains generators quadratic in $x^\mu$, in particular the
generator $T^{(1)}_{\phantom{(1)}\mu\nu\rho} = x_\mu x_\nu \pa_\rho$.
From \Eqnref{clocom} we see that since
$T^{(1)}_{\phantom{(1)}\mu\nu\rho}$ is symmetric in the indices $\mu$
and $\nu$, it is the symmetric part of the generator $R_{\mu\nu}$ that
is important. Generally, we can commute $T^{(n)}$ with $K_\mu$ to
obtain generators of one order higher. Thus any generator of the form
given by \Eqnref{coord_gen} can be obtained as a repeated commutator
of generators from the conformal and the affine algebra.

The reason an infinite dimensional algebra is generated when taking
the closure of two finite dimensional algebras, is the fact that there
is an overlap between the conformal group and the affine group. And
more important, there must exist generators outside the overlap in
both groups, and the commutation relations involving these generators
are non-trivial.
\beq
\underbrace{K_{\mu}}_{conformal~algebra} \hspace{0.3 cm} ,
\hspace{0.3 cm} \overbrace{D \hspace{0.3 cm} , \hspace{0.3 cm}
  M_{\mu\nu} \hspace{0.3 cm} , \hspace{0.3 cm} P_{\mu} \hspace{0.3
    cm}}^{overlap} , \hspace{0.3 cm}
\underbrace{S_{\mu\nu}}_{affine~algebra}
\eeq
The generator $S_{\mu\nu}$ denotes the symmetric and traceless part of
$R_{\mu\nu}$, whereas $M_{\mu\nu}$ denotes the antisymmetric part.

The generators in the conformal algebra which are not included in the
affine algebra are the special conformal ones. The geometric
interpretation of the special conformal transformations is not very
obvious, in comparison with the other transformations. It is
essentially an inversion followed by a translation and yet another
inversion
\beq 
x^\mu \rightarrow \frac{x^\mu}{x^2} \rightarrow y^\mu =
\frac{x^\mu}{x^2} + b^\mu \rightarrow \frac{y^\mu}{y^2} =
\frac{x^\mu + b^\mu x^2}{1+2b^\nu x_\nu+b^2x^2}.
\eeq
We can also consider the special conformal generators directly, by
first defining the coordinate inversion operator $I$
\beq
I[\phi(x^\mu)] = \phi\left(\frac{x^\mu}{x^2}\right).
\eeq
The corresponding sequence of transformations for an arbitrary field
will then be
\beq
\begin{array}{rcl}
\phi(x^\nu) &\rightarrow& I[\phi(x^\nu)] \rightarrow
P_{\mu}(I[\phi(x^\nu)]) \\
&\rightarrow& I[P_{\mu}(I[\phi(x^\nu)])] =
I[P_\mu](I[I[\phi(x^\nu)]]) = I[P_\mu]\phi(x^\nu).
\end{array}
\eeq
The operator $I[P_\mu]$ is then found to be
\beq
\eqnlab{transl_inv}
\begin{array}{rcl}
I[P_\mu] &=& I[\frac{\pa}{\pa x^\mu}] = \frac{\pa
  x^\si}{\pa\left(\frac{x^\mu}{x^\nu x_\nu}\right)}\frac{\pa}{\pa x^\si} \\
  &=& (g^\si_{\phantom{\si}\mu} x^\nu x_\nu - 2x^\si
  x_\mu)\frac{\pa}{\pa x^\si} \\
  &=& -K_\mu.
\end{array}
\eeq 
The conclusion is that the conformal transformations consist of
translations, Lorentz transformations, dilations and inversions. The
inversion itself cannot be written as a generator, since the operation
inversion is discrete and thus contradicts the infinitesimal nature of
the generators for a continuous Lie group. Therefore, inversions appear
in a disguised form as special conformal transformations.

The analysis of the closure between the conformal and the affine group
can easily be generalised to include supersymmetry
\cite{Ivanov:1985nu}.

\chapter{Transformations under the conformal group}
\chlab{transformation}

In this appendix we calculate the transformations of the preferred
fields and an arbitrary field of degree $l$ under a coset
transformation, having the conformal group as global symmetry and the
Lorentz group as local symmetry \cite{Salam:1969qk}. The reducing
matrix 
\beq
g = e^{-x^a P_a}e^{\phi^b K_b}e^{\si D}
\eeq
transforms according to
\beq
g(\zeta) \rightarrow g'(\zeta') = \Lambda g(\Lambda^{-1}\zeta)
h^{-1}(\Lambda^{-1}\zeta,\Lambda)
\eeq
under the coset, $\Lambda$ is a global conformal transformation and
$h$ is a local Lorentz transformation. The coordinate vector in (11,2)
dimensions is given by $\zeta_A$. Conformal transformations in (10,1)
dimensions can be written as Lorentz transformations in 13 dimensions,
and we assume that the local 11-dimensional Lorentz subgroup only acts
on the 11-dimensional subspace spanned by the indices $\mu$. We can
thus form a true 13-vector by
\beq
\Phi_A = \frac{1}{2}\left(g_A^{\phantom{A}12} - g_A^{\phantom{A}13}
\right),
\eeq
which transforms as 
\beq
\Phi_A(\zeta) \rightarrow \Phi'_A(\zeta) = \Lambda_A^{\phantom{A}B}
\Phi_B(\Lambda^{-1}\zeta). 
\eeq
Inverting \Eqnref{phi_matrix} gives
\beq
\eqnlab{linear_pref}
\begin{array}{rcl}
\phi_\mu e^{-\si} &=& \Phi_\mu - x_\mu(\Phi_5 + \Phi_6) \\
\phi^2 e^{-\si} &=& (\Phi_5 + \Phi_6).
\end{array}
\eeq
Knowing the transformation laws for $\Phi_A(\zeta)$ under different
global conformal transformations, we can deduce the transformations of
the preferred fields using \Eqnref{linear_pref}
\benu
\item inhomogeneous Lorentz transformations
\beq
\eqnlab{Lorentz_finite}
\begin{array}{rcccl}
x_\mu &\rightarrow& x'_\mu &=& \Lambda_\mu^{\phantom{\mu}\nu}x_\nu +
\alpha_\mu \\
\si(x) &\rightarrow& \si'(x') &=& \si(x) \\
\phi_\mu(x) &\rightarrow& \phi'_\mu(x') &=&
\Lambda_\mu^{\phantom{\mu}\nu}\phi_\nu(x)
\end{array}
\eeq

\item dilations
\beq
\eqnlab{dila_finite}
\begin{array}{rcccl}
x_\mu &\rightarrow& x'_\mu &=& e^{-\la} x_\mu \\
\si(x) &\rightarrow& \si'(x') &=& \si(x) + \la \\
\phi_\mu(x) &\rightarrow& \phi'_\mu(x') &=& e^{\la} \phi_\mu(x)
\end{array}
\eeq

\item special conformal transformations
\beq
\eqnlab{SCT_finite}
\begin{array}{rcccl}
x_\mu &\rightarrow& x'_\mu &=& \frac{x_\mu+b_\mu x^2}{1+2b \cdot
  x+b^2x^2} \\
\si(x) &\rightarrow& \si'(x') &=& \si(x) + \ln|1+2b \cdot x+b^2x^2| \\ 
\phi_\mu(x) &\rightarrow& \phi'_\mu(x') &=& (1+2b \cdot x+b^2x^2)
  \phi_\mu(x) \\
& & & & + [(1+2x\cdot\phi)(1+\beta\cdot x) - 2x^2\beta\cdot\phi]
  \beta_\mu \\
& & & & - [2\beta\cdot\phi + \beta^2(1+2x\cdot\phi)]x_\mu.
\end{array}
\eeq
\eenu

The local Lorentz group elements in 11 dimensions are obtained as
\beq
h(\zeta,\Lambda) = g'^{-1}\Lambda g = e^{-\si' D} e^{-\phi'\cdot K}
e^{x'\cdot P} \Lambda e^{-x\cdot P} e^{\phi\cdot K} e^{\si D},
\eeq
into which we substitute \Eqsref{Lorentz_finite}, \eqnref{dila_finite}
and \eqnref{SCT_finite} to get 
\benu
\item inhomogeneous Lorentz transformations
\beq
\eqnlab{Lorentz_h}
\begin{array}{rcl}
x'_\mu &=& \Lambda_\mu^{\phantom{\mu}\nu}x_\nu + \alpha_\mu \\
h_{\mu\nu}(\zeta,\Lambda) &=& \Lambda_{\mu\nu}
\end{array}
\eeq

\item dilations
\beq
\eqnlab{dila_h}
\begin{array}{rcl}
x'_\mu &=& e^{-\la} x_\mu \\
h_{\mu\nu}(\zeta,\Lambda) &=& g_{\mu\nu}
\end{array}
\eeq

\item special conformal transformations
\beq
\eqnlab{SCT_h}
\begin{array}{rcl}
x'_\mu &=& x_\mu + (x^2\delta^\nu_\mu - 2x_\mu x^\nu)\beta_\nu +
\ldots \\
h_{\mu\nu}(\zeta,\Lambda) &=& g_{\mu\nu} + 2(\beta_\mu x_\nu -
\beta_\nu x_\mu) + \ldots,
\end{array}
\eeq
\eenu
for different types of $\Lambda$. Note that for pure translations,
i.e., $x'_\mu = x_\mu + \alpha_\mu$, the Lorentz group element will be
$h_{\mu\nu} = \eta_{\mu\nu}$.

We can also compute the infinitesimal variations in an arbitrary field
$\psi$, induced by the conformal transformations. Assume $\psi(\zeta)$
is homogeneous of degree $l$, it will then have the transformation
\beq
\begin{array}{rcl}
\psi(\zeta) \rightarrow \psi'(\zeta') &=& D(h)\psi(\Lambda^{-1}\zeta)
\approx (1+D(\delta h))\psi(\zeta) \\
&=& \left(1+\frac{1}{2}(\delta
  h)_\mu^{\phantom{\mu}\nu}S^\mu_{\phantom{\mu}\nu}\right)\psi(\zeta).
\end{array}
\eeq
We define a function depending on the coordinates in 11 dimensions as 
\beq
\psi(x) = \ka^{-l}\psi(\zeta),
\eeq
where $l \equiv \ka\pa/\pa\ka$ is the degree of homogeneity and $\ka =
\zeta_{12} + \zeta_{13}$. The infinitesimal transformation of a such
field is
\beq
\begin{array}{rcl}
\psi(x) \rightarrow \psi'(x') &=& \ka'^{-l}\psi'(\zeta') \approx
\left(\frac{\ka}{\ka'}\right)^l \left(1+\frac{1}{2}(\delta
  h)_\mu^{\phantom{\mu}\nu}S^\mu_{\phantom{\mu}\nu}\right)\psi(x) \\
&\approx& \left(1-l\frac{\delta\ka}{\ka}+\frac{1}{2}(\delta
  h)_\mu^{\phantom{\mu}\nu}S^\mu_{\phantom{\mu}\nu}\right)\psi(x) \\
&\approx& \psi'(x) + \delta x^\mu\pa_\mu\psi(x),
\end{array}
\eeq
resulting in the infinitesimal variation
\beq
\eqnlab{field_variation}
\delta\psi(x) = \psi'(x) - \psi(x) \approx -\left(\delta x^\mu\pa_\mu +
l\frac{\delta\ka}{\ka} - \frac{1}{2}(\delta
h)_\mu^{\phantom{\mu}\nu}S^\mu_{\phantom{\mu}\nu} \right)\psi(x).
\eeq
$S^\mu_{\phantom{\mu}\nu}$ are the field specified representations of
the generators in the local Lorentz group. The quantity
$\delta\ka/\ka$ is given by  
\beq
\eqnlab{kappa}
\frac{\delta\ka}{\ka} = \left\{ \begin{array}{ll}
0 & \textrm{inhomogeneous Lorentz transformations} \\
\la & \textrm{dilations} \\
2\beta\cdot x & \textrm{special conformal transformations} 
\end{array} \right..
\eeq
Inserting \Eqsref{Lorentz_h}, \eqnref{dila_h}, \eqnref{SCT_h} and
\eqnref{kappa} into \Eqnref{field_variation} yields
\benu
\item inhomogeneous Lorentz transformations
\beq
\eqnlab{Lorentz_infin}
\begin{array}{rcl}
\delta x_\mu &=& \epsilon_\mu^{\phantom{\mu}\nu}x_\nu + \alpha_\mu
\\
\delta h_{\mu\nu}(\zeta,\Lambda) &=& \epsilon_{\mu\nu},
\hspace{1 cm} \epsilon_{\mu\nu} = -\epsilon_{\nu\mu} \\
\delta\si &=& -(\epsilon^{\mu\nu}x_\nu + \alpha^\mu)\pa_\mu\si \\
\delta\phi_\mu &=& -(\epsilon^{\nu\ta}x_\ta + \alpha^\nu)\pa_\nu\phi_\mu
+ \epsilon_\mu^{\phantom{\mu}\nu}\phi_\nu \\
\delta\psi &=& -(\epsilon^{\mu\nu}x_\nu + \alpha^\mu) \pa_\mu\psi +
\frac{1}{2}\epsilon_\mu^{\phantom{\mu}\nu} S^\mu_{\phantom{\mu}\nu}\psi
\end{array}
\eeq

\item dilations
\beq
\eqnlab{dila_infin}
\begin{array}{rcl}
\delta x_\mu &=& -\la x_\mu \\
\delta h_{\mu\nu}(\zeta,\Lambda) &=& 0 \\
\delta\si &=& \la(x^\mu\pa_\mu\si+1) \\
\delta\phi_\mu &=& \la(x^\nu\pa_\nu + 1)\phi_\mu \\
\delta\psi &=& \la(x^\nu\pa_\nu - l)\psi
\end{array}
\eeq

\item special conformal transformations
\beq
\eqnlab{SCT_infin}
\begin{array}{rcl}
\delta x_\mu &=& (x^2\delta^\nu_\mu - 2x_\mu x^\nu)\beta_\nu \\
\delta h_{\mu\nu}(\zeta,\Lambda) &=& 2(\beta_\mu x_\nu - \beta_\nu
x_\mu) \\
\delta\si &=& -\beta^\nu(x^2\delta^\mu_\nu - 2x_\nu x^\mu)\pa_\mu\si +
2\beta^\mu x_\mu \\
\delta\phi_\mu &=& -\beta_\ta(x^2 g^{\ta\nu} - 2x^\ta
x^\nu)\pa_\nu\phi_\mu + \beta_\mu  \\
& & + 2\beta^\nu x_\nu\phi_\mu - 2\beta^\nu\phi_\nu x_\mu +
2x^\nu\phi_\nu\beta_\mu \\ 
\delta\psi &=& -\beta_\nu(x^2 g^{\nu\mu} - 2x^\nu x^\mu)\pa_\mu\psi -
2l\beta^\mu x_\mu\psi + 2\beta_\mu x^\nu S^\mu_{\phantom{\mu}\nu}\psi,
\end{array}
\eeq
\eenu
where we also have expanded \Eqsref{Lorentz_finite},
\eqnref{dila_finite} and \eqnref{SCT_finite}. The non-vanishing of the
vacuum expectation values at the origin, $\delta\si(0) = \la$ and
$\delta\phi_\mu(0) = \beta_\mu$, implies the asymmetry of the vacuum.

\chapter{Real principal subalgebras}
\chlab{principal}

A rank $r$ Lie algebra can always be represented in the form
\beq
\eqnlab{triangulation}
\g = N_{-} \oplus H \oplus N_{+},
\eeq
where $H$ is the Cartan subalgebra, $N_{-}$ and $N_{+}$ are the
triangular subalgebras consisting of the multiple commutators of the
generators $F_i$ and $E_i$, respectively. We will now find the
conditions for a Lie algebra to contain a rank three real principal
subalgebra \cite{Gaberdiel:2002db,Nicolai:2001ir}.

\subsubsection{Principal so(3) subalgebras}

Every finite dimensional semi-simple Lie algebra $\g$ has a principal
so(3) subalgebra, due to the positive definiteness of the inverse
Cartan matrix. The generators of the principal so(3) subalgebra are
constructed using the Weyl vector $\rho = \sum_{i=1}^r
\frac{2}{(\alpha_i,\alpha_i)} \la_i$, with $\alpha_i$ and $\la_i$
being the simple roots and the fundamental weights of $\g$,
respectively. The Cartan generator of the principal so(3) is defined
by
\beq
J_3 = \displaystyle\sum_{i=1}^r \rho^{(i)}H^{Cartan}_{\alpha_i}
\hspace{0.5 cm} \Rightarrow \hspace{0.5 cm} \com{J_3,E_\alpha} =
(\rho,\alpha) E_\alpha,
\eeq
where $H^{Cartan}_{\alpha_i}$ and $E_\alpha$ are the Cartan generators
and the generator associated with the root $\alpha$, respectively,
given in the Cartan-Weyl basis. Also, the combinations of the simple
root generators
\beq
J^{+} = \displaystyle\sum_{i=1}^r k_i E_{\alpha_i} \hspace{0.5 cm} ,
\hspace{0.5 cm} J^{-} = \displaystyle\sum_{i=1}^r k_i E_{-\alpha_i}
\eeq
satisfy
\beq
\com{J_3,J^{\pm}} = \pm J^{\pm} \hspace{0.5 cm} , \hspace{0.5 cm}
\com{J^{+},J^{-}} = J_3
\eeq
with $k_i^2 = p^{(i)}$. The Lie algebra can then be decomposed with
respect to the principal so(3) subalgebra into $r$ irreducible
representations of spin $s_j$
\beq
\g = \displaystyle\bigoplus_{j=1}^r \g^{(s_j)},
\eeq
where $\g^{(s_j)}$ carries the $(2s_j+1)$-dimensional irreducible
representation of so(3), and the $r$ spins $s_j$ are called the
exponents of $\g$. Generally $\g^{(0)}$ is empty, whereas $\g^{(1)}$
is the adjoint representation of the principal so(3) subalgebra. The
exponents $s_j$ contain important informations about the Lie algebra
$\g$. For instance, the orders of the invariant Casimir operators are
given by the numbers $s_j+1$, and the quadratic Casimir invariant is
thus always associated to the representation $s_1=1$. Using the
Chevalley-Serre basis $E_i=E_{\alpha_i}$, $F_i=E_{-\alpha_i}$ and
$H_i=H^{Cartan}_{\alpha_i}$ we can rewrite the generators $J_3$ and
$J^{\pm}$ according to
\beq
J_3 = \displaystyle\sum_{i=1}^r p_i H_i \hspace{0.5 cm},\hspace{0.5
  cm} J^{+} = \displaystyle\sum_{i=1}^r k_i E_i \hspace{0.5 cm} ,
\hspace{0.5 cm} J^{-} = \displaystyle\sum_{i=1}^r k_i F_i,
\eeq
with
\beq
p_i \equiv \rho^{(i)} = \displaystyle\sum_{i=1}^r A^{-1}_{ij} > 0
\hspace{0.5 cm} \textrm{and} \hspace{0.5 cm} k_i \equiv \sqrt{p_i}.
\eeq
The relations $p_i>0$ are due to that $A^{-1}_{ij} \ge 0$ for a finite
dimensional Lie algebra.

\subsubsection{Principal so(2,1) subalgebras}

The concept of principal subalgebras can be generalised to a rank $r$
Kac-Moody algebra $\tilde{\g}$. We can again define a generator $J_3 =
\sum_{i=1}^r \rho^{(i)}H^{Cartan}_{\alpha_i}$, where
$H^{Cartan}_{\alpha_i}$ are the Cartan generators in the Cartan-Weyl
basis, and $\rho = \sum_{i=1}^r \frac{2}{(\alpha_i,\alpha_i)} \la_i$
is the Weyl vector. Expressing the fundamental weights as
\beq
\la_i = \displaystyle\sum_{j=1}^r (\alpha_i,\alpha_i)A^{-1}_{ij}
\frac{\alpha_j}{(\alpha_j,\alpha_j)},
\eeq
allows us to rewrite the Weyl vector as
\beq
\rho = \displaystyle\sum_{i,j=1}^r A^{-1}_{ij}
\frac{2\alpha_j}{(\alpha_j,\alpha_j)}. 
\eeq
The generator $J_3$ then becomes
\beq
J_3 = \displaystyle\sum_{i,j=1}^r A^{-1}_{ij} H_j,
\eeq
where $H_j$ are the Cartan generators of $\tilde{\g}$ in the
Chevalley-Serre basis. The generator $J_3$, together with
\beq
J^{+} = \displaystyle\sum_{i=1}^r k_i E_i \hspace{0.5 cm},\hspace{0.5
  cm} J^{-} = \displaystyle\sum_{i=1}^r l_i F_i,
\eeq
constitute a so(2,1) algebra. The hermiticity property $(J^{+})^{\da}
= J^{-}$ will be satisfied only if $k_i = l_i^{*}$, since $(E_i)^{\da}
= F_i$. Furthermore, demanding $\com{J^{+},J^{-}} = \sum_{i=1}^r k_i
l_i H_i = -J_3$ requires
\beq
0 \le |k_j|^2 = k_j l_j = -\displaystyle\sum_{i=1}^r A^{-1}_{ij}
\equiv p_j. 
\eeq
The remaining Lie bracket of so(2,1), $\com{J_3,J^{\pm}} = \pm
J^{\pm}$ are then satisfied automatically. Thus, a real principal
so(2,1) subalgebra exists if and only if
\beq
\eqnlab{so21_cond}
\displaystyle\sum_{i=1}^r A^{-1}_{ij} \le 0, \hspace{0.5 cm}
\forall~j. 
\eeq
For instance, all the hyperbolic Kac-Moody algebras have a principal
so(2,1) subalgebra, since all their fundamental weights lie in the
forward light-cone, and thus $A^{-1}_{ij} =
\frac{2}{(\alpha_i,\alpha_i)} (\la_i,\la_j) \le 0$ for all the indices
$i$ and $j$ \cite{Ruuska:1991ne}.

The algebra $\tilde{\g}$ can again be decomposed into irreducible
representations of the principal subalgebra. Due to the
non-compactness of the group SO(2,1), all these irreducible
representations will be infinite dimensional and unitary, except the
adjoint representation of so(2,1), which is neither. The unitarity
means that one can define a scalar product $(x,y)$ on the
representation space, being hermitian and positive definite with
the properties
\beq
\eqnlab{repr_scalar}
\begin{array}{rcl}
(\com{E_i,x},y) &=& (x,\com{F_i,y}) \\
(\com{H_i,x},y) &=& (x,\com{H_i,y})
\end{array}, \hspace{0.5 cm} \forall~x,y \in{\tilde{\g}}.
\eeq
The representation space is then the vector space of the algebra
$\tilde{\g}$, with $\tilde{\g}$ acting on itself by the adjoint
action. Assuming that the algebra possesses an invariant bilinear form
$\langle \cdot | \cdot \rangle$, i.e., a generalised Killing form, we
can define the scalar product on the representation space as
\beq
\eqnlab{scalar_Killing}
(x,y) \equiv -\langle x|\theta(y) \rangle,
\eeq
where $\theta$ is the Cartan involution
\beq
\theta(E_i) = -F_i \hspace{0.5 cm},\hspace{0.5 cm} \theta(F_i) = -E_i
\hspace{0.5 cm},\hspace{0.5 cm} \theta(H_i) = -H_i.
\eeq

The generators $J_3$ and $J^{\pm}$ of the principal so(2,1) can be
verified to satisfy \Eqnref{repr_scalar}, with $J_3$ being
self-adjoint and $J^{\pm}$ mutually adjoint, respectively. Taking the
norms
\beq
(J_3,J_3) = \rho^2 = -\displaystyle\sum_{i=1}^r p_i < 0 \hspace{0.5
  cm} \textrm{and} \hspace{0.5 cm} (J^{+},J^{+}) = (J^{-},J^{-}) =
\displaystyle\sum_{i=1}^r p_i > 0
\eeq
shows that the adjoint representation of so(2,1) is indeed not
unitary. Since the representation space is decomposed into orthogonal
subspaces consisting of the Cartan subalgebra and subspaces associated
with each root, all the other irreducible representations in the
decomposition of $\tilde{\g}$ will be unitary. This conclusion is made
by observing that the subspace spanned by all the vectors orthogonal
to the Weyl vector $\rho$ is positive definite with respect to the
scalar product \eqnref{scalar_Killing}.

Because of the adjoint action the spectrum of $J_3$ will be integral,
i.e., $e^{i2\pi J_3}=1$ and the representations arised are called
single valued or unitary. Using the irreducible adjoint representation
of so(2,1) we define the quadratic Casimir operator
\beq
\begin{array}{rcl}
\mathcal{Q} &=& J_3J_3 - J^{+}J^{-} - J^{-}J^{+} \\
&=& J_3(J_3+1) - 2J^{-}J^{+} = J_3(J_3-1) - 2J^{+}J^{-},
\end{array}
\eeq
which acts on a generator adjointly according to
\beq
\textrm{ad}_{\mathcal{Q}}(x) = \com{J_3,\com{J_3,x}} -
\com{J^{+},\com{J^{-},x}} - \com{J^{-},\com{J^{+},x}}.
\eeq
The eigenvalues of this Casimir operator are then used to label the
irreducible representations of so(2,1) occurring in the decomposition
of $\tilde{\g}$. Besides the non-unitary finite dimensional
representations such as the three dimensional one, so(2,1) possesses
two different kinds of unitary infinite dimensional representations.
\benu
\item \textbf{The discrete series representations} \\
The Casimir eigenvalues for the discrete series representations are
\beq
\mathcal{Q} = s(s-1) > 0.
\eeq
These representations contain a lowest (highest) weight state
satisfying $J^{-}|s,s\rangle=0$ (or $J^{+}|-s,-s\rangle=0$), and the
states of a representation are denoted by $|s,m\rangle$ (or
$|-s,-m\rangle$) for $m=s,s+1,\ldots$, where $s=2,3,4,\ldots$. These
representations are entirely contained in the triangular subalgebras
$N_{+}$ or $N_{-}$ defined in \Eqnref{triangulation}. The lowest
weight representations are built on states of the form
\beq
v^{(s)} = \displaystyle\sum_{j_1,\ldots,j_s} c_{j_1\ldots j_s}
\com{E_{j_1},\ldots,\com{E_{j_{s-1}},E_{j_s}}\ldots} \hspace{0.5
  cm},\hspace{0.5 cm} \com{J^{-},v^{(s)}} = 0
\eeq
by repeated application of $J^{+}$. Similarly, the highest weight
states are obtained by acting on
\beq
v^{(-s)} = \displaystyle\sum_{j_1,\ldots,j_s} c_{j_1\ldots j_s}
\com{F_{j_1},\ldots,\com{F_{j_{s-1}},F_{j_s}}\ldots} \hspace{0.5
  cm},\hspace{0.5 cm} \com{J^{+},v^{(-s)}} = 0
\eeq
with $J^{-}$.

\item \textbf{The continuous representations} \\
The continuous representations split into principal and supplementary
series representations with the Casimir eigenvalues 
\beq
\begin{array}{rc}
\textrm{principal series :} & -\infty < \mathcal{Q} < -\frac{1}{4} \\
\textrm{supplementary series :} & -\frac{1}{4} < \mathcal{Q} < 0, \\
\end{array}
\eeq
respectively. The supplementary series can in practical never appear
as unitary representations of so(2,1), since $e^{i2\pi J_3}$ can never
equal unity for these. Thus, we consider only the principal series in
detail. 

Apart from $J_3$, there are $(r-1)$ linearly independent combinations
of the Cartan generators of $\tilde{\g}$, written generally as
$\sum_{i=1}^r c_i H_i$. These are determined by diagonalising the
so(2,1) Casimir operator. Acting with the Casimir operator adjointly
we obtain
\beq
\eqnlab{adQ_lincomb}
\textrm{ad}_{\mathcal{Q}}\left(\displaystyle\sum_i^r c_i H_i\right) =
-2 \displaystyle\sum_{i,j=1}^r c_iA_{ij}p_{j}H_{j}.
\eeq 
Note that using $\sum_{j=1}^rA_{ij}p_j = -\sum_{j,k=1}^r
A_{ij}A^{-1}_{jk} = -1$ and putting $c_i=-p_i$, \Eqnref{adQ_lincomb}
becomes 
\beq
\textrm{ad}_{\mathcal{Q}}(J_3) = 2\displaystyle\sum_{i,j=1}^r
p_iA_{ij}p_{j}H_{j} = 2J_3,
\eeq
which is precisely the result expected for the adjoint representation
of so(2,1). The coefficients of the remaining $(r-1)$ linearly
independent combinations will be orthogonal to $J_3$ and thus satisfy
\beq
\left( \displaystyle\sum_{i=1}^r c_iH_i,-\displaystyle\sum_{j=1}^r
  p_jH_j \right) = -\displaystyle\sum_{i,j=1}^r c_ip_jA_{ij} = 0
\hspace{0.5 cm} \Rightarrow \hspace{0.5 cm} \displaystyle\sum_{i=1}^r
c_i = 0. 
\eeq

The full representations are then generated by multiply commuting
$\sum_{i=1}^r c_iH_i$ with $J^{+}$ and $J^{-}$.  Since the eigenvalues
of $J_3$ are bounded neither from above nor from below, the orthogonal
complement of $J_3$ must belong to $(r-1)$ principal series
representations.  These representations extend simultaneously into
both $N_{+}$ and $N_{-}$. It can also be verified that all the
elements in the principal series have positive norm, i.e., $(x,x)>0$.
\eenu

\chapter{The lattices $\Pi_{D-1;1}$}
\chlab{lattice}

The lattices $\Pi_{D-1;1}$ are the unique even self-dual Lorentzian
lattices, existing only in the dimensions $D=8n+2$,
$n=0,1,2,\ldots$. These lattices are spanned by the vectors
$x=(x^1,\ldots,x^{D-1};x^0) \in{\mathbb R^{D-1,1}}$ satisfying
\beq
x\cdot r \equiv \displaystyle\sum_{i=1}^{D-1} x^ir^i - x^0r^0
\in{\mathbb Z},
\eeq
where $r=(\frac{1}{2},\ldots,\frac{1}{2};\frac{1}{2}) \in{\mathbb
  R^{D-1,1}}$, and the scalar product is defined using the flat
Minkowski metric. In addition, either
\beq
\textrm{all }x^\mu \in{\mathbb Z} \hspace{0.5 cm} \textrm{or}
\hspace{0.5 cm} \textrm{all }(x^\mu-r^\mu) \in{\mathbb Z} 
\eeq
must be obeyed \cite{Conway:1998}.

Concentrating on the lattice $\Pi_{1;1}$, we find it consisting of the
vectors 
\beq
(m;2p+m) \hspace{0.5 cm} \textrm{and} \hspace{0.5 cm}
\left(n+\frac{1}{2};2q+n+\frac{1}{2}\right) \hspace{0.5 cm}
\forall~m,n,p,q \in{\mathbb Z}.
\eeq
It is straightforward to show that $\Pi_{1;1}$ is even
\beq
\left\{ \begin{array}{rcl}
(m;2p+m)^2 &=& -4p(p+m) \\
\left(n+\frac{1}{2};2q+n+\frac{1}{2}\right)^2 &=& -2q(2q+2n+1) 
\end{array}
\in{2\mathbb Z} \right..
\eeq
The self-duality can be shown by explicitly finding the dual
lattice. For convenience we change the notation to $z=(z^{+},z^{-})$,
where
\beq
\left\{ \begin{array}{rcl}
z^{+} &=& x^0 + x^1 \\
z^{-} &=& \frac{1}{2}(x^0 - x^1)
\end{array} \right.
\hspace{0.5 cm} \Rightarrow \hspace{0.5 cm}
\left\{ \begin{array}{rcl}
x^0 &=& \frac{1}{2}(z^{+} + 2z^{-}) \\
x^1 &=& \frac{1}{2}(z^{+} - 2z^{-})
\end{array} \right..
\eeq
The scalar product then becomes
\beq
\begin{array}{rcl}
x\cdot y &=& \frac{1}{4}\left\{ (z^{+} - 2z^{-})(w^{+} - 2w^{-}) -
  (z^{+} + 2z^{-})(w^{+} + 2w^{-}) \right\} \\
&=& -z^{+}w^{-} - z^{-}w^{+}.
\end{array}
\eeq
In the basis given by $(z^{+},z^{-})$ the points in $\Pi_{1;1}$ can be
written simply as
\beq
(n,m) \hspace{0.5 cm} \forall~n,m \in{\mathbb Z},
\eeq
e.g., the vector $r$ will be $r = (1,0)$. Thus, the lattice $\Pi_{1;1}$
is the integer span of the basis vectors 
\beq
k \equiv (1,0) \hspace{0.5 cm} \textrm{and} \hspace{0.5 cm} \bar{k}
\equiv (0,-1),
\eeq
which have be chosen to satisfy $k^2 = \bar{k}^2 = 0$ and
$k\cdot\bar{k}=1$. Also, $\pm(k+\bar{k})$ are the only two points in
the whole lattice with length squared equal two.

%
%
\cleardoublepage
\pagestyle{plain}
\def\href#1#2{#2}
\bibliographystyle{utphysmod2}
\addcontentsline{toc}{chapter}{\sffamily\bfseries Bibliography}
\bibliography{biblio}

%
%

\end{document}